\documentclass[aps,showpacs,floatfix,preprintnumbers,nofootinbib,11pt,prd]{revtex4-2}

\usepackage{graphicx} 
\usepackage{dcolumn} 
\usepackage{bm} 
\usepackage{amsfonts}
\usepackage{amsthm}
\usepackage{amsmath}
\usepackage{amssymb}
\usepackage{epsfig}
\usepackage{color}
\usepackage{textcomp}
\usepackage{hyperref}
\usepackage{titlesec}
\usepackage{slashed}
\usepackage{caption}
\usepackage{subcaption}
\usepackage{natbib}

\newcommand{\kmsMpc}{\, \text{km}\,\text{s}^{-1}\, \text{Mpc}^{-1}}

\def\ee{\end{equation}}
\def\ba{\begin{eqnarray}}
\def\ea{\end{eqnarray}}

\def\bdm{\begin{displaymath}}
\def\edm{\end{displaymath}}

\def\bq{\begin{quote}}
\def\eq{\end{quote}}

 at 10truept

\usepackage{amsmath}
\numberwithin{equation}{section}
\def\ee{\end{equation}}
\def\ba{\begin{eqnarray}}
\def\ea{\end{eqnarray}}

\def\bq{\begin{quote}}
\def\eq{\end{quote}}

 at 10truept

\newcommand{\beq}{\begin{equation}}
\newcommand{\eeq}{\end{equation}}
\newcommand{\beqa}{\begin{eqnarray}}
\newcommand{\eeqa}{\end{eqnarray}}
\newcommand{\bea}{\begin{eqnarray}}
\newcommand{\eea}{\end{eqnarray}}

\newcommand{\al}{\alpha}
 \newcommand{\be}{\beta}

\def\lesssim{~\mbox{\raisebox{-.6ex}{$\stackrel{<}{\sim}$}}~}

\def\ltap{\ \raise.3ex\hbox{$<$\kern-.75em\lower1ex\hbox{$\sim$}}\ }
\def\gtap{\ \raise.3ex\hbox{$>$\kern-.75em\lower1ex\hbox{$\sim$}}\ }
\def\gl{\ \raise.5ex\hbox{$>$}\kern-.8em\lower.5ex\hbox{$<$}\ }
\def\roughly#1{\raise.3ex\hbox{$#1$\kern-.75em\lower1ex\hbox{$\sim$}}}

\newcommand{\bi}{\begin{itemize}}
\newcommand{\ei}{\end{itemize}}

\def\ltap{\ \raise.3ex\hbox{$<$\kern-.75em\lower1ex\hbox{$\sim$}}\ }
\def\gtap{\ \raise.3ex\hbox{$>$\kern-.75em\lower1ex\hbox{$\sim$}}\ }
\def\gl{\ \raise.5ex\hbox{$>$}\kern-.8em\lower.5ex\hbox{$<$}\ }
\def\roughly#1{\raise.3ex\hbox{$#1$\kern-.75em\lower1ex\hbox{$\sim$}}}

\begin{document}

\title{Hot New Early Dark Energy}

\author{Florian Niedermann}
\email{florian.niedermann@su.se}
\affiliation{Nordita, KTH Royal Institute of Technology and Stockholm University\\
Hannes Alfv\'ens v\"ag 12, SE-106 91 Stockholm, Sweden}
\author{Martin S.~Sloth} 
\email{sloth@cp3.sdu.dk}
\affiliation{CP$^3$-Origins, Center for Cosmology and Particle Physics Phenomenology \\ University of Southern Denmark, Campusvej 55, 5230 Odense M, Denmark}

\begin{abstract} 
New early dark energy (NEDE) makes the cosmic microwave background consistent with a higher value of the Hubble constant inferred from supernovae observations. 
It is an improvement over the old early dark energy model (EDE) because it explains naturally the decay of the extra energy component in terms of a vacuum first-order phase transition that is triggered by a subdominant scalar field at zero temperature. With hot NEDE, we introduce a new mechanism to trigger the phase transition. It relies on thermal corrections that subside as a subdominant radiation fluid in a dark gauge sector cools. We explore the phenomenology of hot NEDE and identify the strong supercooled regime as the scenario favored by phenomenology. 
In a second step, we propose different microscopic embeddings of hot NEDE. This includes the (non-)Abelian dark matter model, which has the potential to also resolve the LSS tension through interactions with the dark radiation fluid. We also address the coincidence problem generically present in EDE models by relating NEDE to the mass generation of neutrinos via the inverse seesaw mechanism. We finally propose a more complete dark sector model, which embeds the NEDE field in a larger symmetry group and discuss the possibility that the hot NEDE field is central for spontaneously breaking lepton number symmetry.
\end{abstract}

\maketitle

\newpage

\tableofcontents

\section{Introduction}

The Hubble tension is by now a well-known discrepancy between the value of the Hubble constant today, $H_0$, measured directly using supernovae (SNe) observations and the lower value indirectly inferred from measurements of the cosmic microwave background (CMB) \cite{Aghanim:2018eyx} and baryonic acoustic oscillations (BAO) \cite{Alam:2016hwk} when assuming the $\Lambda$CDM standard model of cosmology (for recent reviews see~\cite{Freedman:2017yms,Verde:2019ivm,DiValentino:2020zio}). It is currently reported to be a $4.1 \sigma$ tension by the SH0ES team~\cite{Riess:2020fzl}, which uses Cepheids to calibrate the second rung of the distance ladder needed to reach out into the Hubble flow. At present, the significance of the tension is contested by the Chicago-Carnegie Hubble program, which, instead of Cepheids, uses stars at the tip of the red giant branch to calibrate the distance ladder and finds a value which is still compatible with the CMB-inferred value~\cite{Freedman:2020dne,Freedman:2021ahq}, although alternative studies using a similar approach find a higher  value of $H_0$, more in line with the SH0ES result~\cite{Yuan:2019npk,Soltis:2020gpl,Anand:2021sum}. Ultimately, the hope is that this debate will be settled by independent local measurements that are not reliant on the difficult calibration of SNe and for example use gravitational waves as standard sirens~\cite{Abbott:2019yzh} or strongly-lensed quasars~\cite{Wong:2019kwg,Birrer:2020tax}. In this work, rather than contributing to the ongoing  (and sometimes heated) controversy about the robustness of the tension (for references see~\cite{DiValentino:2021izs}), we will use it as a guide for finding new physics and rewriting our cosmological standard model (while of course keeping an open mind about future developments in the astrophysics community). In that context, it is by now also rather well-known among cosmologists that a late-time modification of the $\Lambda$CDM model after recombination does not offer much hope of completely resolving the tension \cite{Verde:2016ccp,Bernal:2016gxb,Knox:2019rjx,Aylor_2019,Arendse:2019itb,Efstathiou:2021ocp}. On the other hand, it has been proposed that an extra component of dark energy, which decays just before recombination, shows more promise towards resolving the tension. The first attempt in this direction, called early dark energy (EDE) \cite{Poulin:2018dzj, Poulin:2018cxd,Poulin:2018cxd,Smith:2019ihp,Smith:2020rxx,Murgia:2020ryi,Poulin:2021bjr}, proposed that the decay of the extra EDE component happened through a second order rollover phase transition (see~\cite{Lin:2019qug,Kaloper:2019lpl,Alexander:2019rsc,Hardy:2019apu,Sakstein:2019fmf,Berghaus:2019cls,1798362,Lin:2020jcb,CarrilloGonzalez:2020oac,Freese:2021rjq,Allali:2021azp,Sabla:2021nfy,Karwal:2021vpk,Vagnozzi:2021gjh,Gomez-Valent:2021cbe,Moss:2021obd,Clark:2021hlo} for different studies building up on this idea and~\cite{Escudero:2019gvw,Kreisch:2019yzn,Park:2019ibn,Pandey:2019plg,Sekiguchi:2020teg,Jedamzik:2020krr,Escudero:2021rfi,Bansal:2021dfh,Aloni:2021eaq} for other early-time approaches).

This model, which, for the reminder of this paper, we will refer to as old EDE\footnote{An attempt to be witty by referring to the opposite situation with old and new inflation in the eighties.}, is however not completely without problems. It turns out that data prefers a very fast decay of the EDE component, which is very hard to achieve for a scalar field undergoing a roll-over phase transition without invoking a fine-tuning. Moreover, the phenomenology is sensitive to the perturbation sector which requires a flattening of the potential at high field values, which seems to rule out simple monomial potentials~\cite{Agrawal:2019lmo}.  Finally, the old EDE model cannot resolve the large-scale structure (LSS) tension, although it does not make it significantly worse either (as argued in~\cite{Smith:2020rxx,Murgia:2020ryi} as an answer to the bleaker picture drawn in~\cite{Ivanov:2020ril,Hill:2020osr}). This problem, also referred to as the $\sigma_8$ tension, is with a significance of $2 $ to $3 \sigma$ less severe but still gaining attention in the observational community.  It tells us that the  $\Lambda$CDM model predicts too much power in the linear matter power spectrum on small scales~\cite{DiValentino:2020vvd}.
 
For these reasons, we have proposed the New Early Dark Energy (NEDE) model~\cite{Niedermann:2019olb,Niedermann:2020dwg,Niedermann:2020qbw}, which differs from the old EDE model in crucial ways. In the NEDE model, the early dark energy component is carried by a scalar field $\psi$ that decays in a triggered first-order phase transition trough vacuum tunneling. This realizes an almost instantaneous transition and  naturally leads to a fluid that, after the transition, is at least as stiff as radiation. This is because bubbles of true vacuum when they collide will create anisotropic stress in the form of a scalar field condensate, which on small scales acts as a source of gravitational waves (and potentially other types of radiation) and on large scales gives rise to a homogenous and isotropic fluid. It is characterized by a time-dependent equation of state parameter $w_\mathrm{NEDE}(t)$ that controls the decay of NEDE~\cite{Niedermann:2020dwg} and asymptotes to $1/3$ (corresponding to a complete dissipation of the condensate).\footnote{Ultimately, $w_\mathrm{NEDE}(t)$ will be related to the microscopic parameters of the tunneling field.} 
Approximating the value of $w_\mathrm{NEDE}(t)$ right after the phase transition as a constant with value $2/3$, a combination of CMB, BAO and SNe data (including a late-time prior on $H_0)$ shows that NEDE is favored at the 4$\sigma$  level over $\Lambda$CDM reducing the tension down to $2.5 \sigma$~\cite{Niedermann:2020dwg,Niedermann:2020qbw}. Moreover, a recent comparison with other early-time modifications ranked NEDE high among its often more phenomenological competitors~\cite{Schoneberg:2021qvd} (treating $w_\mathrm{NEDE}$ as a constant phenomenological parameter). With regard to the LSS tension, it was recently showed that with this rather simple implementation of NEDE, the tension is not significantly worsened in NEDE \cite{Niedermann:2020qbw} (but also not relieved). Beyond that, in a few recent studies~\cite{Hill:2021yec,Poulin:2021bjr,Moss:2021obd}, additional evidence for EDE-type models (including NEDE~\cite{Poulin:2021bjr}) was reported when using ACT (rather than Planck) data, although a joint analysis of Planck and ACT data raises concerns about internal inconsistencies within both datasets.  Since the phase transition occurs in a vacuum at zero temperature, we will refer to this first version of NEDE as \textit{cold} NEDE.

In cold NEDE, the first-order vacuum phase transition is triggered by a second, subdominant field once it becomes heavy compared to the Hubble constant and starts evolving. This additional trigger field must be very light compared to the NEDE tunneling field whose mass is set by the eV energy scale of the transition or equivalently the temperature around recombination. While this may be natural in the landscape or for UV completions of NEDE in the axiverse and/or monodromy models \cite{Niedermann:2020dwg}, another possibility, which we will consider here, is that the mass scales are tied to known standard model (SM) physics. To be precise, we will introduce the \textit{hot} NEDE model where the first-order phase transition happens at finite, dark-sector temperature $T_d \sim \mathrm{eV}$.\footnote{Recently, the authors in \cite{Allali:2021azp} proposed a confinement phase transition as a possible microscopic realization of new early dark energy. It is similarly triggered by temperature corrections. They then used the cold NEDE Boltzmann code {\tt TriggerCLASS}, reliant on an ultralight trigger field, as an approximation to test their model against data. Instead, we will consider a weakly coupled, thermal phase transition and work out the phenomenological difference between hot and cold NEDE. } The potential of the NEDE field $\Psi$, which we assume to be charged under a dark gauge group, will then acquire temperature corrections, which restore the symmetry of the vacuum state at high temperatures. In this situation, the temperature can replace the ultralight field as the trigger of a rapid phase transition. This is very similar to the SM Higgs field, although important differences remain: Unlike the Higgs field, we are in a regime where the transition is first-order and can be described perturbatively. Moreover, our transition happens at much lower energies during the CMB epoch which introduces a new set of phenomenological constraints (and signatures). In any event, having a thermal trigger disposes the need of a mass scale much below the eV scale, while the (sub-)eV mass of the NEDE field itself can be related to the neutrino sector. Our belief is that by doing this, we can make the occurrence of the NEDE energy scale as much (or little) a coincidence as the scale of neutrino masses in the SM.\footnote{The idea to tie the physics of early dark energy to neutrino physics has been discussed in a different context in \cite{Sakstein:2019fmf,CarrilloGonzalez:2020oac}, where a conformal coupling to the neutrino sector (similar to \cite{DAmico:2018hgc}) kicks off the early dark energy epoch when neutrinos become non-relativistic. In \cite{Escudero:2019gvw,Escudero:2021rfi,Fernandez-Martinez:2021ypo,DiBari:2021dri}, on the other hand, the neutrino mass generation has been discussed in the context of the Hubble tension yet without relating it to an EDE/NEDE phase transition. } In fact, we will outline a version of hot NEDE where the phase transition provides the mechanism for generating the mass of the SM neutrinos along with a spontaneous breaking of lepton number symmetry.

In Sec.~\ref{sec:hotNEDE}, as the main part of this work, we will study the phenomenology of hot NEDE for a class of simple  $\mathrm{SU(N)}$ and $\mathrm{U(1)}$ gauge theories with coupling parameter $g_\mathrm{NEDE}$ {under which $\Psi$ is charged}. To be precise, we will express different phenomenological parameters such as the fraction of NEDE at decay time $f_\mathrm{NEDE}$, the  temperature of the dark sector $T_d$,  the inverse duration $\bar{\beta}$ and strength $\alpha$ of the phase transition, and the wall-thickness parameter $\delta_\mathrm{eff}$ in terms of $g_\mathrm{NEDE}$ as well as the field's vacuum mass scale $\mu$ and quartic coupling parameter $\lambda$.   This in turn singles out  $\gamma = \lambda / (4 \pi g_\mathrm{NEDE}^4) \lesssim 1$ as the most promising  parameter region for hot NEDE, which includes the strong supercooled regime where the energy difference between the false and true vacuum dominates over the dark radiation (DR) fluid (in agreement with the discussion in~\cite{Allali:2021azp} as well as \cite{Barreiro:1996dx} in an inflationary context). In particular, this regime allows for a dark sector significantly colder than the visible sector, avoiding cosmological bounds on the number of relativistic degrees of freedom, which otherwise tend to constrain dark sector model building efforts. As part of this section, we also propose two decay scenarios for the hot NEDE fluid component. One where the colliding bubble wall condensate shows a stiff behavior, which makes it dilute quickly, and another one where  it dissipates almost instantaneously through the decay of $\Psi$ quanta into a lighter particle species that subsequently turns non-relativistic and contributes to DM. This second \textit{mixed DM} scenario offers an entirely new possibility for early dark energy model building.

While the phenomenology discussion starts from the finite-temperature effective potential, we work out possible microphysical scenarios in Sec.~\ref{sec:microphysics}.  This includes a model for heating up the dark sector, which makes use of the (non-)Abelian dark matter model [(N)ADM]~\cite{Buen-Abad:2015ova,Lesgourgues:2015wza,Buen-Abad:2017gxg}. Here, the idea is that a $\mathrm{TeV}$ mass fermionic field which is simultaneously charged under a dark and SM gauge group is responsible for heating up the dark sector and providing DM. The resulting dark sector temperature is then used to inform and constrain the phenomenological discussion of hot NEDE. Moreover, the model introduces an interaction between the DR fluid and dark matter (DM), which has been shown in a different context to help with the LSS tension. We will therefore argue that incorporating the (N)ADM mechanism can be naturally incorporated in hot NEDE, providing a simple way of heating up the dark sector and addressing both the Hubble and LSS tension simultaneously.

Having established the viability of the hot NEDE phenomenology, we next turn to neutrino physics. We argue that the NEDE phase transition is responsible for generating the Majorana mass scale $m_s \gtrsim \mathrm{eV}$ of a set of sterile neutrinos that through the inverse seesaw mechanism~\cite{Abada:2014vea} give mass to the active neutrinos. We will derive an explicit relation between the NEDE parameters and $m_s$, showing that this mechanism can be naturally embedded in a supercooled phase transition. {We also discuss the possibility that a fourth neutrino mass eigenstate $\nu_4$, which has been claimed to resolve short-baseline oscillation anomalies~\cite{Kopp:2013vaa,Boser:2019rta,Dasgupta:2021ies}, is present in the low-energy mass spectrum. Our NEDE scalar, which gives mass to $\nu_4$ (and the active neutrino eigenstates), is then also responsible for the ``secret interaction'' needed to make the presence of $\nu_4$ compatible with cosmology~\cite{Hannestad:2013ana,Dasgupta:2013zpn,Archidiacono:2014nda,Archidiacono:2015oma,Archidiacono:2016kkh}. }

Motivated by these positive findings, we finally raise the bar and outline a more inclusive neutrino model dubbed the dark electroweak model (DEW). Here, the sterile  neutrinos carry lepton number and are embedded along with the NEDE field in a dark gauge group.
To be precise, we consider an additional phase transition in the dark sector operative above the $\mathrm{TeV}$ scale. It breaks a  $\mathrm{SU(2)}_\mathrm{D} \times \mathrm{U(1)_\mathrm{Y_D}} $ down to a dark electromagnetism $\mathrm{U(1)}_\mathrm{DEM}$  and generates the mass mixing between a set of right-handed and the sterile neutrinos required by the inverse seesaw mechanism. The NEDE scalar field can then be identified as the neutral component $\Psi_0$ of a field $\Psi = (\Psi_1,\Psi_2,\Psi_3)^T$ that transforms in the triplet representation of  $\mathrm{SU(2)}_\mathrm{D} $. As before, its potential receives thermal corrections through its coupling to a subdominant DR plasma {consisting of charged $\Psi$ quanta}. Finally, when the NEDE phase transition takes place at the $\mathrm{eV}$ temperature scale, the sterile neutrinos acquire their (super-)$\mathrm{eV}$ Majorana mass, which, in turn, completes the inverse seesaw mass matrix while breaking lepton number spontaneously.  Moreover, this model naturally incorporates different fermionic and bosonic decay channels for the bubble wall condensate, which can realize the proposed mixed DM scenario needed for a successful NEDE phenomenology. In summary, we link the hot NEDE phase transition to the neutrino mass generation and spontaneous lepton number breaking through the inverse seesaw mechanism. 
Adopting a more general point of view, the DEW model exemplifies how, in the future, cosmological tensions will guide our search for a more complete description of the dark sector. This point will be made more prominently in our companion paper~\cite{Letter}.

\section{Cold NEDE}\label{sec:coldNEDE}
 
The first realization of NEDE relied on a phase transition that occurred in a dark sector at zero temperature. We will therefore refer to the previous model as \textit{cold} NEDE to distinguish it from its \textit{hot} counterpart. Here, we briefly review cold NEDE (see \cite{Niedermann:2020dwg} for more details) by discussing its field theoretic model and cosmological signatures. This then serves as a blueprint for the discussion of hot NEDE.

\subsection{Effective field theory model}
Cold NEDE uses an ultralight scalar field $\phi$ to trigger the phase transition in the {(real-valued)} NEDE field $\psi$. Its action reads\footnote{An high-energy version of this potential has been considered in \cite{Linde:1990gz,Adams:1990ds,Copeland:1994vg} to end inflation through a first-order phase transition.}
\begin{align}\label{potential_cold_NEDE}
V(\psi, \phi) = \frac{\lambda}{4} \, \psi^4  + \frac{1}{2}\beta M^2 \psi^2 - \frac{1}{3} \alpha M \psi^3 
+ \frac{1}{2} m^2 \phi^2 +\frac{1}{2} \tilde{\lambda} \, \phi^2 \psi^2 \,,
\end{align}
where $\lambda$, $\beta$, $\alpha$, and $\tilde \lambda$ are dimensionless constants, and the potential is controlled by the two mass scales $m \ll M \sim \mathrm{eV}$. For understanding the mechanism, it is convenient to rewrite $V(\psi, \phi)$ in terms of dimensionless variables
\begin{subequations}
\begin{align}\label{eq:pot2d}
\bar{V}(\bar{\psi},\bar{\phi}) = \frac{1}{4} \bar{\psi}^4 - \bar{\psi}^3  + \frac{\delta_\text{eff}(\bar{\phi})}{2} \, \bar{\psi}^2 + \frac{1}{2} \kappa^2 \bar{\phi}^2\,,
\end{align}
where we introduced
\begin{align}\label{dimless_vars}
\bar{V} = \frac{81 \lambda^3}{\alpha^4 \, M^4} \, V \,, && \bar{\psi} = \frac{3 \lambda}{\alpha M} \psi \,, && \bar{\phi} = \frac{3 \sqrt{\lambda \tilde{\lambda}}}{\alpha M} \phi 
\end{align}
and
\begin{align}\label{eq:kappa}
\kappa = \frac{3 \lambda}{\alpha\, \sqrt{ \tilde \lambda}} \frac{m}{M} \;.
\end{align}
\end{subequations}
With these choices, the $\psi$-dependent part of the potential is  characterized by a single parameter $\delta_\mathrm{eff}(\bar \phi)$, which is a function of the trigger field $\bar{\phi}$
\begin{align}\label{eq:delta_phi}
\delta_\text{eff}(\bar{\phi}) = 9 \, \frac{\lambda \beta }{\alpha^2} + \bar{\phi}^2 \,.
\end{align}
The $\psi$-dependent part of the potential for different values of $\delta_\mathrm{eff}$ is depicted in Fig.~\ref{fig:Potential}. A global minimum with $\bar{\psi} \neq 0$ exists if $\delta_0 \equiv 9 \lambda \beta / \alpha^2 < 2$ and lies at
\begin{align}\label{eq:psi_true}
(\bar \phi,\bar \psi)_\mathrm{true} = \left(0, \frac{1}{2} \left[ 3 + \sqrt{9-4 \delta_0 }\right] \right)\,.
\end{align}

As initial conditions, we choose  $(\bar{\psi}, \bar{\phi})=(0, \bar \phi_\mathrm{ini}=\textrm{const})$, where $\phi_\mathrm{ini} \ll M_\mathrm{pl} $ ensures that $\phi$ gives a sub-dominant contribution to the energy budget.  Then, irrespective of the value of $\phi$, there is always a non-vanishing probability $\Gamma$ for the field to tunnel to the true minimum, provided $\delta_0 < 2$. In \cite{Niedermann:2020dwg}, it was shown that,  for $\tilde \lambda / \lambda \ll 1$, the corresponding two-field tunneling, described in terms of the coupled system of $\bar \phi$ and $\bar \psi$, can be well approximated as a one-field process, controlled solely by the effective potential for $\bar{\psi}$. To be explicit, the tunneling rate is 
\begin{align}
\Gamma(t) \sim M^4 \exp\left[ - S_4(t) \right]\,.
\end{align}
where $S_4$ is the Euclidian action evaluated at the $O(4)$-symmetric solution of the Euclidian equations of motion
\begin{align}\label{eq:eom_psi1}
\bar{\psi}''(\bar r) +  \frac{3}{\bar r}\, \bar{\psi}'(\bar r) = \bar{\psi}^3 - 3 \, \bar{\psi}^2 + \delta_\mathrm{eff}(\bar{\phi}) \, \bar{\psi } \,,
\end{align}
 subject to the boundary conditions $ \frac{\mathrm{d}\bar \psi}{\mathrm{d}\bar r}|_{\bar r=0} = 0 $ and $\bar \psi|_{\bar r \to \infty}=0$. Here, $\bar r$ is a four-dimensional radial coordinate in Euclidian space, and $\bar\phi(\bar r) \simeq \mathrm{const}$ (valid for $\tilde \lambda / \lambda \ll 1$). Two numerical solutions of \eqref{eq:eom_psi1} for different values of $\delta_\mathrm{eff}$ are provided as the blue curves in Fig.~\ref{fig:profile}. A semi-analytic formula then reads~\cite{Adams:1993zs}
\begin{align}\label{SE4}
S_4 \simeq \frac{4 \, \pi^2}{3 \lambda} \left( 2 - \delta_\mathrm{eff}\right)^{-3} \left(\alpha_1 \delta_\mathrm{eff} + \alpha_2 \delta_\mathrm{eff}^2 + \alpha_3 \delta_\mathrm{eff} ^3  \right)\,,
\end{align}  
where $\alpha_1 = 13.832$, $\alpha_2 = -10.819$, and $\alpha_3 = 2.0765$
are numerically determined coefficients, and $\delta_\mathrm{eff}(t)$ is a function of time through its dependence on $\bar \phi(t)$. The product $\lambda S_4$ is plotted as the blue curve in Fig.~\ref{fig:S_El}  alongside a set of numerically determined values depicted as blue dots.  For our phenomenological discussion, we introduce the percolation parameter~\cite{Anderson:1991zb}
\begin{align}\label{eq:p}
p(t) = \int^t_0 \mathrm{d}t' \frac{\Gamma(t')}{H(t')^3} \;,
\end{align}
which measures the number of nucleation events per causal volume $\sim 1/H^3$ at time $t$. Tunneling is strong enough to compete with the Hubble expansion when $p(t)$ reaches order unity, which defines the time of the phase transition $t_*$ through $p(t_*) \simeq 1$.  We can further evaluate it by approximating $\Gamma(t)$ as a linear exponential $\Gamma(t) \sim M^4 \exp\left[-S(t_*)+\bar\be (t-t_*)\right]$, where
\begin{align}\label{beta}
\bar{\beta} \equiv - \frac{d S_4}{dt}\Big|_{t=t_*} \simeq \frac{\dot \Gamma (t_*)}{\Gamma(t_*)}  \,
\end{align}
has the interpretation of the inverse duration of the phase transition (because it sets the time it takes to increase $p$ beyond order unity and therefore complete the transition). This is justified because the integral in \eqref{eq:p} only picks up its value in a small time window around $t_*$. In the limit of a fast phase transition, $\bar \beta^{-1} H(t_*) \equiv \bar \beta^{-1} H_*  \ll1$, and we obtain
\begin{align}\label{percolation_parameter}
p(t_*) \sim \frac{M^4}{\bar \beta H_*^3} \mathrm{e}^{-S_4(t_*)} \sim1\, ,
\end{align}
which requires\footnote{We will see that the main contribution to $S_4(t_*)$ arises from the first term, whereas the second term can be neglected for order of magnitude statements (as done, for example, in \cite{Niedermann:2020dwg}).} 
\begin{align}\label{S_4_star}
S_4(t_*) \simeq  \ln(M^4/H_*^4)+\ln(\bar{\beta}^{-1}H_*) \,.
\end{align}

We define NEDE as the vacuum energy released trough the tunneling,
\begin{align}\label{def:rho_NEDE}
{\rho}_\text{NEDE}^* \equiv \rho_\mathrm{NEDE} (t_*) = V(0, \phi_*)-V(\psi_\mathrm{true},0)\,.
\end{align}
Using \eqref{dimless_vars} and \eqref{eq:psi_true}, it  evaluates to
\begin{align}\label{eq:def_EDE}
\rho_\mathrm{NEDE}^* = \frac{c(\delta_0)}{12}  \frac{\alpha^4 \, M^4}{\lambda^3}  + \frac{1}{2} m^2 \phi_*^2\,,
\end{align}
where 
\begin{align}\label{eq:c_delta}
{c(\delta_0) = \frac{1}{216} \left( 3 + \sqrt{9 -4 \delta_0}\right)^2 \left( 3 - 2 \delta_0 + \sqrt{9 -4 \delta_0}\right)}\,,
\end{align}
which is of order unity for the relevant regime where $\delta_0 < 1.5$. It is straightforward to show that the second term in \eqref{eq:def_EDE} is subdominant because $\phi_* < \phi_\mathrm{ini} \ll M_\mathrm{pl}$. We also see that if we want NEDE to be sizable around matter-radiation equality, we need $M \sim \mathrm{eV}$ unless $\alpha^4/\lambda^3$ introduces a hierarchy.

\subsection{Phenomenology} \label{Phenomenology_coldNEDE}

We now discuss in more detail how the evolution of $\bar{\phi}$ controls the different phases of the NEDE phase transition.

\begin{enumerate}
\item[(a)]
\textbf{Initial regime [$\delta_\mathrm{eff}(\bar{\phi}) \geq 2$]:} At early times, $m \ll H$, and the trigger field is frozen $\bar{\phi} \simeq \bar{\phi}_\mathrm{ini}$. The tunneling is highly suppressed (at least for our preferred choice $\tilde \lambda / \lambda \ll 1$) and hence $p \ll 1$. As a consequence, the NEDE field resides in its false vacuum at $\bar{\psi} =0 $, providing the early dark energy needed to resolve the Hubble tension. The Hubble drag is released when $m \sim H$, and $\bar \phi$ starts to roll down the potential, scanning different values of $\delta_\mathrm{eff}(\bar{\phi})$ and hence various shapes of $\bar{V}$.

\item[(b)]
\textbf{Fostering regime [$2 > \delta_\mathrm{eff}(\bar{\phi})  \geq  \delta_\mathrm{eff}(\bar{\phi}_*)$]:} 
Here, the tunneling probability builds up until $\bar \phi$ drops to the critical value $\bar \phi_* \equiv \bar{\phi}(t_*)$ where $p \simeq 1$. For phenomenological reasons, we want this to happen close to matter-radiation equality, which fixes $m \sim H_\mathrm{eq} \sim 10^{-27}\mathrm{eV}$. From \eqref{S_4_star}, we then obtain that for $M\sim \mathrm{eV}$ and $H_* \sim H_\mathrm{eq}$, the Euclidian action needs to fall below $S_4(t_*)\simeq 250$. The corresponding value  $\delta_\mathrm{eff}^* \equiv \delta_\mathrm{eff}(\bar{\phi}_*)$ depends on $\lambda$ but, as a consequence of \eqref{SE4}, is bounded from above by $\simeq 1.5$ (when assuming a weak quartic coupling $\lambda < 1$). Eq.~\eqref{eq:delta_phi} then implies that we need $\delta_0 < \delta_\mathrm{eff}^* (< 1.5)$ to ensure that the critical value of $\delta_\mathrm{eff}$ is actually scanned as $\bar{\phi} \to 0$.

\item[(c)]
\textbf{Tunneling regime [$\delta_\mathrm{eff}(\bar{\phi}_*) > \delta_\mathrm{eff}(\bar{\phi})  \geq  \delta_0 >0]$:} 
As $\bar \phi \to 0$, tunneling becomes more and more efficient, filling the space with bubbles of true vacuum as $p$ increases beyond unity. The corresponding time scale was found to be $\bar{\beta}^{-1} \propto \lambda^{3/2} H^{-1}_*$; in particular, we have $ H_* \bar{\beta}^{-1} \lesssim 10^{-3}$ for $\lambda \lesssim 0.01 $. This is important as it prevents bubbles from growing to cosmological sizes, which would lead to potentially large anisotropies observable in the CMB.\footnote{A different way to avoid these anisotropies has been discussed recently in \cite{Freese:2021rjq} in terms of a sequence of phase transitions, each corresponding to a relatively small energy release.} To be more precise, structures introduced by the bubbles subtend an angle~\cite{Niedermann:2020dwg}
\begin{align}
\theta_\mathrm{NEDE} \simeq 0.4^\circ \times H_* \bar{\beta}^{-1} \left( \frac{5000}{z_*} \right) \left( \frac{h}{0.7}\right) \left( 1 - f_\mathrm{NEDE}\right)^{1/2}\,.
\end{align}
While the absence of NEDE structures in the CMB demands  $\theta_\mathrm{NEDE} < 0.1^\circ $, the authors in~\cite{Freese:2021rjq} argued for an even stronger bound arising from LSS data probed through the Lyman-$\alpha$ forest, which yields $\theta_\mathrm{NEDE} < 0.002^\circ $ and can be satisfied for $H_* \bar{\beta}^{-1} < 0.005$. In any event, this bound is easily satisfied for $\lambda \lesssim 0.01$, and, therefore, bubbles collide on scales $\ll 1/H_*$, forming a homogeneous and isotropic fluid on cosmological scales. We expect this fluid to be dominated by the kinetic energy of the bubble wall condensate and to be described by an equation of state $1/3 < w_\mathrm{NEDE}(t) < 1$.\footnote{If we deviate strongly from the thin-wall limit, e.g.~$\delta^*_\mathrm{eff}=1$, we also expect an admixture of a pressureless component due to coherent field oscillations around the true vacuum (see Sec.~\ref{sec:decay}).} The tunneling is complete before $\bar{\phi} \to 0$ because the trigger evolution takes place on a cosmological time scale $1/H_* (\gg \bar{\beta}^{-1}) $. \\
\end{enumerate}

In summary, cold NEDE provides a scenario where an ultralight field $\phi$ initiates a phase transition of an $\mathrm{eV}$ scalar $\psi$. As argued in \cite{Niedermann:2020dwg}, provided the coupling $\tilde \lambda <  10^3 m^2/(\beta M^2) \ll 1$, $\psi$-induced loop corrections to the trigger mass scale $m$ are under control, avoiding a fine-tuning. In contrast with old EDE, this model does not rely on oscillations in a tuned, anharmonic potential to explain the quick decay of early dark energy after the phase transition.  Instead, the quick decay is  a manifestation of the relativistic bubble wall condensate on large scales.

This type of triggered first-order phase transition is a promising candidate for resolving the Hubble tension. There are two mechanisms at play, one at background and one at perturbation level. 

The background mechanism is similar to the one used by other early-time approaches to the Hubble tension and has been discussed extensively in the literature~\cite{Poulin:2018cxd}: The false vacuum energy of the scalar field leads to an energy injection in the cosmic fluid around the decay time $t_*$. This raises $H(t)$ for $t < t_* < t_\mathrm{rec}$ relative to $\Lambda$CDM and, in turn, lowers the comoving sound horizon $r_s[H]$, which is a linear functional of $1/H(t)$. As the CMB measurement fixes the angular scale of the sound horizon $\theta_\mathrm{rec} = r_s/D_\mathrm{rec}$ at the sub-permille level, a reduced value of $r_s$ needs to be compensated by also decreasing the comoving distance to the surface of last scattering  $D_\mathrm{rec}$. Since $D_\mathrm{rec} \propto 1/H_0$, this is achieved by raising $H_0$, which then reconciles the early and late measurements of $H_0$. Raising $H(t)$ prior to recombination also leads to an excess diffusion damping on small scales, which can be countered through raising the spectral tilt and amplitude. 

On perturbation level, the important observation is that the decaying NEDE fluid supports its own acoustic oscillations. They source the gravitational potential and, due to their positive pressure, lead to a quicker decay of the Weyl potential (this point was made in the context of acoustic early dark energy \cite{Lin:2019qug}, but equally applies to other early dark energy models~\cite{Niedermann:2020dwg}). This is then countered by increasing the (dimensionless) DM energy density $\omega_\mathrm{cdm} $ (and slightly increasing $\omega_b $ too). For this to work, it is important that the acoustic NEDE oscillations are not too large as, otherwise, it becomes more difficult to mitigate their effects on the gravitational potential.  As they are seeded by adiabatic perturbations of the trigger field $\delta \phi(t_*, \mathbf{x})$ at decay time, this constrains the trigger sector. There is an intuitive reason for that sensitivity:  fluctuations $\delta \phi(t_*, \mathbf{x})$ lead to spatial variations of the decay time because the threshold value $\phi_*$ at two different locations is reached at slightly different times. Correspondingly, the NEDE fluid $\rho_\mathrm{NEDE}$ starts its decay at different times, which then induces perturbations $\delta \rho_\mathrm{NEDE}(t_*, \mathbf{x})$. For sub-horizon scales they depend on the preceding dynamics of the trigger field. This, for example, disfavors the parameter range where $\phi$ has entered too deep into the oscillatory regime at decay time, which would enhance the perturbations too strongly.  This results in a lower phenomenological bound on $H(t_*)/m$, which is in remarkable agreement with the theoretical bounds. Beyond the perturbation dynamics, another unique feature of NEDE is the transition's abruptness. Both background and perturbation effects lead to distinct signatures in the CMB power spectra that distinguish it from other early dark energy models.

Using CMB, BAO, and SNe data, it was shown in \cite{Niedermann:2020dwg} that the tension is reduced to $2.5\, \sigma$ (down from $>4 \sigma$) corresponding to $H_0 = 69.6^{+1.0}_{-1.3} \, \kmsMpc$ $(68 \%$  C.L.).  Further including the local measurement from SH$0$ES, a $>4 \sigma$ evidence for a non-vanishing fraction of NEDE was reported alongside $H_0 = 71.4 \pm 1.0 \, \kmsMpc$ $(68 \%$  C.L.). In both cases, the transition takes place at redshift $z_* \simeq 5000$ and $w_\mathrm{NEDE}=2/3$ was fixed. Similarly encouraging results have been obtained recently in a broader comparison of different early-time modifications of $\Lambda$CDM, singling out EDE and NEDE as particularly promising directions~\cite{Schoneberg:2021qvd}.  A recent study, more focused on the difference between EDE and NEDE, finds that both models react differently to ACT data, although important questions about the internal consistency of ACT and Planck data within early dark energy models remain~\cite{Poulin:2021bjr}. Despite this phenomenological success, there have been concerns about the viability of early dark energy proposals in the light of recent LSS data~\cite{Hill:2020osr,Ivanov:2020ril,DAmico:2019fhj}. They were first addressed in the case of NEDE in \cite{Niedermann:2020qbw}. To be specific, it was shown that NEDE in its simplest form predicts matter to be too clumpy on short scales -- quantified by the parameter $S_8$. However, this so-called $S_8$ tension is already present within the $\Lambda$CDM model at the $2$ to $3 \sigma$ level~\cite{DiValentino:2020vvd}, and it was found that NEDE is not making it worse. {In fact, even when including LSS data, NEDE maintains its strong statistical evidence of $\simeq 4 \sigma$}. This statement relies on adding the effective field theory of LSS~\cite{DAmico:2019fhj} applied to BOSS data~\cite{Alam:2016hwk,6dF} to further constrain the model. These results were later confirmed {also for the old early dark energy, single-field, second-order} model~\cite{Murgia:2020ryi,Smith:2020rxx}. Of course, a rather pessimistic reading of these findings would be that NEDE failed at resolving the $S_8$ tension. Instead, in this paper, we argue that this conclusion is premature and describe how we think hot NEDE has the potential to address both tensions simultaneously.\footnote{Also cold NEDE offers a possibility to address both tensions. Here, the idea is to go away from the thin-wall limit where coherent field oscillations around the true vacuum introduce an interacting DM component, which is known to relieve the $S_8$ tension if its interacting partner is relativistic~\cite{Archidiacono:2019wdp}. }

\section{Hot NEDE}\label{sec:hotNEDE}

\subsection{Effective field theory model} \label{sec:hotNEDE_EFT}
We will consider the spontaneous breaking of a gauge group with coupling $g_\mathrm{NEDE}$ in a hidden sector with temperature $T_d$, induced by a complex scalar field $\Psi$ acquiring a vacuum expectation value (vev), $\Psi \to v_\Psi/\sqrt{2}$, below a critical temperature~$T_c$. This can be described in terms of a temperature-dependent effective potential, which is typically derived in the limit where the temperature is larger than the gauge boson and scalar mass after the breaking. We will start by exploring in detail this \textit{small-mass/high-temperature} regime before moving on to the \textit{large-mass/low-temperature} scenario. In both cases, we derive the phenomenological parameters needed to discuss the phenomenology of hot NEDE in the next section.

\subsubsection{Small-mass/high-temperature potential}

\begin{figure}[t]
    \centering
    \includegraphics[width=0.65\textwidth]{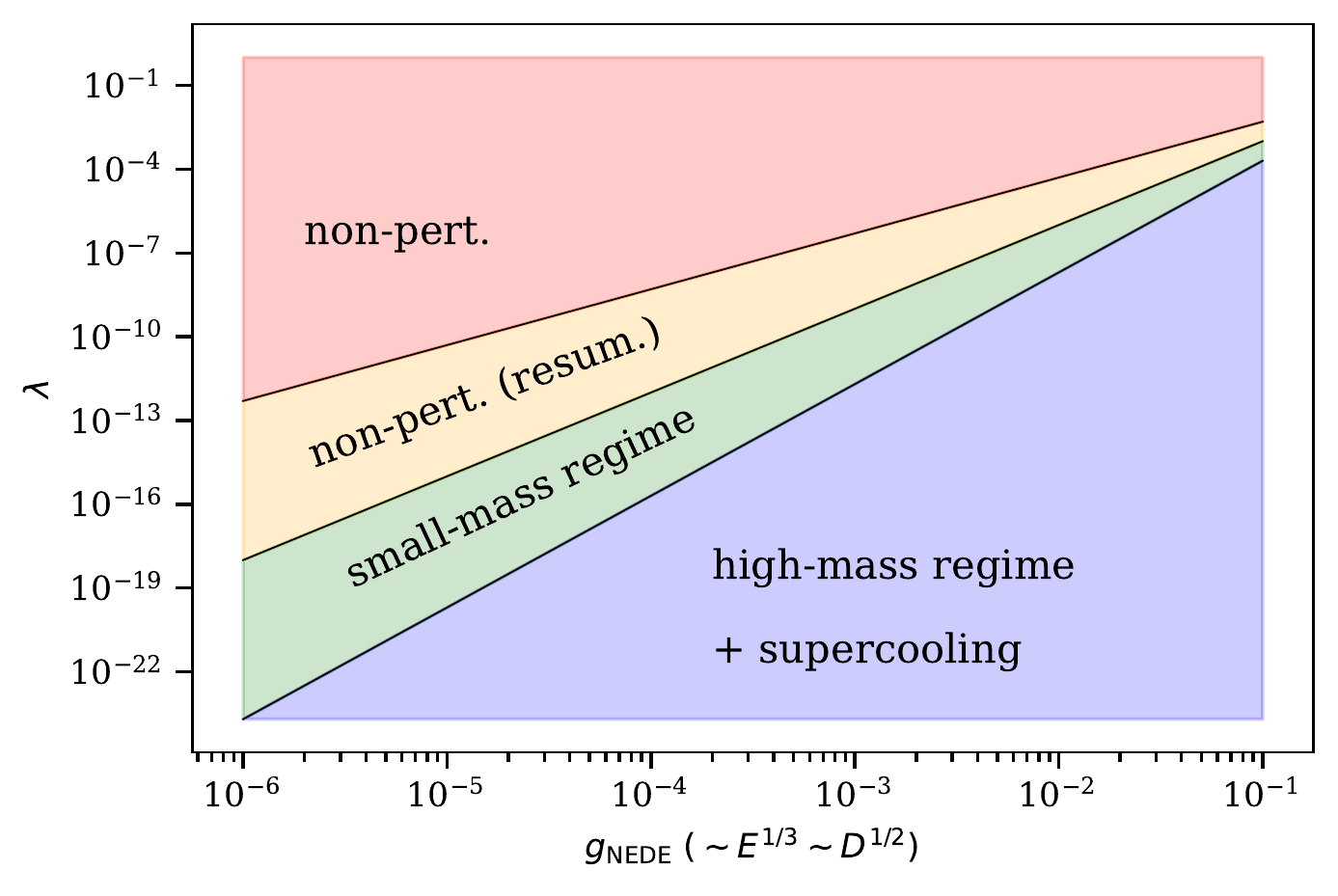}
    \caption{Regime of validity of the finite temperature potential for a theory with a single gauge coupling $g_\mathrm{NEDE}$. The small-mass expansion leading to \protect\eqref{eq:effective_T_pot} is applicable in the green region (where $\gamma = 4 \pi \lambda / g_\mathrm{NEDE} \gg 1$). The daisy-resummed formula in \protect\eqref{eq:effective_T_pot_gen} is needed to access the orange region. Above that (red), perturbation theory breaks down, making non-perturbative techniques necessary. The small-mass expansion breaks down in the blue region (where $\gamma \lesssim 1$), requiring the use of \protect\eqref{eq:thermal_corr_gen}. This region accommodates strong supercooling and provides the best conditions for hot NEDE.}
    \label{fig:coupling_regions}
\end{figure}

At this stage, we remain mostly agnostic about the microscopic details, although in  Sec.~\ref{sec:microscopic_model} we will introduce a specific candidate model. In any case, this situation is very similar to the well studied electro-weak phase transition, 
and -- within the perturbative regime -- the typical form of the finite temperature effective potential for $\psi \equiv \sqrt{2} |\Psi| $ is~\cite{Linde:1981zj,Dine:1992wr,Arnold:1992rz}
\beq\label{eq:effective_T_pot}
V(\psi;T_d) = D(T_d^2-T_\circ^2)\psi^2 -E T_d \psi^3 +\frac{\lambda}{4}\psi^4 + V_0(T_d)
\eeq
where $D > 0$ and $E>0$ are dimensionless constants, and $\lambda(T_d) > 0$ in general has a weak logarithmic temperature dependence. $V_0$ comprises the field-independent thermal corrections and is therefore not relevant for describing the dynamics of the phase transition (yet contains information about the subsequent evolution of the true vacuum).  Besides the dark temperature $T_d$, the potential is controlled by $T_\circ$, which sets the energy scale of the radiation plasma at the time of the phase transition and can, alongside $D$ and $E$ be related to the fundamental parameters of the microscopic theory. Specifically, we show in the Appendix~\ref{appendix_potential} that in the case of a simple U(1) gauge theory (we use the Abelian Higgs model as our working example), we have $E \simeq g_\mathrm{NEDE}^3/(4 \pi)$, $D \simeq g_\mathrm{NEDE}^2/8$ (although the scaling with the gauge coupling parameter is expected to hold in more complicated gauge theories; see for example~\cite{Dine:1992wr,Arnold:1992rz}). We further show that $T_\circ = 2 \mu / g_\mathrm{NEDE} $, where $\mu$ sets the mass scale of the zero-temperature potential; in particular, $\mu \ll T_\circ \lesssim \mathrm{eV}$ for small couplings $g_\mathrm{NEDE}$. We also demonstrate that for the effective description in \eqref{eq:effective_T_pot} to be valid 
\begin{align}\label{lambda_E_cond}
E^4 \ll \lambda^3 \lesssim E^3 && \text{and} && D^6 \ll \lambda^3 \lesssim D^{9/2}  \,
\end{align}
have to hold, which corresponds to the green shaded region in Fig.~\ref{fig:coupling_regions} and slightly tightens the result in~\cite{Arnold:1992rz} (after identifying $E\sim D^{3/2} \sim g_\mathrm{NEDE}^3 $). In particular, the lower bounds implement the small-mass (or high-temperature) condition. In order to distinguish different regimes, we introduce the parameter
\begin{align}\label{def:gamma}
\gamma=\frac{\lambda}{\left( 4 \pi E^4 \right)^{1/3}}\,,
\end{align}
which in the specific case of the Abelian Higgs model becomes $\gamma = 4 \pi \lambda/g_\mathrm{NEDE}^4$.
Then, the form \eqref{eq:effective_T_pot} is valid only if $\gamma \gg 1$, whereas the regime $\gamma \lesssim 1$ (blue shaded region) requires a generalized form of the potential, which we will introduce in Sec.~\ref{sec:large_mass}. 

\begin{figure}[t]
    \centering
    \includegraphics[width=0.65\textwidth]{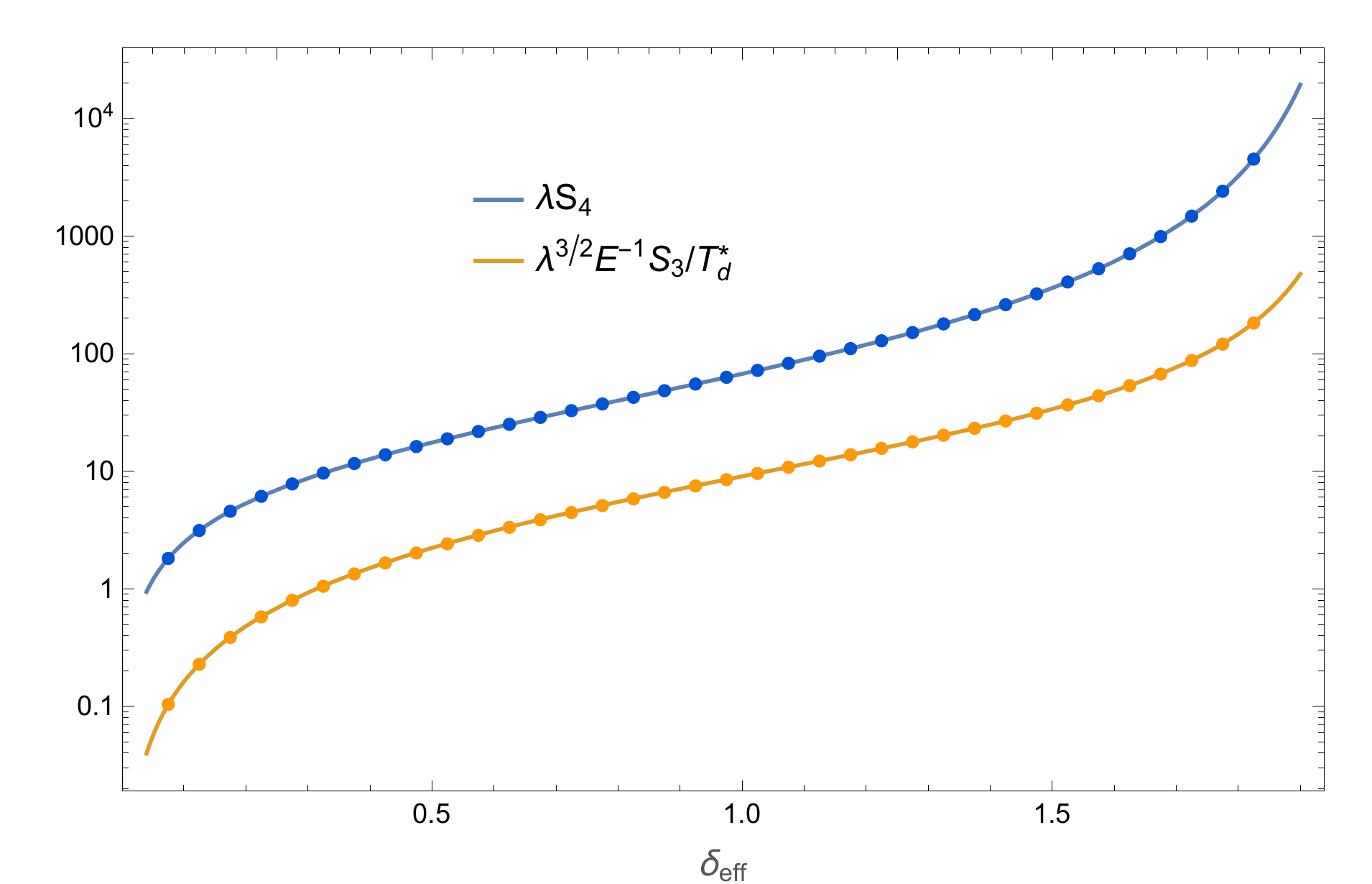}
    \caption{Euclidian action in the $O(4)$ (blue) and $O(3)$ (orange) symmetric case. The solid solid lines correspond to the semi-analytic expression in \protect\eqref{SE4} and \protect\eqref{SE3}. Each dot corresponds to a numerical value obtained by using a shooting method. The Euclidian action diverges in the thin-wall limit $\delta_\mathrm{eff} \to 2$ and vanishes for $\delta_\mathrm{eff} \to 0$. }
    \label{fig:S_El}
\end{figure}

\begin{figure}
     \centering
     \begin{subfigure}[b]{0.99\textwidth}
         \centering
         \includegraphics[width=\textwidth]{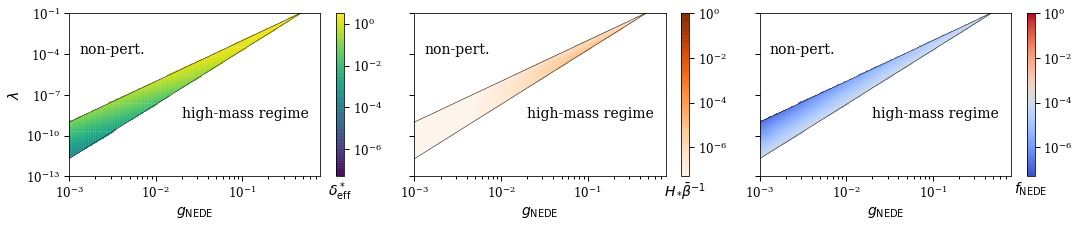}
         \caption{Small-mass/high-temperature regime ($\gamma \gg 1$) }
         \label{fig:params}
     \end{subfigure}
\\
     \begin{subfigure}[b]{0.99\textwidth}
         \centering
         \includegraphics[width=\textwidth]{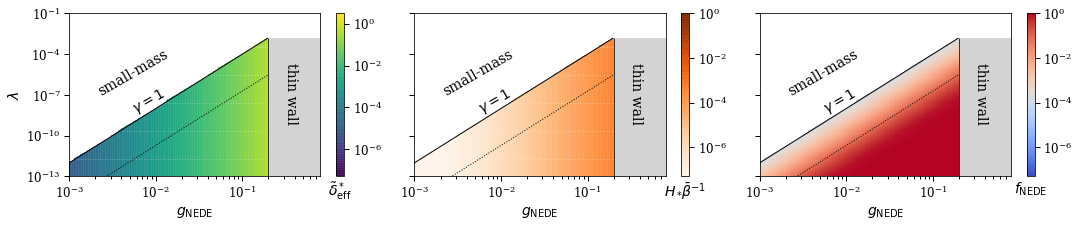}
          \caption{Large-mass/low-temperature regime ($\gamma \lesssim 1$) }
         \label{fig:params_large_mass}
     \end{subfigure}
          \caption{Phenomenological parameters as functions of $\lambda$ and $g_\mathrm{NEDE}$ in the case of the Abelian Higgs model [$E$ and $D$ as in~\protect\eqref{D_E_abelian}]. The colored bands in (a) and (b) correspond to the green and blue band in Fig.~\protect\ref{fig:coupling_regions}. To derive the values of $\delta_\mathrm{eff}^*$ (upper left) and $\tilde \delta_\mathrm{eff}^*$ (lower left), we used the percolation condition $S_{3}/T_d^*\simeq 250$ (from $p\sim 1$). The middle plots detail the parameter dependence of the duration $H_*\bar{\beta}^{-1}$ as in \protect\eqref{beta} and \protect\eqref{beta_large_mass}. The right plots show the achievable NEDE fraction $f_\mathrm{NEDE}$ (for $\xi_*=0.34$) inferred from \protect\eqref{fNEDE}. We can easily achieve a sizable fraction of NEDE with the dotted line corresponding to $f_\mathrm{NEDE} = 10\%$ provided we are in the low-temperature regime. A discussion of the thin-wall limit for $\gamma \lesssim 1$ (light dark region) is left to future work.} 
        \label{fig:params_both}
\end{figure}

Having established the regime of validity of \eqref{eq:effective_T_pot}, we can now study the corresponding tunneling probability.
Comparing with the cold NEDE potential \eqref{potential_cold_NEDE}, we identify
\beq\label{eq:dict}
M=T_\circ\,,\, \quad \al(T_d) = 3 E \frac{T_d}{T_\circ}\,, \quad \text{and} \quad \be = -2 D \,.
\eeq
We also find that the trigger-field coupling in \eqref{potential_cold_NEDE} is replaced with the temperature dependence via  $\tilde \lambda \phi^2 \to 2 D T_d^2$. So instead of $\phi$, here, the dark temperature triggers the phase transition. 
With these identifications, the dimensionless, temperature-dependent potential $\bar{V}$, defined in \eqref{dimless_vars}, becomes (dropping the field-independent term)
\beq\label{dimless_potential}
\bar V(\bar \psi; T_d) = \frac{1}{4} \bar \psi^4-\bar\psi^3+ \frac{\delta_{\textrm{eff}}(T_d)}{2}\bar\psi^2\,,
\eeq
where $\bar\psi\equiv 3\lambda \psi/[\alpha(T_d) M] = \lambda \psi / (E T_d)  $ and  
\beq\label{delta_eff}
\delta_{\textrm{eff}}(T_d)= 2\frac{\lambda D}{E^2}\left(1-\frac{T_\circ^2}{T_d^2}\right)~.
\eeq
\begin{figure}[t]
    \centering
    \includegraphics[width=0.65\textwidth]{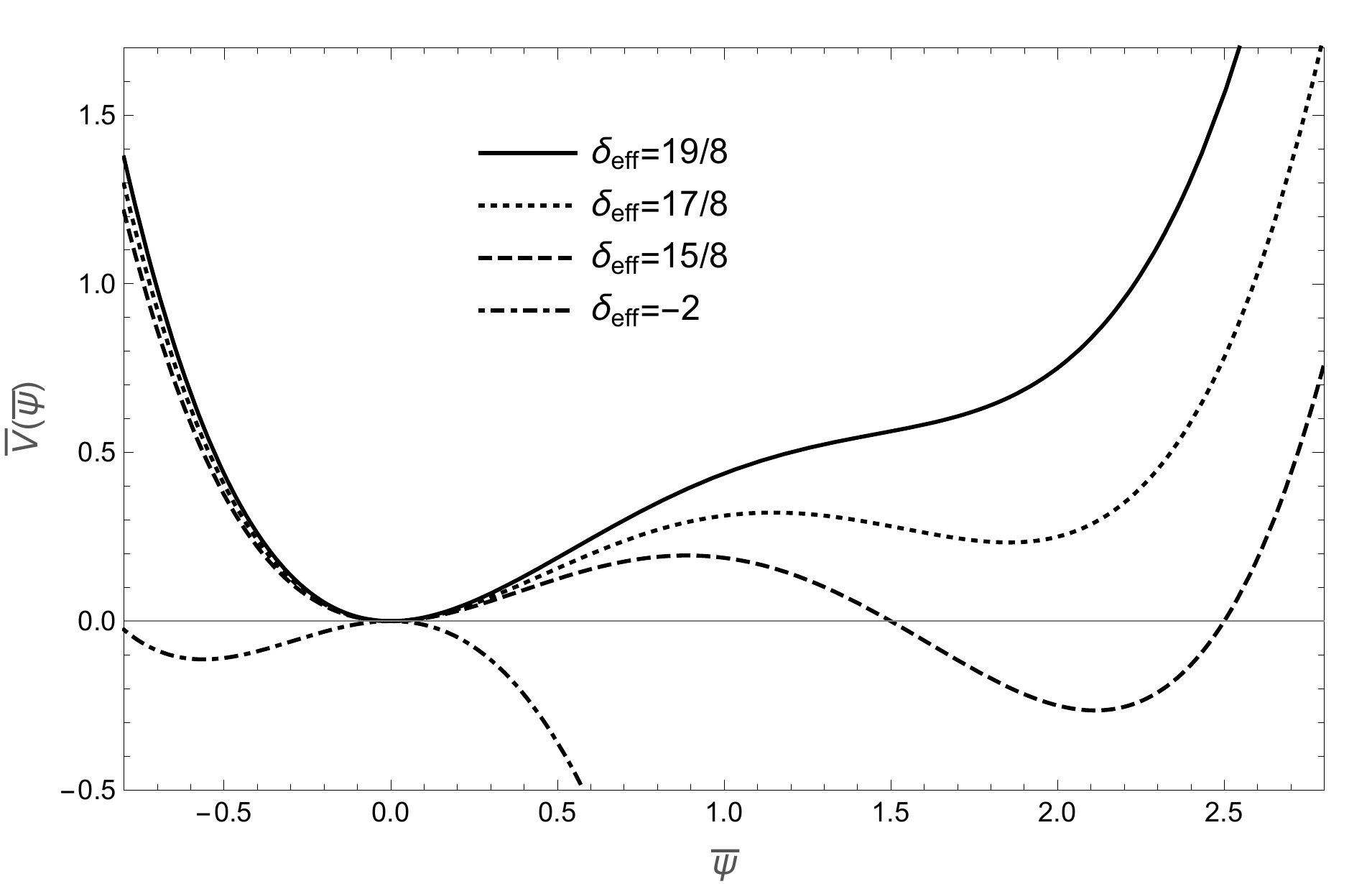}
    \caption{Dimensionless potential $\bar V$ as defined in \protect\eqref{dimless_potential} [or \protect\eqref{eq:pot2d}] for different choices of $\delta_\mathrm{eff}$. A second minimum develops for $\delta_\mathrm{eff} <9 / 4 $ (dotted) which becomes global for  $\delta_\mathrm{eff} < 2 $ (dashed). For $\delta_\mathrm{eff} < 0$ (dash-dotted), the local minimum at $\bar \psi =0$ turns into a maximum.}
    \label{fig:Potential}
\end{figure}
In order to restore the symmetric phase at high temperatures, corresponding to $\delta_\mathrm{eff} > 2$,  we require
\beq\label{param_condition_1}
\frac{\lambda D}{E^2} >1~,
\eeq
which is always fulfilled within the regime of validity of the small-mass expansion \eqref{lambda_E_cond}.
For $\delta_\mathrm{eff} < 2$, on the other hand, there is a global minimum at  
\begin{align}\label{psimin}
\bar \psi_\mathrm{true}(T_d) = \frac{1}{2} \left[ 3 + \sqrt{9-4 \delta_\mathrm{eff}(T_d) }\right] \,,
\end{align}
which renders the false vacuum at $\bar\psi_\mathrm{false} = 0$ unstable against tunneling (see the dashed line in Fig.~\ref{fig:Potential}).
The decay rate per volume then becomes~\cite{Linde:1981zj}
\begin{align}\label{Gamma_3}
\Gamma \sim T_d^4 \exp{\left( - S_3 / T_d\right)}\,,
\end{align}
where $S_3$ is the Euclidian action in the $O(3)$-invariant case. Denoting the three-dimensional radial coordinate with $r$, it is defined as
\begin{align}\label{SE3}
\frac{S_3}{T_d}	&= \frac{4\pi}{T_d} \int \mathrm{d}r r^2 \left[ \frac{1}{2} \psi'(r)^2 + V(\psi;T_d) \right] \nonumber \\
			&\equiv \frac{E}{\lambda^{3/2}}  \bar{F}_3(\delta_\mathrm{eff})\,,
\end{align}
where we have introduced the rescaled, dimensionless action
\begin{align}\label{F_3}
 \bar{F}_3(\delta_\mathrm{eff}) = 4\pi \int \mathrm{d}\bar r \bar r^2 \left[ \frac{1}{2} \bar \psi'(\bar r)^2 + \bar V(\bar \psi; T_d) \right] \,,
\end{align}
together with the coordinate $\bar{r} = E T_d r / \sqrt{\lambda}$. It only depends on $\delta_\mathrm{eff}$ and can be approximated in terms of the semi-analytic function \cite{Dine:1992wr}\footnote{We note that this agrees within the claimed accuracy with the semi-analytic formula provided in \cite{Adams:1993zs} (up to a presumably lost factor of $8$: $S_{E3}^\mathrm{here} =  8S_{E3}^\mathrm{there}$).}
\begin{figure}[t]
    \centering
    \includegraphics[width=0.65\textwidth]{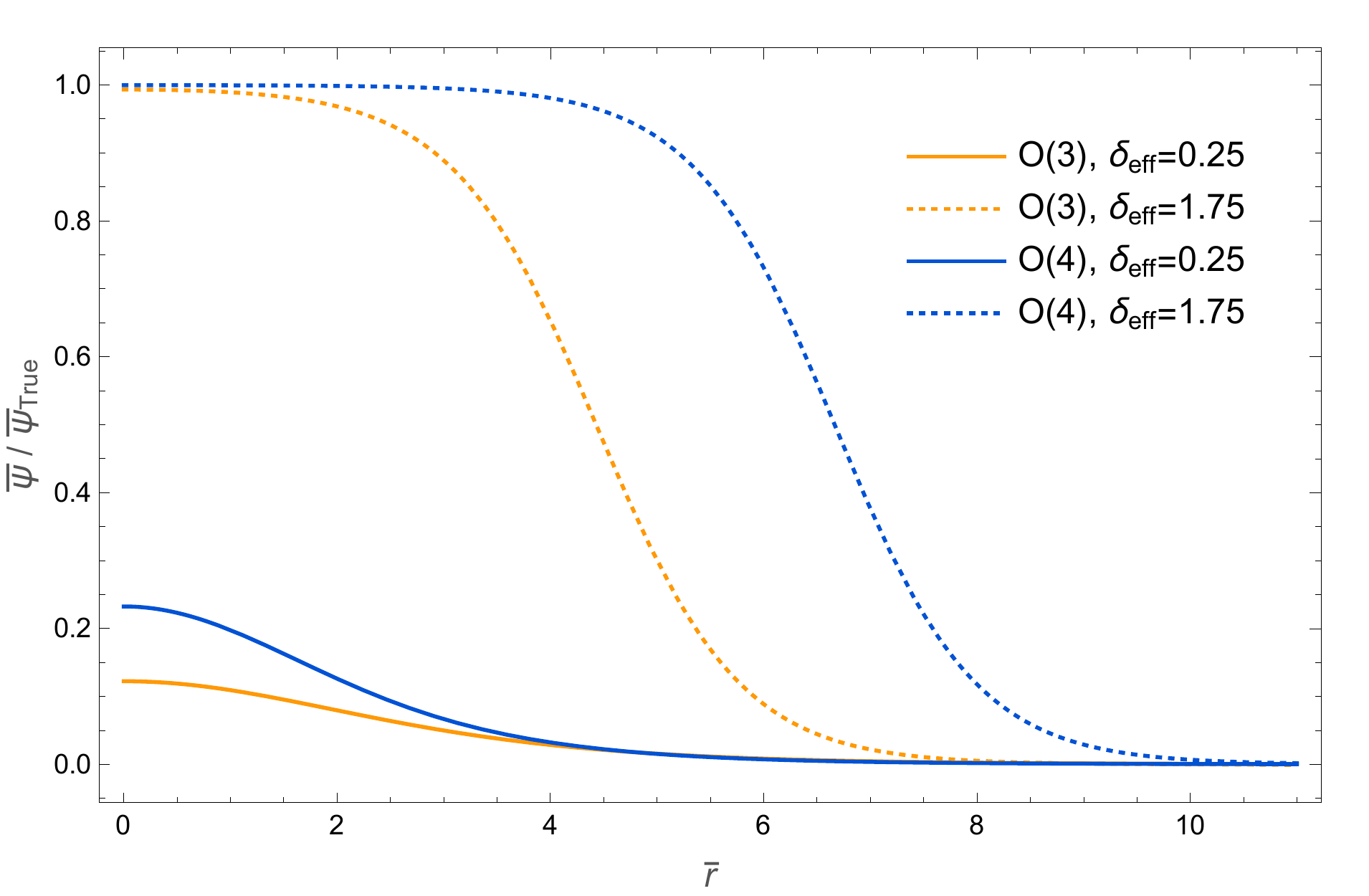}
    \caption{Bounce solution $\bar{\psi}(\bar r) / \bar{\psi}_\mathrm{min}$  in the $O(4)$ (blue) and $O(3)$ (orange) symmetric case for $\delta_\mathrm{eff}=0.25$ (solid) and $\delta_\mathrm{eff}=0.75$ (dotted). For $\delta_\mathrm{eff} < 1.5$, the field inside the bubble $\bar{\psi}(0) \equiv \bar \psi_-$ is displaced from the true vacuum at $\bar \psi_\mathrm{True}$. This effect is slightly more pronounced in the $O(3)$ symmetric case. }
    \label{fig:profile}
\end{figure}
\beq\label{F_3_explicit}
 \bar{F}_3(\delta_\textrm{eff}) = 0.61 \delta_{\textrm{eff}}^{3/2} (2-\delta_{\textrm{eff}})^{-2} (32-17.36 \, \delta_{\textrm{eff}}-0.8 \, \delta_{\textrm{eff}}^2+\delta_{\textrm{eff}}^3)~.
\eeq
To be more precise, it is obtained from numerically evaluating $S_3$ at the $O(3)$-symmetric solution of the Euclidian equation of motion
\begin{align}\label{eq:eom_S3}
\bar{\psi}''(\bar r) +  \frac{2}{\bar r}\, \bar{\psi}'(\bar r) = \bar{\psi}^3(\bar r) - 3 \, \bar{\psi}^2(\bar r) + \delta_\mathrm{eff}(T_d) \, \bar{\psi }(\bar r) \,,
\end{align}
subject to the same boundary conditions as in the  $O(4)$-symmetric case. We depict the bounce solutions for two different values of $\delta_\mathrm{eff}$ and $\lambda^{3/2} E^{-1}S_3(\delta_\mathrm{eff})/T_d^*$ as the orange curves in Fig.~\ref{fig:profile} and Fig.~\ref{fig:S_El}, respectively. The applicability of \eqref{Gamma_3} requires $S_3/ T_d < S_4$ (otherwise the O($4$)-symmetric bounce solution should be used). From \eqref{SE4} and \eqref{SE3}, we derive the sufficient condition $E / \lambda^{1/2} < 7.3$\,, which is always satisfied when \eqref{lambda_E_cond} holds [this might no longer be true if we drop the high-temperature/small-mass approximation used in deriving \eqref{eq:effective_T_pot}]. 
The phase transition occurs at temperature $T_d^*$ (or time $t_*$ equivalently) once the percolation parameter in \eqref{eq:p} has risen to $p(T_d^*) \sim 1$. Analogous to cold NEDE, this defines $S_{3}/T_d^*\simeq  \ln(T_d^{*4}/H_*^4)+\ln(\bar{\beta}^{-1}H_*)$, which, for an $\mathrm{eV}$-scale transition, amounts to 
\begin{align}\label{eq:perc_cond}
S_{3}/T_d^* \simeq  \ln(M_\mathrm{pl}^4/\mathrm{eV}^4) \simeq 250\,,
\end{align} 
provided there are no extreme parameter hierarchies (which could lead to $T_d^* \ll T_\mathrm{vis}^*$ or $\ln(\bar{\beta}^{-1}H_*) <  -10$).  We can then infer $\delta_\mathrm{eff}^* \equiv \delta_\mathrm{eff}(T_d^*)$ as a function of $\lambda^{3/2}/E$ by inverting $ \bar{F}_3(\delta_\mathrm{eff})$ (or reading it off from Fig.~\ref{fig:S_El}), which finally fixes $T_d^*$ through \eqref{delta_eff}. We find that $\delta_\mathrm{eff}^* < 1.86$ when we use that $\lambda^{3/2}/E \lesssim \sqrt{\lambda} < 1$, which marginally includes the thin-wall regime (whereas in cold NEDE $\delta_\mathrm{eff}^* < 1.5$). The parameter dependence of  $\delta_\mathrm{eff}^*$ is explored more systematically in the first plot of Fig.~\ref{fig:params}. In general, we see that for smaller values of $\lambda$, we are driven further away from the thin-wall limit at $\delta_\mathrm{eff}=2$.

Like with cold NEDE, the duration of the phase transition can again be quantified in terms of $1/ \bar{\beta}$ where we use the definition in \eqref{beta} subject to the replacement $S_4 \to S_3/T_d$. Substituting \eqref{SE3} and using \eqref{delta_eff} alongside $T_d \propto 1/a(t)$, we obtain

\begin{align}\label{Hbeta}
H_*\bar{\beta}^{-1}  = \frac{\sqrt{\lambda}E}{4D} \left(1 - \frac{\delta_\mathrm{eff}^*}{2} \frac{E^2}{\lambda D}\right)^{-1} \left[\frac{\mathrm{d} \bar{F}_3}{\mathrm{d}\delta_\mathrm{eff}}\right]^{-1}_{\delta_\mathrm{eff}^*}\,,
\end{align}
where the last factor is order unity. The transition is fast compared to a Hubble time if $\bar\beta/H_* \gg1$. In Fig.~\ref{fig:params} (orange plot), it can be seen that this is always satisfied within the accessible parameter range, thus not imposing any further constraints.

We can then characterize the model's phenomenology in terms of the fraction of hot NEDE $f_\mathrm{NEDE}$ and the dark sector temperature $T_d^*$ at decay time. In the next step, we will relate these phenomenological parameters to the EFT parameters in the hot NEDE potential.  Analog to~\eqref{def:rho_NEDE}, we define
\beq\label{eq:rho_NEDE}
f_{\textrm{NEDE}} \equiv \frac{\rho_{\textrm{NEDE}}(t_*)}{\rho_\mathrm{tot} (t_*)}= \frac{V(\psi_\mathrm{false}; T_d^*)-V(\psi_\mathrm{true}(T_d);T_d)\big|_{T_d \to 0}}{\rho_\mathrm{tot} (t_*)}\,,
\eeq
where we normalized with respect to the potential energy in vacuum, which can be obtained in the limit where $T_d \to 0$. Note that this limit simply corresponds to the trivial case where any thermal corrections are absent, and, as a result, it is not affected by the breakdown of the small-mass expansion.  
After using \eqref{dimless_vars}, \eqref{eq:dict}, \eqref{psimin}, and $\psi_\mathrm{false}=0$, \eqref{def:rho_NEDE} evaluates to
\beq\label{fNEDE0}
f_{\textrm{NEDE}} =   \frac{27}{4} \frac{E^4}{\lambda^3}\frac{c\left[\delta_\mathrm{eff}(T_d)\right]T_d^{4}}{\rho_\mathrm{tot}(t_*)}\Big|_{T_d \ll T_\circ}\,,
\eeq
where $c[\delta_\mathrm{eff}]$ has been defined in \eqref{eq:c_delta}. In the limit $T_d \ll T_\circ$, we obtain from \eqref{delta_eff} and \eqref{param_condition_1}
\begin{align}\label{eq:delta_late}
-\delta_\mathrm{eff}\big|_\mathrm{today} \simeq 2 \frac{\lambda D}{E^2} \frac{T^2_\circ }{T^2_d}  \gg 1\,,
\end{align}
which, in turn, implies that $c[\delta_\mathrm{eff}] \simeq 8 \delta_\mathrm{eff}^2/216 \gg 1$. After substituting this back into \eqref{fNEDE0}, the $T_d$ dependence drops out, and we have
\begin{align}
f_{\textrm{NEDE}} 	=   \frac{D^2}{\lambda}\frac{T_\circ^{4}}{T_d^{*4}}\frac{T_d^{*4}}{\rho_\mathrm{tot}(t_*)}\;.
\end{align}
The same expression could have been obtained directly from minimizing \eqref{eq:effective_T_pot} for $T_d=0$. It can be further evaluated by expressing $T_\circ$ in terms of the dark decay temperature $T_d^*$ using  \eqref{delta_eff}:
\begin{align}
\label{fNEDE}
f_{\textrm{NEDE}} 	=   \frac{D^2}{\lambda}  \left( 1- \frac{\delta_\mathrm{eff}^*}{2}\frac{E^2}{\lambda D} \right)^{2}  \frac{T_d^{*4}}{\rho_\mathrm{tot}(t_*)}
\end{align}
 If we assume that the transition happens during radiation domination -- recall that the cold NEDE transition occurs around $z_* \simeq 5000$, which is still within radiation domination -- we can substitute $\rho_\mathrm{tot}(t_*) \simeq \rho_\mathrm{rad}(t_*)/(1-f_\mathrm{NEDE})$, where the radiation density at decay time is
\begin{align}\label{rad}
\rho_\mathrm{rad}(t_*)= \rho_\mathrm{rad,vis}(T_\mathrm{vis}^*)+ \rho_\mathrm{rad,d}(T_d^*)= \frac{\pi^2}{30} \left( g_\mathrm{rel, vis}^* T^{*4}_\mathrm{vis} + g_\mathrm{rel, d}^* T^{*4}_d \right).
\end{align} 
Here $g_\mathrm{rel, vis}^*\simeq 3.4$ and $g_\mathrm{rel, d}^*$ are the effective number of relativistic degrees of freedom in the visible and dark sector, respectively. Introducing the fraction $\xi = T_d / T_\mathrm{vis}$, we derive from \eqref{fNEDE}
\begin{align}\label{xi}
\xi^4_*\equiv \xi^4(t_*) \simeq 0.11 \times \frac{\lambda}{D^2} \left( 1- \frac{\delta_\mathrm{eff}^*}{2}\frac{E^2}{\lambda D} \right)^{-2} \left[\frac{f_{\textrm{NEDE}} /(1-f_{\textrm{NEDE}} )}{0.1}\right]  \;,
\end{align}
where we neglected the second term in \eqref{rad}, which is valid for a sufficiently cold dark sector with $\xi_* < 1$.  In fact, in Sec.~\ref{sec:NADM}, we will see that a typical value is $\xi_* \simeq 0.34 $ when the dark sector decouples between $10$ and $100 \, \mathrm{GeV}$, which is compatible with a subdominant DR component (as required for hot NEDE) if $g_{\mathrm{rel, d}}$ remains order unity. At this point, we encounter a phenomenological challenge. For a generic gauge theory, we have~\cite{Linde:1981zj,Arnold:1992rz}) $D \sim E^{2/3}\sim g_\mathrm{NEDE}^2 $ (for a detailed discussion of the Abelian Higgs model see the  Appendix \ref{appendix_potential}). Demanding $\xi_*<1$ along with $f_\mathrm{NEDE} \simeq 0.1$ therefore requires $\lambda < 2 E^{4/3}$ (or $\gamma \lesssim 1$ equivalently), which is incompatible with the regime of validity of the high-temperature/small-mass approximation~\eqref{lambda_E_cond} used to derive \eqref{eq:effective_T_pot}. This is also illustrated by the third plot in Fig.~\ref{fig:params}, which depicts  $f_\mathrm{NEDE}$ as a function of $\lambda$ and $g_\mathrm{NEDE}$ in the case of the Abelian Higgs model. However, this is only a limitation of the approximation, and we will drop it all-together in the next section.  As a result, we will see that it is possible to realize a large fraction of NEDE with  $f_\mathrm{NEDE} \gtrsim 0.1$.

Finally, we introduce the parameter 
\begin{align}\label{def:alpha}
\alpha = \frac{\Delta V}{\rho_\mathrm{rad,d}(T_d^*)}\,,
\end{align}
which measures what is commonly referred to as the \textit{strength} of the phase transition. Here, $\Delta V = V(\psi_\mathrm{false}, T^*_d)-V(\psi_\mathrm{true}(T_d^*),T^*_d)$ is the vacuum energy released in the phase transition. This quantity is different from $\rho_\mathrm{NEDE}(T_*)$, defined in \eqref{eq:rho_NEDE}, because it does not take into account the subsequent evolution of the true vacuum.  We derive \
\begin{align}\label{alpha}
\alpha \simeq 21  \frac{c(\delta^*_\mathrm{eff})}{g^*_\mathrm{rel,d}} \frac{E^4}{\lambda^3} \,,
\end{align}
where we used \eqref{dimless_vars}, \eqref{psimin}, \eqref{eq:dict}, and $\rho_{\mathrm{rad,d}}(T_d^*) = \pi^2 g_\mathrm{rel,d}^* T_d^{*4}/30$.
As $c(\delta_\mathrm{eff}^*)/g_\mathrm{rel,d}^* < 1$, we see that $\alpha > 1$ requires
\begin{align}
\frac{E^4}{\lambda^3} = \frac{1}{4 \pi \gamma^3}> 1 \,,
\end{align}
where we used used \eqref{def:gamma}. This is incompatible with the lower bound in \eqref{lambda_E_cond} (or $\gamma \gg 1$ equivalently).  As before, this means that we cannot rely on the small-mass/high-temperature approximation to analyze a strong first-order transition as favored by the hot NEDE phenomenology. Motivated by these finding, we will explore the large-mass/small-temperature regime in the next section.

\subsubsection{Large-mass/low-temperature potential}\label{sec:large_mass}

\begin{figure}[t]
    \centering
    \includegraphics[width=0.65\textwidth]{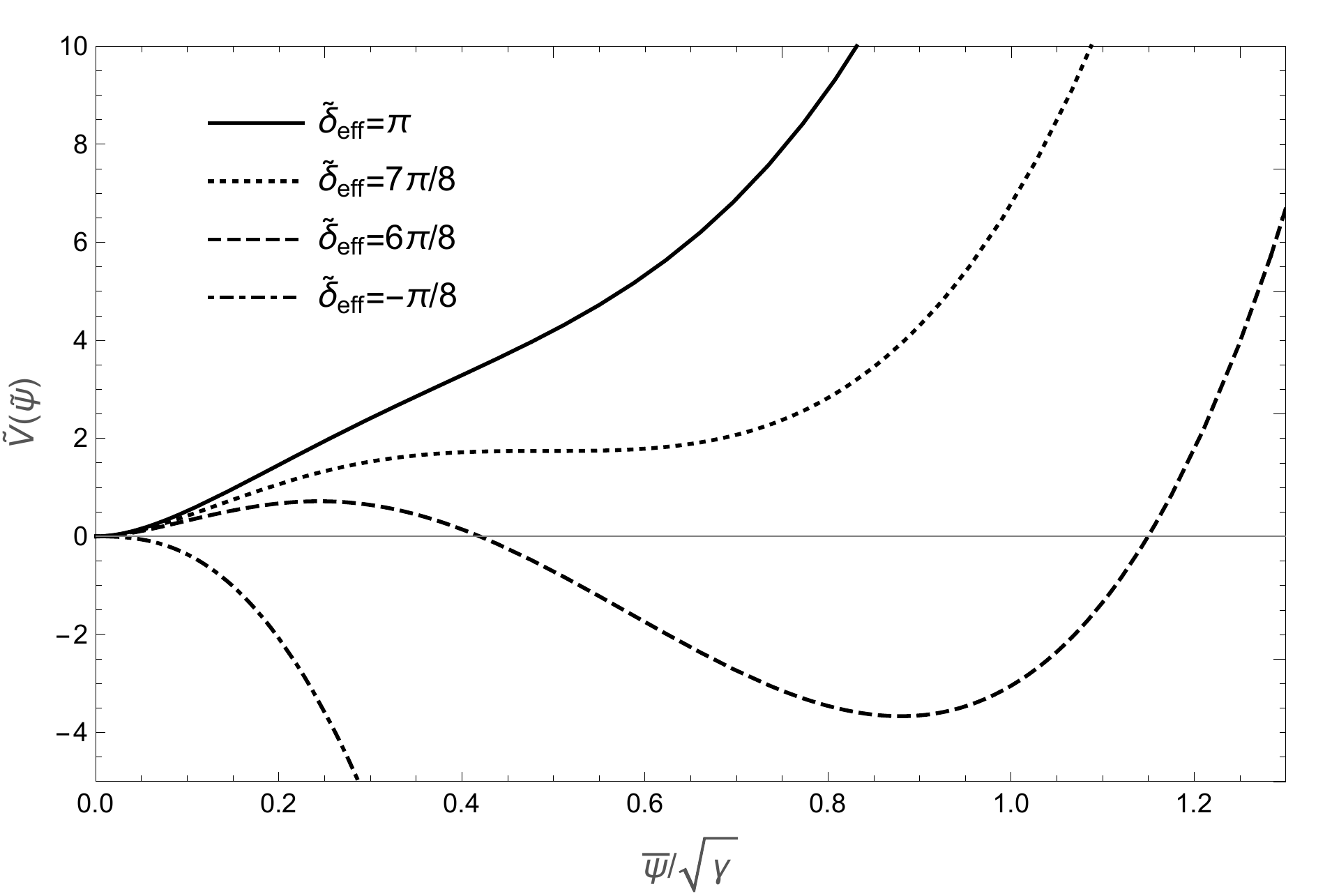}
    \caption{Dimensionless potential $\tilde V = \bar V_\gamma/\gamma^3$ as defined in \protect\eqref{eq:high_mass_pot} for $\gamma=0.02$ and different choices of $\tilde \delta_\mathrm{eff}= \delta_\mathrm{eff}/\gamma$. At high temperatures $\tilde \delta_\mathrm{eff} \to \pi $ and $\bar \psi=0$ is the global, symmetric vacuum (solid). A second minimum develops for $\tilde{\delta}_\mathrm{eff} < \pi $ (dotted) which becomes the true vacuum  $\psi_\mathrm{True}$ shortly after~(dashed).}
    \label{fig:Potential_large_m}
\end{figure}

Here, we study the case where $\gamma \lesssim 1$ (including $\gamma \ll 1$ ) corresponding to the blue region in Fig.~\ref{fig:coupling_regions}. As we will argue later, this regime very naturally accommodates strong supercooling. It requires a more careful analysis because $T_d$ no longer exceeds the gauge boson mass, which in the broken phase becomes $ g_\mathrm{NEDE} \,\psi_\mathrm{True} \sim T_d\, \bar{\psi}_\mathrm{True} / \gamma \gtrsim T_d $. As a result, we have to go beyond the approximation leading to \eqref{eq:effective_T_pot}. In the Appendix \ref{appendix_potential}, we derive
\beq\label{eq:effective_T_pot_low_T}
V(\psi;T_d) =  - D T_\circ^2 \psi^2 +\frac{\lambda}{4}\psi^4 + 3 T_d^4 K\left( \sqrt{8 D} \psi / T_d\right) \mathrm{e}^{-\sqrt{8 D} \psi / T_d} + V_0(T_d)
\eeq
where $K(a)$ only varies slowly in the range $0.1 < |K(a)| \lesssim 10$ for $0< a \equiv \sqrt{8 D} \psi / T_d < 20 $; we provide a semi-analytic approximation in \eqref{eq:K}. While the discussion of the previous section was mostly model-independent, we will focus here on the case of the Abelian Higgs model for which (similar relations are expected to hold for more complicated gauge theories)
\begin{align}\label{def:D_E_Abelian_Higgs}
8 D \simeq (4 \pi E)^{2/3} =g_\mathrm{NEDE}^2 \,.
\end{align}
Using the definitions in \eqref{dimless_vars} and \eqref{eq:dict}, we derive the dimensionless potential (suppressing $V_0$)
\begin{align}\label{eq:high_mass_pot}
\bar V_\gamma(\bar \psi, T_d) = \frac{1}{4} \bar \psi^4 - \frac{1}{2} \left[\pi \gamma - \delta_{\textrm{eff}}(T_d)\right] \bar\psi^2 + 12  \pi \gamma^3 K\left(\bar \psi/\gamma \right) \mathrm{e}^{- \bar\psi/\gamma} \,.
\end{align}
It depends on the two parameters $\delta_\mathrm{eff}$ and $\gamma$ as defined in \eqref{delta_eff} and \eqref{def:gamma}, respectively. Using~\eqref{def:D_E_Abelian_Higgs}, they evaluate to
\begin{align}\label{eq:gamma}
\gamma = \frac{4 \pi \lambda}{g_\mathrm{NEDE}^4} \,,
\end{align}
and
\begin{align}\label{eq:delta_large_mass}
\delta_\mathrm{eff}(T_d) = \pi \gamma\left( 1 - \frac{T_\circ^2}{T_d^2}\right) \,.
\end{align}
The previous potential can be recovered in the limit where $\gamma \gg 1$ (or $\lambda \gg g_\mathrm{NEDE}^4$ equivalently); explicitly, $\bar{V} = \lim_{\gamma \to \infty} \bar{V}_{\gamma}$. For $\gamma \lesssim 1$ (including $\gamma \ll 1$),  there is still a local, metastable minimum around $\bar \psi = 0$ if $0 < \delta_\mathrm{eff}(T_d)\lesssim \pi \gamma$. This can be easily demonstrated by expanding the last term in~\eqref{eq:high_mass_pot} as $12 \pi \gamma^3 K(\bar \psi / \gamma) \mathrm{e}^{-\bar \psi / \gamma} \simeq \mathrm{const} +\pi  \gamma \bar{\psi}^2/2 $, which indeed combines with the second term to the (metastable) mass term  $\delta_\mathrm{eff}(T_d) \bar \psi^2 >0 $. However, when $\bar \psi \gtrsim \gamma$ that expansion breaks down and the last term becomes exponentially suppressed. Since $\delta_\mathrm{eff}(T_d) < \pi \gamma$ due to \eqref{eq:delta_large_mass}, the quadratic term turns negative leading to a local maximum around $\bar \psi \sim \gamma$ . For even larger field values, the potential decreases until it reaches the true minimum. Minimizing the first and the second term (the third term can be neglected in this regime), we obtain
\begin{align}\label{psi_true_large_mass}
\bar \psi_\mathrm{True} \simeq \sqrt{\pi \gamma - \delta_\mathrm{eff}} \,.   
\end{align}
In particular, right after the transition $ \bar \psi_\mathrm{True} \lesssim \sqrt{\pi \gamma}$ (using that $\delta_\mathrm{eff}^* < \pi \gamma$). This has to be contrasted with the previous result  in \eqref{psimin} where $\bar{\psi}_\mathrm{True} = \mathcal{O}(1)$ (valid for $\gamma \gg 1$). A corresponding example is depicted as the dashed line in Fig.~\ref{fig:Potential_large_m}. We also show that the second minimum disappears at high temperatures when $\delta_\mathrm{eff} \to \pi \gamma$ (solid line).
Substituting \eqref{psi_true_large_mass} back into \eqref{eq:high_mass_pot}, we find 
\begin{align}
\bar{V}_\gamma(\bar{\psi}_\mathrm{True}) \simeq -\frac{\pi^2 \gamma^2}{4} \left(1- \frac{\delta_\mathrm{eff}^*}{\pi \gamma}\right)\,.
\end{align}
With this we can derive a generalized expression for the strength parameter defined in \eqref{def:alpha}, which becomes 
\begin{align}\label{alpha_large_mass}
\alpha \simeq \frac{15}{8 \pi} \frac{1}{ g_\mathrm{rel,d}} \frac{1}{ \gamma}  \left(1- \frac{\delta_\mathrm{eff}^*}{\pi \gamma}\right)\,.
\end{align}
This demonstrates that we can indeed have a strong first-order phase transition, characterized by $\alpha > 1$, for $\gamma \ll 1$ (as opposed to the previous case with  $\gamma \gg 1$). Moreover, the formula for $f_\mathrm{NEDE}$ in~\eqref{fNEDE} is still applicable, and we rewrite it in therms of $\gamma$ with \eqref{def:D_E_Abelian_Higgs} and \eqref{eq:gamma} as 
\begin{align}\label{f_NEDE_large_mass}
f_{\textrm{NEDE}} 	=   \frac{\pi}{16 \gamma}  \left( 1- \frac{\delta_\mathrm{eff}^*}{\pi \gamma} \right)^{2}  \frac{T_d^{*4}}{\rho_\mathrm{tot}(t_*)}\,.
\end{align}
As before, $\delta_\mathrm{eff}^*$
is determined implicitly through $S_{3}/T_d^* \simeq 250$  [see \eqref{eq:perc_cond}]. What has changed, however, is the expression for the Euclidian action in \eqref{SE3}. It can now be parametrized as
 \beq\label{S_3_large_mass}
\frac{S_{3}}{T_d} \simeq  \frac{\sqrt{4 \pi}}{g_\mathrm{NEDE}^3 \gamma^{3/2}} \bar{F}_3(\delta_\mathrm{eff},\gamma)~,
\eeq
where $ \bar{F}_3(\delta_\mathrm{eff},\gamma)$ is obtained by evaluating the rescaled Euclidian action in \eqref{F_3} for the bounce solution obtained from the generalized (dimensionless) potential \eqref{eq:high_mass_pot}. In the limit where $\delta_\mathrm{eff} \to 0$ and/or $\gamma \to 0$, the potential barrier shrinks to zero implying $\bar  F_3(\delta_\mathrm{eff},\gamma) \to 0$ (thick-wall/deep-well limit). If $\delta_\mathrm{eff}$ increases on the other hand, also the potential barrier increases, and, eventually, for  $\delta_\mathrm{eff} \simeq \pi \gamma$, the second minimum is lifted above the symmetric one at $\bar \psi =0$, which in turn implies that $ \bar{F}_3(\delta_\mathrm{eff},\gamma) \to \infty$ (thin-wall limit). This case is represented by the dotted line in Fig.~\ref{fig:Potential_large_m}. While it is difficult to derive the functional form of $ \bar{F}_3(\delta_\mathrm{eff},\gamma)$ in full generality, we will argue in the following that it scales as 
\begin{align}\label{tilde_F3}
  \bar{F}_3(\delta_\mathrm{eff},\gamma) \sim \gamma^{3/2} \delta_\mathrm{eff}
\end{align}
in the thick-wall limit where $\delta_\mathrm{eff} \ll \pi \gamma$. In that case, we can neglect the quartic term in the potential because it is never probed by the bounce solution, which exits the potential closer to the maximum at $\bar \psi \sim \gamma$ than the minimum at $\bar \psi_\mathrm{True}$.\footnote{The same argument is used in \cite{Linde:1981zj} when studying the thick-wall limit.} This allows us to scale out $\gamma$ by introducing 
\begin{align}
\bar \psi = \gamma \tilde \psi\,, && \delta_\mathrm{eff} = \gamma \tilde \delta_\mathrm{eff}\,, &&\bar V_\gamma = \gamma^3 \tilde V\,,&&\bar r =  \tilde r/\sqrt{\gamma}\,.
\end{align}
Substituting back into \eqref{F_3} (with $V$ generalized to $V_\gamma$) then yields $ \bar{F}_3( \delta_\mathrm{eff},\gamma) =  \gamma^{3/2} \tilde F_3(\tilde \delta_\mathrm{eff}) $, where
\begin{align}\label{F_3_tilde}
\tilde  F_3(\tilde \delta_\mathrm{eff}) = 4\pi \int \mathrm{d} \tilde r \tilde r^2 \left[ \frac{1}{2} \tilde \psi'(\tilde r)^2 + \tilde V(\tilde \psi; T_d) \right] \,.
\end{align}
\begin{figure}[t]
    \centering    \includegraphics[width=0.65\textwidth]{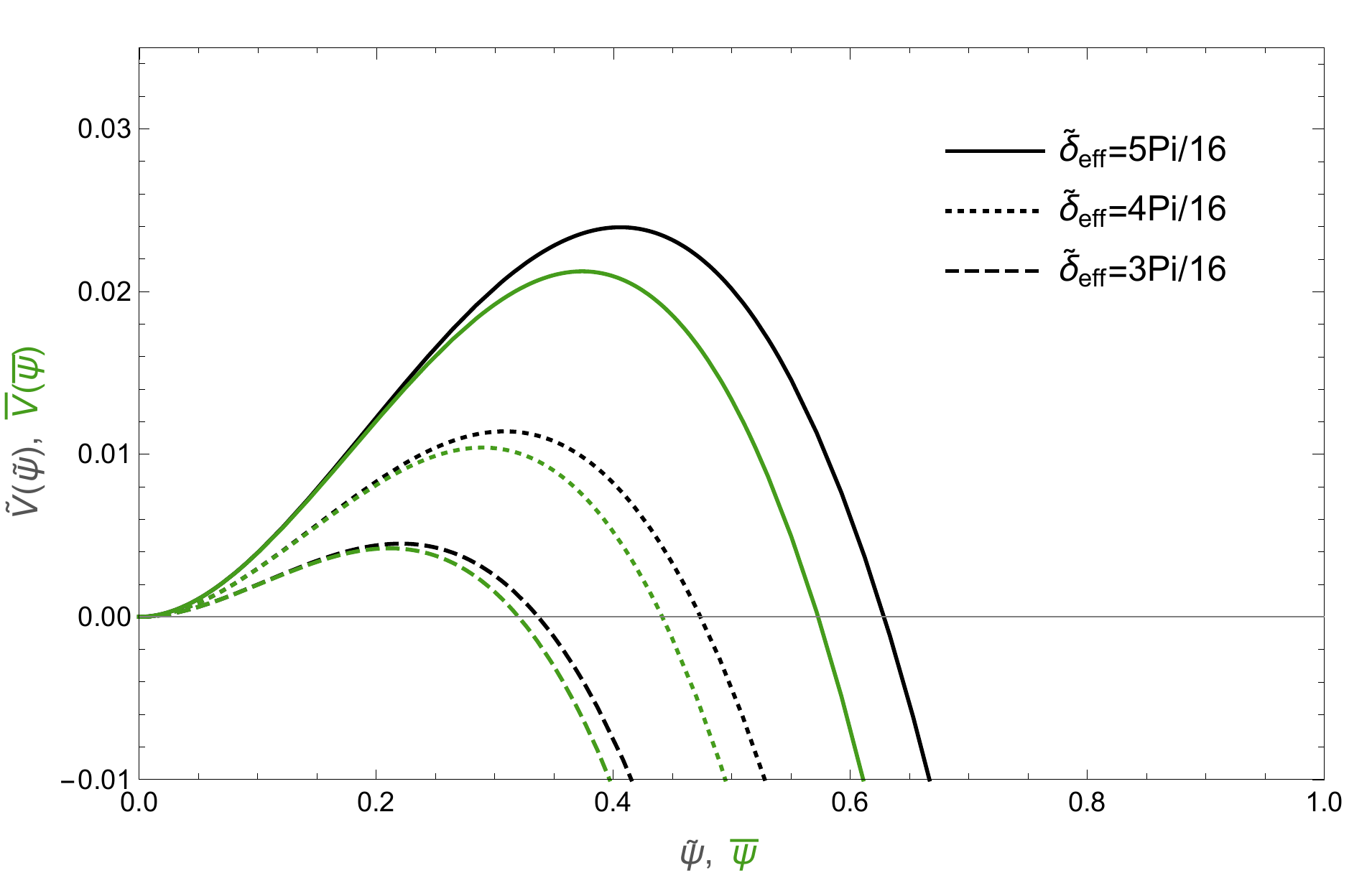}

    \caption{The rescaled large-mass/low-temperature potential $\tilde V(\tilde \psi) = \bar V_\gamma / \gamma^3 $ (black) defined in \protect\eqref{eq:high_mass_pot} can be roughly approximated around the maximum by the potential $\bar V(\bar \psi)$ (green) defined in \protect\eqref{dimless_potential}. Both potentials differ strongly around the true minimum though.}
    \label{fig:Potential_approx}
\end{figure}
Now, we expect the scaling with $\tilde \delta_\mathrm{eff}$ to be approximated by the functional form in \eqref{F_3_explicit} since the (rescaled) potential as a function of $\tilde \delta_\mathrm{eff}$ has a very similar shape around the maximum as the small-mass potential in \eqref{dimless_potential}, explicitly $\tilde V(\tilde \psi) \sim \bar V(\tilde \psi)$, which then implies $\tilde F_3(\tilde \delta_\mathrm{eff})\sim \bar{F}_3(\tilde \delta_\mathrm{eff})$. We demonstrate this explicitly in Fig.~\ref{fig:Potential_approx} by ploting $\tilde V(\tilde \psi)$ (black) and $\bar{V}(\bar \psi)$ (green) on the same axes.\footnote{We note that this approximation only provides the right order of magnitude; a numerical analysis is required for achieving a higher precision or accessing the thin-wall limit .} Expanding  \eqref{F_3_explicit} for small $\tilde \delta_\mathrm{eff}$ then yields the expression in \eqref{tilde_F3}. Further using the percolation condition \eqref{eq:perc_cond} together with \eqref{S_3_large_mass} fixes
\begin{align}\label{delta_eff_star}
\delta^*_\mathrm{eff} \sim 10 \gamma g_\mathrm{NEDE}^2\,,
\end{align}
which imposes the very weak parameter constraint $g_\mathrm{NEDE} ^2 \ll \pi / 10 $ to make sure that $\delta_\mathrm{eff}^* \ll \pi \gamma$. It also shows \textit{a posteriori} that  using the thick-wall approximation was justified for a sufficiently weak gauge coupling parameter $g_\mathrm{NEDE} \lesssim 0.1$. For larger couplings we have to solve the thin-wall limit instead. Moreover, performing the analogue calculation in the $\mathrm{O(4)}$ invariant case, we find that the thermal tunneling dominates over the vacuum tunneling. We attribute this to the fact that in our model there is no tunneling barrier in the vacuum potential that could dominate over the thermally induced barrier.

{In any event, having an expression for $\delta_\mathrm{eff}^*$ enables us to further evaluate $f_\mathrm{NEDE}$.} Substituting \eqref{delta_eff_star} back into \eqref{f_NEDE_large_mass} yields
\begin{align}
f_{\textrm{NEDE}} 	\simeq   \frac{\pi}{16} \frac{1}{\gamma}  \frac{T_d^{*4}}{\rho_\mathrm{tot}(t_*)} \quad\quad (\text{for}\,\,\gamma \lesssim 1\,\, \text{and} \,\, g_\mathrm{NEDE} \lesssim 0.1 )~.\end{align}
This demonstrates that for a small dark sector temperature $T_d \ll T_\mathrm{vis}$, we can still have a large fraction of NEDE by demanding $\gamma \ll 1$. Again, we can solve for $\xi_*$ (neglecting the {subdominant} DR contribution to $\rho_\mathrm{tot}$): 
\begin{align}\label{xi_large_mass}
\xi^4_* \simeq 0.56 \times \gamma \left[\frac{f_{\textrm{NEDE}} /(1-f_{\textrm{NEDE}} )}{0.1}\right] \,.
\end{align}
In particular, it shows that we can have $\xi_* \simeq 0.34 $ (as later required in the context of the interacting DM model in Sec.~\ref{sec:NADM}.) for $\gamma \simeq 0.024$ while being compatible with a sizable amount of NEDE and realizing a strong transition dominated by vacuum energy (with $\alpha \gg 1$). This scenario is represented by the dotted line in Fig.~\ref{fig:params_large_mass}.
We can use \eqref{xi_large_mass} to express the dark sector temperature $T_d^*$ in terms of the decay redshift as
\begin{align}\label{eq:T_d_star_large_mass}
T_d^{*4} \simeq (0.7 \mathrm{eV})^4 \gamma \left[\frac{f_{\textrm{NEDE}} /(1-f_{\textrm{NEDE}} )}{0.1}\right]\left[ \frac{1+z_*}{5000}\right]^4\,.
\end{align} 
We therefore see that similar to cold NEDE the energy scale of the phase transition is of the order of $\mathrm{eV}$ (or below) for a typical transition at redshift $\sim 5000$. 
Finally, the duration of the phase transition in \eqref{Hbeta} evaluates to
\begin{align}\label{beta_large_mass}
H_*\bar{\beta}^{-1}  \sim 10^{-2} g_\mathrm{NEDE}^2   \quad\quad (\text{for}\,\,\gamma \lesssim 1\,\, \text{and} \,\, g_\mathrm{NEDE} \lesssim 0.1 )~,
\end{align}
where we used $\mathrm{d}\bar{F}_3 / \mathrm{d} \delta^*_\mathrm{eff}= \sqrt{\gamma}\mathrm{d} \tilde{F}_3 / \mathrm{d} \tilde \delta^*_\mathrm{eff} \sim \sqrt{\delta_\mathrm{eff}^*}$ along with the identification \eqref{def:D_E_Abelian_Higgs}. As a result, for $\gamma \lesssim 1$, we always have a very quick phase transition. 

We summarize our findings in Fig.~\ref{fig:params_large_mass}, which depicts the different phenomenological quantities as a function of the microscopic parameters. We stress that this exploration is limited so far to the thick-wall limit with $\delta^*_\mathrm{eff} \ll \pi \gamma$ (requiring $g_\mathrm{NEDE} \lesssim 0.1$). We leave a more complete discussion of the complementary regime, including the thin-wall limit where $\delta^*_\mathrm{eff} \simeq \gamma \pi $, to future work as it will require a more extensive, independent analysis of the full two-parameter potential in \eqref{eq:high_mass_pot}.

\subsection{Phenomenology}

We are now prepared to summarize the different cosmological stages of hot NEDE.  We will discuss the high ($\gamma \gg 1$) and low-temperature ($\gamma \lesssim 1$) scenario in parallel. We will define four characteristic points in the evolution of the thermally corrected potentials in \eqref{dimless_potential} and \eqref{eq:high_mass_pot}, corresponding to the temperatures $T_1>T_c> T_d^* >T_\circ$. \\

\begin{enumerate}

\item[(a)]
\textbf{Initial regime ($T_d>T_1$):}
For $\gamma \gg 1$, if we are in a situation where \eqref{param_condition_1} holds, then initially, for $T_d > T_1 \gg T_\circ$, we have  $\delta_{\textrm{eff}} >2$, and the field is in the lowest energy state at $\bar \psi = 0$, where tunneling is prohibited.  This corresponds to the solid and dotted lines in Fig.~\ref{fig:Potential}. Moreover, the bound \eqref{param_condition_1} gives rise to the upper edge of the colored parameter  regions in Fig.~\ref{fig:params}. For $\gamma \lesssim 1$, we have $  \delta_\mathrm{eff} \simeq \pi \gamma$, which also corresponds to the unbroken phase represented by the solid and dotted lines in Fig.~\ref{fig:Potential_large_m}.  In both cases, as the temperature drops, $\delta_{\textrm{eff}}$  will decrease below a value $ \delta_{\textrm{eff}}(T_1)$ where a second minimum appears. For $\gamma \gg 1$, $\delta_{\textrm{eff}}(T_1)\equiv 9/4$,
which defines 
\beq
T_1^2 = \frac{8\lambda D}{8\lambda D-9 E^2}T_\circ^2 \quad\quad (\text{for}\,\,\gamma \gg 1)~.
\eeq
For $\gamma \lesssim 1$, we can estimate the value of $ \delta_\mathrm{eff}(T_1)$ by equating the first and second term in~\eqref{eq:high_mass_pot} near the local maximum at $\psi \sim \gamma$. Only if the (negative) mass term dominates a second minimum appears, otherwise the only minimum is at $\bar{\psi}=0$. This reasoning fixes $ \delta_\mathrm{eff}(T_1) \simeq \gamma (\pi - \gamma/2)$, which in turn implies
\begin{align}\label{T1}
T_1^2 \sim \frac{2 \pi}{\gamma} T_\circ^2  \quad\quad (\text{for}\,\,\gamma \lesssim 1)~.
\end{align}
 \\
 
 \item[(b)]
\textbf{Fostering regime ($T_1>T_d>T_c$):}
In this regime a barrier between the two minima is developing, but there is still no tunneling as the minimum at the origin, $\bar \psi=0$, is still the global minimum (see the dotted lines in Fig~\ref{fig:Potential} and~\ref{fig:Potential_large_m}). For $\gamma \gg 1$, when $\delta_{\textrm{eff}}(T_c)\equiv 2$, the two minima become degenerate, which defines the critical temperature
\beq\label{Tc}
T_c^2 =\frac{\lambda D}{\lambda D -E^2}T_\circ^2\quad\quad (\text{for}\,\,\gamma \gg 1)~.
\eeq
In a generic situation $T_c \sim T_1$, which also holds for $\gamma \lesssim 1$ where it proves more difficult to derive an exact expression.

\item[(c)]
\textbf{Tunneling regime ($T_c>T_d>T_\circ$):}
In this regime  $\delta_{\textrm{eff}}<\delta_{\textrm{eff}}(T_c)$, and tunneling becomes possible as $\bar \psi=0$ no longer is the global minimum (dashed line). This regime ends, when the barrier disappears, and the origin becomes a maximum again. The phase transition completes before that when the percolation parameter $p $, as defined in \eqref{eq:p}, reaches order unity, implying a unity probability for a bubble to have nucleated in each Hubble volume. As explained above, this defines the temperature $T_d^*$ through $p(T_d^*) \sim 1$. The process completes within the time $\bar{\beta}^{-1} < 1/H_* $. The temperature range $T_d^*< T_d <T_c$ is referred to as the supercooled phase. We derive from \eqref{Tc} and \eqref{delta_eff}
\begin{align}\label{supercool}
\frac{T_c^2}{T_d^{*2}} = \frac{2 \lambda D- \delta_\mathrm{eff}^* E^2 }{2 \lambda D- 2 E^2}\quad\quad (\text{for}\,\,\gamma \gg 1)~.
\end{align}
which is always $\geq 1$.
Strong supercooling refers to a situation with $T_c/T_d^* \gg 1$ and requires $0 < \delta^*_\mathrm{eff} \ll 2$ and $E^2/(\lambda D) \to 1 $. It corresponds to the band right below the black, solid line in Fig.~\ref{fig:params}. As we have argued before and is illustrated in the third plot in Fig.~\ref{fig:params}, this is incompatible with having a sizable amount of NEDE. However, this conclusion does not apply to the regime where $\gamma \lesssim  1$. Combining \eqref{T1} and \eqref{eq:delta_large_mass}, we derive
\begin{align}
\frac{T_c^2}{T_d^{*2}} \simeq \frac{T_1^2}{T_d^{*2}} \sim \frac{2 \pi}{\gamma} \left( 1-\frac{\delta_\mathrm{eff}^*}{\pi \gamma} \right) \simeq \frac{2 \pi}{\gamma} \quad \quad(\text{for}\,\,\gamma \lesssim 1)~,
\end{align}
where we used that $\delta_\mathrm{eff}^* / \gamma \sim g_\mathrm{NEDE}^2 \ll 1$ (for $g_\mathrm{NEDE} \lesssim 0.1$) due to \eqref{delta_eff_star}. This shows that $\gamma \ll 1$ automatically implies strong supercooling. As we have seen in Eq.~\eqref{alpha_large_mass}, this regime also leads to a strong first-order phase transition with $\alpha \gg 1$, where the released vacuum energy $\Delta V$ dominates over DR $\rho_{\mathrm{rad,d}}$. An order unity fraction $\kappa_\psi = \rho_\mathrm{grad}/\Delta V$ is then converted into gradient energy (and kinetic energy) carried by $\psi$; and a negligible fraction $\kappa_v$ goes into the bulk motion of the surrounding radiation plasma. These quantities are typically used to characterize the phenomenology of the wall-plasma fluid after the decay (see for example the discussion in \cite{Binetruy:2012ze,Caprini:2018mtu}).

\item[(d)]
\textbf{Rollover regime ($T_\circ>T_d$):}
Finally, if the first-order phase transition has not already completed, in this regime, it will complete in a second-order roll-over transition. Due to \eqref{delta_eff} and \eqref{eq:p}, $p \gg 1$ is always satisfied before $T_d$ drops below $T_\circ$. Therefore, it depends on $\bar \beta$ if this regime can ever be reached before the phase transition completes. A sufficient condition for avoiding a rollover is $\bar{\beta}^{-1} \ll t_\circ - t_c$, where the times $t_\circ$ and $t_c$ correspond to temperatures $T_\circ$ and $T_c$, respectively. Performing an explicit numerical analysis, we find that for $\gamma \gg 1$ the previous condition is always fulfilled within the accessible parameter space. This can be understood intuitively as a consequence of the large hierarchy of scales in \eqref{percolation_parameter}, which leads to a very quick increase of the percolation parameter $p$.\footnote{This discussion relies on the validity of the perturbative description and would fail in the case of the electroweak phase transition, which indeed corresponds to a rollover~\cite{Kajantie:1996mn,Laine:1998jb}.} However, if the transition happens in the thick-wall limit, i.e.\ for $
\delta_\mathrm{eff} \ll 2$ or $\delta_\mathrm{eff} \ll \pi \gamma$, we will see that the phenomenology of the phase transition is dominated by oscillations around the true minimum and hence resembles very much a rollover. In fact, this becomes a temperature-triggered version of the hybrid NEDE scenario introduced in \cite{Niedermann:2020dwg}. Also, this scenario will be realized later in terms of our microscopic embedding in Sec.~\ref{sec:microphysics}, although other models where the decay happens in the thin-wall limit can be envisioned, too.
\end{enumerate}

Having discussed the different stages of the phase transition, we can now formulate a set of phenomenological conditions that have to be fulfilled for hot NEDE to work as a resolution of the Hubble tension:

\begin{figure}[t]
    \centering
    \includegraphics[width=0.65\textwidth]{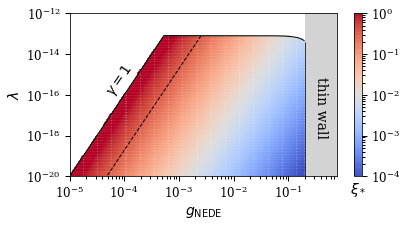}
    \caption{Relative dark sector temperature $\xi_* = T_d^*/T_\mathrm{vis}^*$ compatible with a sizable fraction of NEDE with $f_\mathrm{NEDE}= 10\%$ in the large-mass/small-temperature regime with $\gamma \lesssim 1$ (left edge). As the radiation fluid only serves as a trigger, we can accommodate a hierarchically wide range of dark sector temperatures which is a distinctive feature of hot NEDE. The dashed line corresponds to $\xi_*=0.34$ (or $\gamma=0.024$ equivalently) which is the preferred value in the interacting DM model presented in Sec.~\protect\ref{sec:NADM}. The thin-wall regime (light gray) remains to be explored in detail. }
    \label{fig:xi}
\end{figure}

\begin{enumerate}
\item There has to be sizable fraction of NEDE of order $f_\textrm{NEDE} \simeq 10 \%$ that decays before matter-radiation equality. As we have argued before this singles out the regime with $\gamma \lesssim 1$. The dark sector temperature $T_d^* \lesssim \mathrm{eV}$ is then determined via \eqref{xi_large_mass}. For example, if we demand $\xi_* = 0.34$ (and $g_\mathrm{NEDE} \lesssim 0.1$), we are in the strong supercooled regime with $\gamma \simeq 0.024 \ll 1$.  As one of the main results of this work, we depict the relative dark sector temperature as a function of the microscopic parameters in Fig.~\ref{fig:xi}. It demonstrates that  a sizable fraction of NEDE with $f_\mathrm{NEDE}= 10\%$ can be realized for $\xi_* \lesssim 1$. In particular, $\xi$ can be hierarchically small. This is a crucial feature of hot NEDE, and shows that the model does not rely on tuning the dark sector temperature close to $T_\mathrm{vis}$.\footnote{In fact, other proposals relying on the dark sector radiation plasma to provide the early energy injection, like the step model in~\cite{Aloni:2021eaq}, typically require $T_d \simeq T_\nu$ . Whether this can be achieved in a minimal dark sector model without fine-tuning remains to be seen.}

\item The phase transition has to be short on cosmological time scales in order to prevent bubbles from growing to cosmological sizes and leave (potentially large) imprints in the CMB or LSS data. As argued in Sec.~\ref{Phenomenology_coldNEDE}, a sufficient condition is $H_* \bar{\beta}^{-1} < 0.005$, which due to \eqref{beta_large_mass} is always satisfied for a gauge coupling $g_\mathrm{NEDE} \lesssim 0.1$, although the upper bound can be probed in the thin-wall regime where $g_\mathrm{NEDE} \gtrsim 0.1$.

\item
The crucial feature of EDE-type models is their ability to provide a short energy injection. This is best achieved in terms of a cosmological constant source that relative to the decreasing (yet dominant) radiation fluid grows like $\propto a^4$. In the case of hot NEDE, this type of source is realized through the presence of the false vacuum energy present before the phase transition.  However, this requires that the vacuum energy dominates over the DR fluid; otherwise, the model's phenomenology would resemble that of DR, which, at least in its simple implementation, is known not to help with the Hubble tension~\cite{Aghanim:2018eyx}  (see~\cite{Lesgourgues:2015wza,Raveri:2017jto,Blinov:2020hmc,Blinov:2020uvz} for more sophisticated interacting scenarios). Therefore, we demand that the strength parameter defined in \eqref{def:alpha} fulfills $\alpha \gg 1$, corresponding to a strong first order transition. As we have seen, this can be guaranteed in the regime of strong supercooling where $\gamma \ll 1$ and $g_\mathrm{NEDE} \lesssim 0.1$. This also implies that $\kappa_\psi \simeq 1$, corresponding to a situation where most of the released vacuum energy is carried by $\psi$ (rather than being transferred to the plasma). After the phase transition, we {expect} the NEDE fluid to decay quickly, which has been shown to be {also} a crucial phenomenological requirement. As we will discuss in Sec.~\ref{sec:decay}, our model accommodates different decay scenarios: One where most energy is carried either by coherent oscillations of $\psi$ around its true {vacuum} or by the kinetic energy of the bubble walls (scenario A). And another one where the $\psi$ condensate quickly dissipates through the microscopic decay of $\psi$ quanta (scenario B).

\item As argued in the previous section, acoustic oscillations in the decaying NEDE fluid are a vital ingredient to our setup. Their positive pressure is needed to counter the effect of an increased DM density $\omega_\mathrm{cdm}$. This requires sub-horizon modes to be initialized after the decay with the right amplitude $\delta \rho_\mathrm{NEDE}(t_*,\mathbf{k})$. For cold NEDE, $\delta \rho_\mathrm{NEDE}(t_*,\mathbf{k})$ is determined in terms of adiabatic perturbations of the trigger field $\delta \phi(t_*,\mathbf{k}) $, the slow-roll dynamics of which turned out to provide the phenomenologically preferred scale, specifically~\cite{Niedermann:2020dwg}: $\delta \rho_\mathrm{NEDE}(t_*,\mathbf{k})/  \rho_\mathrm{NEDE}(t_*)= -3 [1+w_\mathrm{NEDE}(t_*)]H_* \delta \phi(t_*,\mathbf{k}) / \dot \phi(t_*)  $. On the other hand, for hot NEDE, the amplitude will be controlled by dark sector temperature fluctuations $\delta T_d(t_*,\mathbf{k})$. Adapting the derivation in~\cite{Niedermann:2020dwg} to this case, we find\footnote{To be specific, we identify the transition surface as the surface of constant temperature. In the notation of~\cite{Niedermann:2020dwg}, this amounts to identifying $q(t,\mathbf{x}) \equiv T_d(t,\mathbf{x})$.}
\begin{align}
\frac{\delta \rho_\mathrm{NEDE}(t_*,\mathbf{k})}{ \rho_\mathrm{NEDE}(t_*)}= -3 [1+w_\mathrm{NEDE}(t_*)] H_* \frac{\delta T_d(t_*,\mathbf{k})}{ \dot T_d(t_*) }\,.
\end{align}
This novel trigger mechanism makes the hot and cold NEDE phenomenology different on perturbation level. Whether it can give the required amplitude as in the case of cold NEDE needs to be investigated through an explicit Boltzmann code implementation, which we postpone to future work.

\end{enumerate}

\subsection{The decay of NEDE}\label{sec:decay}

A crucial feature of early dark energy, independent of the underlying model, is its quick dilution as a function of the scale factor. It is necessary in order to preserve the dynamics leading to the CMB and LSS formation.  Here, we discuss two scenarios how such a decay might be realized. \textit{Scenario A} relies on the same mechanism as in cold NEDE, where small-scale anisotropic stress comprised of colliding bubble walls gives rise to an excess pressure on large scales. \textit{Scenario B} discusses a novel possibility where the bubble wall condensate  decays into relativistic particles that turn non-relativistic shortly after. We will discuss both scenarios in the context of cold \textit{and} hot NEDE.

\subsubsection{Scenario A  (colliding bubble wall fluid)}

In order to discuss this first scenario, it is useful to decompose the released vacuum energy $\Delta V = \Delta \rho_\mathrm{wall}(t_*) +  \Delta \rho_\mathrm{osc}(t_*) $. The physical reason for this split is that the field after the transition is initially displaced from the true vacuum and starts to oscillate around it. The first term measures the energy released directly through the tunneling, which is stored in the bubble walls,\footnote{We note that we are in a regime where $\alpha \gg 1$ and thus the plasma/ DR component can  be neglected.}
\begin{align}
 \Delta \rho_\mathrm{wall}(t_*) =V(\psi_\mathrm{false}, T_d^*)-V(\psi_-,T_d^*)\,,
\end{align} 
and the second one corresponds to the energy that is subsequently dissipated through oscillations of $\psi$ around its true minimum $\psi_\mathrm{true}(T_d^*)$,
\begin{align}
 \Delta \rho_\mathrm{osc}(t_*) =V(\psi_{-},T_d^*)-V(\psi_\mathrm{true}(T_d^*),T_d^*)\,,
\end{align} 
where $\psi_{-} \equiv \lim_{r \to 0} \psi(r) \neq \psi_\mathrm{true} $ is the initial value of $\psi$ at the center of the bubble (following the notation in \cite{Coleman:1977py}). The displacement is demonstrated in Fig.~\ref{fig:profile} for $\gamma \gg 1$, although the definition is equally applicable for  $\gamma \lesssim 1$.
 For $\Delta \rho_\mathrm{wall}(t_*) / \Delta V \ll 1$, we are in a regime where most of the energy is dissipated via oscillations around the true minimum,  which phenomenologically resembles a roll-over. For $ \Delta \rho_\mathrm{osc}(t_*) / \Delta V  \ll 1$, on the other hand, the dissipation happens through the redshift and decay of the bubble wall condensate. The first and second situations correspond to the thick and thin-wall limit, respectively. We derive
\begin{align}\label{eq:ratio}
\frac{\Delta \rho_\mathrm{wall}(t_*) }{ \Delta V}  = \frac{\bar{V}\left(\bar \psi_{-},  T_d^*\right)}{ \left|\bar V(\bar \psi_\mathrm{true}(T^*_d),T^*_d) \right|} \equiv -\bar{V}_-(\delta_\mathrm{eff}^*)
\end{align}
where $\bar{V}_-$ is the (dimensionless) potential energy at the center of the bubble normalized with respect to the true vacuum at time $t_*$.  

\begin{figure}[t]
    \centering
    \includegraphics[width=0.65\textwidth]{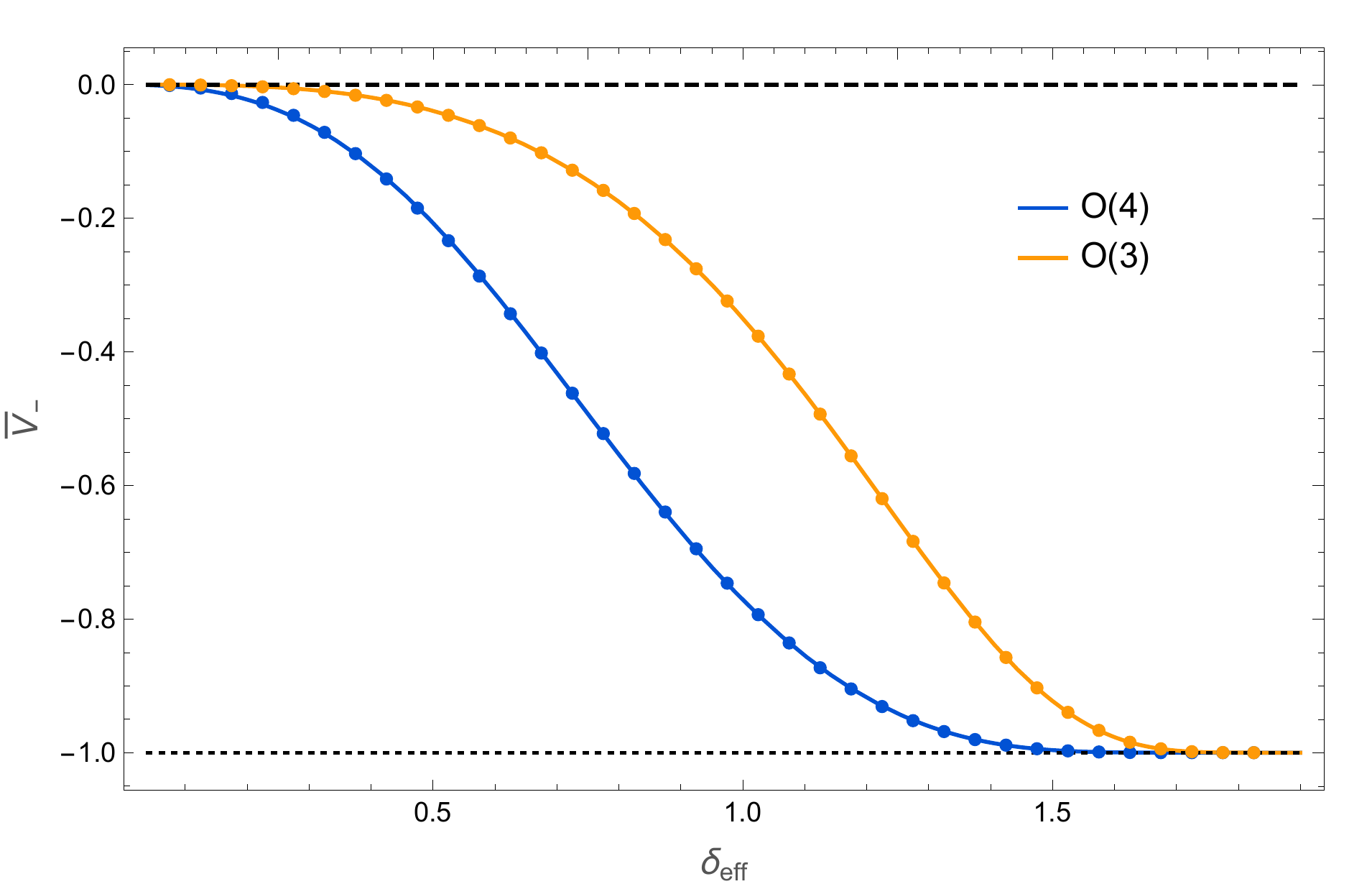}
    \caption{The value of $\bar{V}$ at the center of the the bubble right after the decay. Each dot corresponds to one numerically determined bounce solution. The blue and orange curves depict the semi-analytic approximations in \protect\eqref{eq:V_minus_analytic}. For small values of $\delta_\mathrm{eff}^*$ the field transitions to a configuration which is energetically close to the false vacuum (dashed line). In the thin-wall limit, as $\delta_\mathrm{eff}^* \to 2$, the initial bubble vacuum quickly approaches the true vacuum (dotted line). }
    \label{fig:V_minus}
\end{figure}

As before, we will first discuss the case $\gamma \gg 1$, which can be analyzed more straightforwardly, and then discuss implications for $\gamma \lesssim 1$. The ratio in \eqref{eq:ratio} can be determined numerically as a function of $\delta_\mathrm{eff}^*$ by solving the differential equation in \eqref{eq:eom_S3} or \eqref{eq:eom_psi1} in the O$(3)$ or O$(4)$ symmetric case, respectively. As demonstrated in Fig.~\ref{fig:V_minus}, it can be well approximated in terms of
\begin{subequations}
\label{eq:V_minus_analytic}
\begin{align}
\bar{V}_-(\delta_\mathrm{eff}^*) = \frac{1-\mathrm{e}^{P(\delta_\mathrm{eff}^*)/(2-\delta_\mathrm{eff}^*)} }{\mathrm{e}^{P(\delta_\mathrm{eff}^*)/(2-\delta_\mathrm{eff}^*)} + c_0} \quad \text{and} \quad P(\delta_\mathrm{eff}^*) = c_2 \delta_\mathrm{eff}^{*2}+c_3 \delta_\mathrm{eff}^{*3} +c_4 \delta_\mathrm{eff}^{*4}
\end{align}
with numerical coefficients:
\begin{align}
c_0 &= -0.484\,, & c_2 &= 0.340\,,  & c_3 &=1\,, & c_4&=-0.334\,, &[\mathrm{O}(4)\, \mathrm{symmetry}] \\
c_0 &= 0.227\,, & c_2 &= -0.031\,,  & c_3 &=0.758\,, & c_4&=-0.220\,, &[\mathrm{O}(3)\, \mathrm{symmetry}] 
\end{align}
\end{subequations}
This indeed confirms that $|\bar{V}_- |\to 0$ in the thick-wall limit ($\delta_\mathrm{eff}^* \to 0$)  and $|\bar{V}_- |\to 1$ in the thin-wall limit  ($\delta_\mathrm{eff}^* \to 2$). 

\begin{figure}[t]
    \centering
    \includegraphics[width=0.7\textwidth]{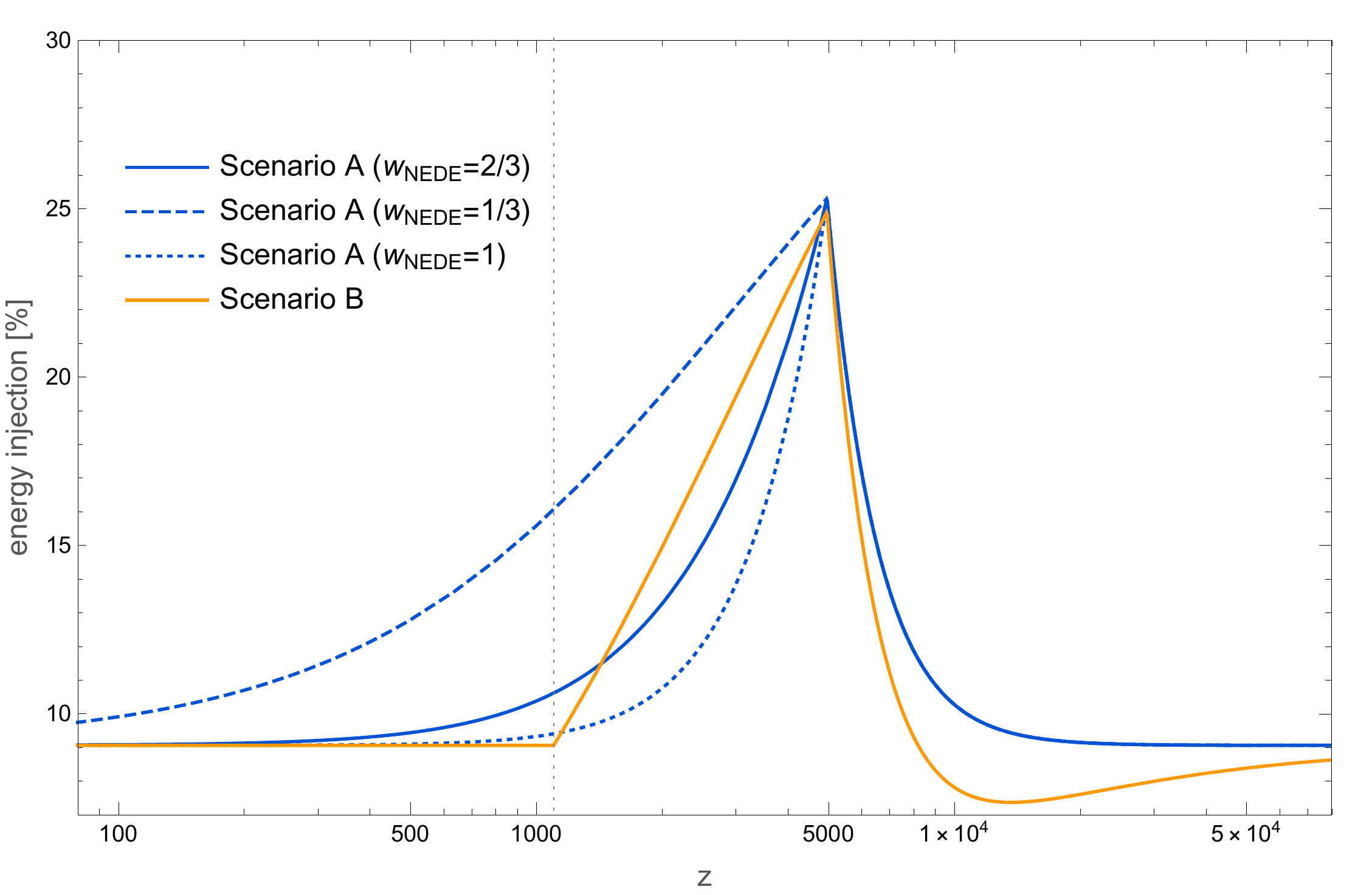}
    \caption{Relative energy injection compared to $\Lambda$CDM. Scenario A (blue curves) relies on a fluid with constant equation of state $w_\mathrm{NEDE}$ between $1/3$ (dashed) and $1$ (dotted). The best fit to data is achieved for $w_\mathrm{NEDE} \simeq 2/3$ (solid line) with parameters taken from Tab.~1 in~\protect\cite{Niedermann:2020dwg}. Scenario B (orange) relies on a mixed DM model where all of the NEDE condensate decays into relativistic particles that become non-relativistic shortly after at $z_{rel}$ (dotted vertical line). For our numerical example, the value of $z_\mathrm{rel}$ is adjusted to achieve the same reduction in $r_s$ while keeping the other parameters fixed. This leads to an energy injection which is similar to the $w_\mathrm{NEDE} \simeq 2/3$ fluid.}
    \label{fig:injection}
\end{figure}

Now, for scenario A, we assume that $\Delta \rho_\mathrm{wall}(t_*) \gg \Delta \rho_\mathrm{osc}(t_*) $, which is fulfilled for $\delta_\mathrm{eff}^*  \gtrsim 1.1$ and  $\delta_\mathrm{eff}^*\gtrsim 1.5$ in the $\mathrm{O}(4)$  and $\mathrm{O}(3)$ symmetric case, respectively. In particular, $\Delta \rho_\mathrm{osc}(t_*)$ becomes hierarchically suppressed in the thin-wall limit $\delta_\mathrm{eff} \to 2$. In any case, we can neglect the contribution from the oscillations and describe NEDE as a single fluid with a time-dependent equation of state parameter $\omega_\mathrm{NEDE}(t)$, sound speed $c_s(t)$ and viscosity parameter characterizing the colliding bubble wall condensate on large scales. The physical reason is that bubbles are small on cosmological length scales when they start to collide with typical size $\bar \beta^{-1} \ll 1/H_*$ (in the plasma-dominated regime where bubble walls are moving with non-relativistic speeds they would remain even smaller). As a result, there is no preferred direction or location on large scales, and the system admits a description in terms of a homogenous and isotropic fluid. On background level, the cosmological model becomes
\begin{subequations}
\begin{align}
\rho_\mathrm{tot} = \rho_\Lambda + \rho_\mathrm{m}+ \rho_\mathrm{rad} +  \rho_\mathrm{NEDE}\,,
\end{align}
where the first three terms denote the cosmological constant $\rho_\Lambda$, matter $ \rho_\mathrm{m}$ and radiation component $\rho_\mathrm{rad} $, respectively. The NEDE component can be parametrized in general as
\begin{align}
\rho_\mathrm{NEDE}(a) = \rho_\mathrm{NEDE}^* 
\begin{cases}
1 & \text{for} \quad a< a_* \,,\\
\exp\left\{ -3 \int_{a_*}^a \frac{\mathrm{d}a}{a} \left[1+w_\mathrm{NEDE}(a)\right]\right\} & \text{for} \quad a> a_* \,,
\end{cases}
\end{align}
\end{subequations}
where at early times the false vacuum energy acts like a cosmological constant term $\rho_\mathrm{NEDE}^*  = \mathrm{const}$. For example, assuming that  $\omega_\mathrm{NEDE}(a) \simeq \mathrm{const}$ in the first efolds ensuing the decay, it was found that  the cold NEDE decay provides the best data fit for $w_\mathrm{NEDE} \simeq 2/3$ corresponding to a decay $\rho_\mathrm{NEDE} \propto 1/a^5$.\footnote{Fitting to Planck, BAO, DR and SH$_0$ES, it was found that $w_\mathrm{NEDE} = 0.70^{+0.12}_{-0.16}$~\cite{Niedermann:2020dwg}. This seems to be a fairly robust statement among different single-fluid EDE models, although it has been challenged by recent analyses including ACT data~\cite{Poulin:2021bjr,Moss:2021obd}.} Using the results from the cosmological parameter extraction in \cite{Niedermann:2020dwg}, we compare the bestfit cosmologies within cold NEDE and $\Lambda$CDM by plotting $[\rho_\mathrm{tot}(\Lambda\mathrm{CDM})-\rho_\mathrm{tot}(\mathrm{NEDE})]/\rho_\mathrm{tot}(\Lambda \mathrm{CDM})$ in Fig.~\ref{fig:injection}  as the blue solid line. We see that $\rho_\mathrm{tot}$ is universally shifted to higher values accounting for a $5 \%$ increase in $H_0$. The distinctive feature of NEDE, however, is the energy injection before matter radiation equality that peaks rather sharply around $z_*\simeq 5000$. It is needed for a sufficient reduction of the sound horizon $r_s$ while allowing to preserve (or even improve) the CMB fit. For comparison, we also plot the energy injection corresponding to a decay like radiation ($w_\mathrm{NEDE} = 1/3$; blue dashed) and a stiff fluid ($w_\mathrm{NEDE} = 1$; blue dotted), which both yield a worse fit.

In the case of hot NEDE, we have seen that the regime with $\gamma \lesssim 1$ is able to provide a large enough fraction of NEDE that dominates over the DR fluid. Although we could not study that regime in detail in Fig.~\ref{fig:V_minus} (which was derived for $\gamma \ll 1$), due to the similarity of the potentials in Fig.~\ref{fig:Potential} and \ref{fig:Potential_large_m}, we expect a very similar phenomenology to emerge when the thin-wall limit is approached as $ \tilde \delta_\mathrm{eff}^* \to \pi $ (the precise numerical threshold on $\tilde \delta_\mathrm{eff}^*$ will depend on $\gamma$ though). In other words, the wall tension will again dominate over the oscillatory component. {While we expect the small scale anisotropic stress of the bubble walls to give the stiff behavior of the NEDE fluid after the transition, for both hot and cold NEDE it remains an important problem to find a precise description of the effective fluid parameters in terms of the microscopic parameters characterizing the bubble walls} in order to further constrain the model and better understand the conditions required for realizing a fluid with $w_\mathrm{NEDE} \simeq 2/3$. In fact, in \cite{Niedermann:2020dwg} it was argued that values $w_\mathrm{NEDE} > 1/3$ can be expected from integrating out small-scale anisotropies.

An obvious generalization of this scenario, which was also pointed out in \cite{Niedermann:2020dwg}, consists in considering a two-fluid model. This would enable us to go away from the thin-wall limit and accommodate a sizable fraction of $ \Delta \rho_\mathrm{osc} $, which could be modeled as a DM component that arises at decay time. We  leave a more detailed investigation of this possibility to future work.  A further assumption underlying scenario A is that the microscopic decay of bubble walls into gravitational waves,  gauge bosons (in the case of hot NEDE) or other decay produces (if present) is not completed within the time window where the fluid leaves its phenomenological imprint. Otherwise, the decaying NEDE fluid would require a different description in terms of a more complicated fluid.  The next scenario can be understood as a novel approach that accommodates one of these additional decay channels (and is independent of the distinction between $\Delta \rho_\mathrm{wall}$ and $ \Delta \rho_\mathrm{osc} $). 

\subsubsection{Scenario B (mixed DM model)}

Within this scenario, we make the assumption that the bubble wall condensate very quickly dissipates through decays of $\psi$ quanta into decoupled particles with masses $m_\mathrm{light} < m_\psi $. This is a non-thermal process that produces particles with  kinetic energies of order of $m_\psi/2 $. For a decay rate $\Gamma \gg H$, we can model this as an instantaneous event taking place simultaneously with the bubble percolation at $z=z_*$. Interestingly, this can lead to a background evolution of $\rho_\mathrm{tot}$ that is very similar to scenario A with $w_\mathrm{NEDE} = 2/3$ (which led to a successful resolution of the Hubble tension within cold NEDE). The crucial idea is that the light particles produced through the decay turn non-relativistic at a later time $z_\mathrm{rel} > z_*$. This happens if they have a mass difference of order of $ (m_\psi - m_\mathrm{light}) / m_\psi \simeq   (z_*-z_\mathrm{rel} )/ z_*$. As a consequence, NEDE will be ultimately converted into pressureless matter that makes up a fraction $x_\mathrm{NEDE}$ of the overall DM density.  In a first implementation, we describe the transition from the relativistic to the non-relativistic regime as a discontinuity in the equation of state parameter.\footnote{We stress that this simple model is only used to illustrate the mechanism. A more complete implementation will track the Boltzmann evolution of the corresponding particle distribution.}  The corresponding fluid model becomes
\begin{align}
\rho_\mathrm{NEDE}(a) = \rho_\mathrm{NEDE}^* 
\begin{cases}
1 & \text{for} \quad a_* > a \\
\left( \frac{a_*}{a} \right)^4  & \text{for} \quad a_\mathrm{rel} > a >  a_* \\
\frac{a_*}{a_\mathrm{rel}}\left( \frac{a_*}{a} \right)^3  & \text{for} \quad a >  a_\mathrm{rel} 
\end{cases}
\end{align}
where $a_*=a(z_*)$ and $a_\mathrm{rel}=a(z_\mathrm{rel})$. Now, we require the overall amount of DM for $z < z_\mathrm{rel}$ to be the same as in scenario A. After taking $z_*\simeq 5000$, we are then left with $z_\mathrm{rel}$ and $f_\mathrm{NEDE}$ as the only undetermined parameters. We fix them by demanding the same reduction of $r_s$ relative to the $\Lambda$CDM cosmology, which is needed for the increase in $H_0$ (the remaining $\Lambda$CDM parameters simply match their bestfit values from before). A numerical example with $z_\mathrm{rel} = 1000$ and $f_\mathrm{NEDE} = 16.3 \%$ is provided by the orange curve in Fig.~\ref{fig:injection}. We  indeed see that this model leads to a pronounced energy injection with a similar fall-off as the one found in scenario A with $w_\mathrm{NEDE} = 2/3$ (blue solid). In particular, it is steeper than the fall-off in the pure radiation model with $w_\mathrm{NEDE} =1/3$ (blue dashed). Physically, this can be understood as a consequence of a reduced cold DM fraction for $z>z_\mathrm{rel}$. In fact, for our numerical example, we find that NEDE makes a $x\simeq 13 \%$ contribution to the DM density.\footnote{Incidentally, this is roughly the difference between the DM density within cold NEDE and $\Lambda$CDM. In scenario B the increase in $\omega_\mathrm{cdm}$ typical for EDE type models is therefore caused by the NEDE fluid at late times. It is intriguing to speculate that this in itself might help with the $S_8$ tension as short scales will not feel this increase (because it occurs when they have already entered the horizon). } In other words, the microscopic model underlying scenario B is providing a potential explanation for the stiff fluid behavior favored within scenario A and observed throughout the literature. A decay of the scalar field condensate works with both cold and hot NEDE irrespective as to whether the condensate is dominated by scalar field oscillations or the wall tensions. We note that this mechanism could also be implemented in old EDE as a way of avoiding an anharmonic potential. In that case, however, one would encounter the problem of achieving $\Gamma \gg H$ without having the EDE condensate decay too early. In contrast, the NEDE condensate only forms during the phase transition and thus it can decay immediately. In fact, in the next section, we will argue that such a decay is very naturally born out of a first order phase transition.  To that end, we will provide two explicit candidates for such a rapid decay channel. While both scenarios yield a similar background modification as compared to $\Lambda$CDM, the perturbation sector is going to be different. It is therefore a priority of our future work, to implement this idea in a Boltzmann code to test it against data.

\section{Microphysical description}\label{sec:microphysics}

Here, we study explicit microscopic models that incorporate hot NEDE and offer a more complete description of the dark sector.  In Sec.~\ref{sec:NADM}, we introduce a dark gauge sector, providing the thermal fluid needed for the hot NEDE phase transition. At the same time, this model introduces an interacting DM candidate, which has the potential to resolve the $S_8$ tension. In Sec.~\ref{sec:neutrino}, we tie the hot NEDE phase transition to the mass generation of a sterile neutrino. Finally, in Sec.~\ref{sec:microscopic_model} we start a first attempt at marrying hot NEDE with a more complete dark sector model that explains the active neutrino masses through a spontaneous breaking of lepton number symmetry. 

\subsection{Interacting dark matter and dark radiation}\label{sec:NADM}

We consider a Higgs model charged under an $\mathrm{SU}(N)$ [or U($1$)] dark gauge group along with a DM field $\chi$; explicitly,
\begin{subequations}
\label{NADM_model}
\begin{align}
\mathcal{L} = - \frac{1}{4} F^2 - \bar{\chi}\left(\mathrm{i} \slashed{D} + M_\chi\right)\chi - |D \Psi |^2 - V(|\Psi|^2)  
\end{align}
with tree-level potential
\begin{align}
V(|\Psi|^2) = -\mu^2 |\Psi|^2 +  \lambda |\Psi|^4\,,
\end{align}
where $\Psi$ is an $\mathrm{SU}(N)$ multiplet and $D_\mu$ the gauge-covariant derivative

\begin{align}
{D}_\mu \Psi = \left( \partial_\mu - \frac{\mathrm{i}}{2} g_d A_\mu \cdot T  \right)\Psi
\end{align}
\end{subequations}
with dark gauge coupling parameter $g_d$ and generators $T^a$.
The microscopic parameters $\mu$, $\lambda$ and $g_\mathrm{d}$ can be related to the parameters of the finite-temperature effective potential \eqref{eq:effective_T_pot} or \eqref{eq:effective_T_pot_low_T} by calculating the thermal corrections to the model's  action in \eqref{NADM_model} and identifying  $g_d = g_\mathrm{NEDE}$. We also recall that $ \psi = \sqrt{2} |\Psi|$. An explicit perturbative calculation in the case of $\mathrm{U}(1)$ is provided in the Appendix \ref{appendix_potential}. This dark sector has all the ingredients needed for a thermal first-order phase transition as needed for hot NEDE.

This gauge sector can be naturally incorporated in the (N)ADM model~\cite{Buen-Abad:2015ova,Lesgourgues:2015wza,Buen-Abad:2017gxg}, which contains a mechanism to  heat the dark sector to temperatures of the order of $T_d \sim T_\mathrm{vis}$. The general idea is to introduce a fermionic DM field $\chi$ that is charged under the $\mathrm{SU}(N)$  (with $N\geq2$) and transforms in its fundamental representation. It also transforms as the neutral component of an $\mathrm{SU}(2)_\mathrm{SM}$ weak triplet, making it a WIMP DM candidate. To be precise, it is a ``wino''-like DM candidate, although additional multiplicity factors change the sensitivity of experimental searches~\cite{Buen-Abad:2015ova}. For $N=2,3,4$ it was shown that DM will be produced with the right abundance from thermal freeze-out for a DM mass of the order of $M_\chi = 1.2,1.0,0.9~\mathrm{TeV}$, respectively. As argued in \cite{Lesgourgues:2015wza}, this model can also be reduced to the Abelian case with an $\mathrm{U}(1)$ symmetry (N=1) giving rise to a single dark photon field. Moreover, different $\mathrm{SU}(2)_\mathrm{SM}$ representations could lead to equally suitable DM candidates~\cite{Cirelli:2005uq}.

In these cases, the dark gluons/photons will be in thermal equilibrium with the SM at high temperatures above the DM mass. After the DM freeze-out, the gluons/photons decouple and evolve with lower temperature $T_d(T_\mathrm{vis}) = [g_\mathrm{rel, vis} (T_\mathrm{vis})/g_\mathrm{rel, vis}(T_\mathrm{vis}^\mathrm{dec})]^{1/3} T_\mathrm{vis}$, where $T^\mathrm{dec}_\mathrm{vis}$ is the visible sector temperature at decoupling time. With the assumption that the DR fluid decouples with a temperature between $10$ and $100~\mathrm{GeV}$, we have 
\begin{align}
\frac{T_d}{T_\nu} = \frac{T_d}{T_\mathrm{vis}}\Big|_{T_\mathrm{vis} \sim \mathrm{Mev}} = \left[\frac{2+ \frac{7}{8} \times 10}{18+ \frac{7}{8} \times 90}\right]^{1/3}\,
\end{align}
where $T_\nu$ is the temperature of the neutrino sector, which decouples at $T_\nu = T_\mathrm{vis}\sim \mathrm{Mev}$.
Before the hot NEDE phase transition, this translates to $\xi = T_d/T_\mathrm{vis} =T_d/T_\mathrm{\nu} (4/11)^{1/3}\simeq 0.34$ and was used as a benchmark value in Fig~\ref{fig:params} and~\ref{fig:params_large_mass} to infer the value of $f_\mathrm{NEDE}$. It is also conventionally expressed as a contribution to the effective equivalent number of neutrino species 
\beq\label{Delta_Neff}
\Delta N_\mathrm{eff} = N_d \frac{8}{7} \left( \frac{11}{4}\right)^{4/3} \xi^4\simeq 0.06 N_d,
\eeq
where $N_d $ is the number of dark, massless gauge bosons. For $N \geq 2 $, it amounts to $N^2-1$ before and $(N-1)^2-1$ after the transition, corresponding to the emergence of $2N-1$ massive gauge bosons. For $N \geq 3$, they decay into the remaining massless gauge bosons, leading to a dark temperature increase by a factor $[(N^2-1)/(N^2-2N)]^{1/3}$. The case $N=2$ has to be treated separately. Here, only one massless boson remains, corresponding to a residual $\mathrm{U}(1)$ and the emergence of two massive gauge bosons. Finally, for $N=1$ no massless boson remains (unless $\Psi$ is not charged under the gauge group, which corresponds to the scenario considered later in Sec.~\ref{sec:microscopic_model} and requires a different origin of the temperature corrections).  
Citing from \cite{Lesgourgues:2015wza}, a fit of the (N)ADM model to CMB, BAO  and LSS data yields an upper bound $\Delta N_\mathrm{eff}  < 0.67 $ at $95\%$ C.L., which singles out  $N=1$ (with $\Delta N_\mathrm{eff} \simeq 0.06$), $N=2$ (with $\Delta N_\mathrm{eff} \simeq 0.18$ before and $\Delta N_\mathrm{eff} \simeq 0.26 $ after the transition), and $N=3$ (with $\Delta N_\mathrm{eff} \simeq 0.48$ and $\Delta N_\mathrm{eff} \simeq 0.67 $ after the transition) as the relevant cases.\footnote{The phenomenological bounds might change within hot NEDE, although we still expect $N > 3$ to be excluded.} Although the value of $\Delta N_\mathrm{eff}$ after the transition seems to marginally violate the bound above for $N=3$, we expect it to have no phenomenological consequences as the transition occurs close to matter-radiation equality where observables are less sensitive to the exact radiation density. For $N=2,3$, the dark sector temperature after the decay is increased to $\xi \simeq 0.5$, while it remains constant for $N=1$.

The DM particles will experience a drag force as they move through the dark gluon/photon soup. The drag coefficient can be computed to be~\cite{Buen-Abad:2015ova,Lesgourgues:2015wza} 
\begin{subequations}
\begin{align}
\Gamma^\mathrm{DM-DR} = N_d \Gamma^\mathrm{DM-DR}_0 \, \frac{ T^2_\mathrm{vis}}{T_{\mathrm{vis},0}^2} \left[\frac{g_{\mathrm{rel},d}(T_\mathrm{vis})}{g_{\mathrm{rel},d}(T_{\mathrm{vis},0})}\right]^{2/3}
\end{align}
where 
\begin{align}
\label{eq:Gamma_0}
\Gamma_0^\mathrm{DM-DR} 	&=\frac{\pi}{9}\al_d^2\log{\al_d^{-1}}\left.\frac{T_d^2}{M_X}\right|_\mathrm{today} \nonumber\\
			&= 1.2 \times 10^{-7} \mathrm{Mpc}^{-1} \left[ \frac{\alpha_d^2 \log\alpha_d^{-1}}{2.0 \times 10^{-16}}\right] \left[ \frac{1.2 \mathrm{TeV}}{M_\chi}\right]
\end{align}
\end{subequations}

is the coefficient today for a single gauge boson, and we slightly generalized the formula to account for its dependence on the number of dark, relativistic degrees of freedom $g_{\mathrm{rel},d}$ and the number of dark, massless gauge bosons $N_d$. The drag force is discontinuous across the transition due to the change in $g_{\mathrm{rel},d}$ and $N_d$. For $N \geq 3$ it gets reduced by a factor $\left[(N^2-2N)/(N^2-1) \right]^{2/3}$, which evaluates to $\simeq 0.72$ for $N=3$ (a modified calculation yields a similar decrease for $N=2$). For the numerical estimate in \eqref{eq:Gamma_0} we used $\alpha_d = (g_d)^2 / (4 \pi) \simeq 3 \times 10^{-9}$, which is the value suggested by previous analyses studying the (N)ADM model alone. In general, LSS observations constrain $g_d < 10^{-3}$~\cite{Buen-Abad:2015ova}. As a result, the dark gauge sector is always in the deconfined, weakly-coupled phase, which sets this implementation of hot NEDE apart from other proposals that study a dark confinement phase transition. In fact, this latter possibility was touched upon in~\cite{Allali:2021azp} and might provide another interesting scenario for realizing hot NEDE.

The model's potential to resolve the $S_8$ tension derives from the DM-DR drag force. To be more specific, the positive radiation pressure leads to a damping of matter density perturbations  and thus lowers $S_8$. Crucially, the momentum transfer rate between DM and DR scales as $T_\mathrm{vis}^2$, which is the scaling of $H$ during radiation domination. This implies that the damping effect acts equally on all scales that enter the horizon during radiation domination, but turns off shortly after when $H$ decreases slower as $T_\mathrm{vis}^{3/2}$ during matter domination. This avoids potential problems that would otherwise arise by changing the integrated Sachs-Wolfe effect. 

While the phenomenology of (N)ADM has been studied in a series of papers~\cite{Buen-Abad:2015ova,Lesgourgues:2015wza,Buen-Abad:2017gxg,Archidiacono:2019wdp}, showing promise as a resolution to the $S_8$ tension, these studies need to be updated to include the effect of the hot NEDE phase transition. In particular, the $\simeq 30 \%$ decrease of the drag coefficient from before to after the phase transition is expected to significantly change the damping dynamics for modes that have entered the horizon after the phase transition. Beyond that, as the NEDE field also couples to the gauge field, we expect DR and NEDE perturbations to interact. This can be described in terms of another interaction coefficient $\Gamma^\mathrm{NEDE-DR}$, which will affect the acoustic oscillations carried by the NEDE fluid. 

After fixing $N=1, 2, 3$ the model will be described in terms of four phenomenological parameters: the drag force parametrized through $\Gamma_0$, the fraction of hot NEDE $f_\mathrm{NEDE}$, the decay redshift $z_*$ and the equation of state parameter of the decaying NEDE fluid $w_\mathrm{NEDE}$.  The dark sector temperature, on the other hand, is completely fixed as detailed above. While this model assumes that all of DM is interacting with the dark sector, this could be generalized by allowing for a fraction of non-interacting DM. Moreover, as the self-scattering rate of the DR is large compared to the Hubble rate during radiation domination, it can be modeled as an ideal (rather than free-streaming) fluid with vanishing viscosity.  The corresponding system of DR-DM perturbation equations can be found in \cite{Lesgourgues:2015wza}. 

In short, NEDE naturally accommodates (N)ADM and thus provides a simple way of using the same dark physics needed to resolve the $S_8$ and $H_0$ tension to account for DM.

\subsection{NEDE and neutrino mass generation} \label{sec:neutrino}
It has been argued that the inverse seesaw mechanism can explain the observed neutrino oscillation data and mass spectrum~\cite{Mohapatra:1986bd,Gonzalez-Garcia:1988okv,Deppisch:2004fa,Abada:2014vea}. It introduces a mass mixing between the active left-handed neutrinos $\nu_{L} = \left( \nu_\mathrm{e},\, \nu_\mu,\, \nu_\tau \right)^T$, right-handed neutrinos $(\nu_{R})_\alpha$ , and a set of sterile fermions $(\nu_s)_i$ lurking in the dark sector. The corresponding action contains a light Majorana mass  $ \mathrm{eV} < m_s <\mathrm{GeV}$, which is protected by a broken lepton symmetry that gets restored in the limit $m_s \to 0$. Here, we show how this low-energy sector of the inverse seesaw mechanism could dynamically arise in the hot NEDE phase transition when $\Psi $ acquires its $\mathrm{vev}$.

Writing $N \equiv (\nu_L, \nu_R^c, \nu_s)^T$, the neutrino mass term takes the form
\beq
\mathcal{L}_\mathrm{\nu} = - \frac{1}{2} N^T C M N + \mathrm{h.c.}
\eeq
where $C=i\gamma^2\gamma^0$ is the charge conjugation matrix and
\beq\label{massmatrix}
M =\left(\begin{matrix}{}
  0 & d & 0\\
  d & 0 & n \\
 0 & n & m_s
\end{matrix}\right)
\eeq
specifies the mass mixing.
It includes the matrices $d$ and $n$, mixing the right-handed with the active left-handed and the sterile neutrino, respectively. Their components are set at or above the TeV scale, specifically $n \gtrsim d\sim \mathrm{TeV}$. $M$ further contains a symmetric Majorana mass matrix $(m_s)_{ij}$ for the sterile neutrinos. Here, we will consider the case of three sterile neutrinos  $\nu_{s} = \left( \nu_{s,1},\, \nu_{s,2},\, \nu_{s,3}\right)^T$ and a family of either two or three right-handed neutrinos $(\nu_{R})_\alpha$ with $\alpha= 1,2(,3)$ . Both configurations have been identified as phenomenologically promising models~\cite{Abada:2014vea} that can explain the observed mass spectrum and mixing pattern without a fine-tuning of $M$.\footnote{There is also a more minimal version of the inverse seesaw mechanism with only two right-handed and sterile neutrinos; however, it requires a fine-tuning to fulfill the phenomenological constraints~\cite{Abada:2014vea}.} After diagonalization, this model gives rise to three light active states with masses 
\begin{align} \label{active_masses}
m_i \lesssim \mathcal{O}(m_s) \kappa^2/(1+\kappa^2)\,,
\end{align}
where $\kappa = \mathcal{O}(d)/\mathcal{O}(n$). In particular, their mass spectrum and mixing pattern is compatible with phenomenological constraints.
In addition, the model predicts the presence of heavy pairs of pseudo-Dirac neutrinos with masses $\mathcal{O}(n)$ that do not participate in the low-energy dynamics. Finally, in the case of only two right-handed steriles there is an additional  light mass eigenstate $\nu_4$  with (super-)eV mass  $m_4=\mathcal{O}(m_s)$. 

In our work, we propose that the low-energy mass matrix $m_s$ is generated through the hot NEDE phase transition. To that end, we assume that the sterile neutrinos $(\nu_s)_i$ couple to the NEDE field through a dimension-four coupling of the form\footnote{Fifth-force constraints are of no concern as the sterile is only coupled to the visible sector via gravity and the active neutrinos, which does not give rise to stringent bounds.}
\beq\label{vertex}
\mathcal{L} \supset -\frac{1}{\sqrt{2}} \sum_{ij}(g_s)_{ij} \Psi \overline{{(\nu_s)_i}^c} {(\nu_s)_j^{\phantom{c}}} + \mathrm{h.c.}~.
\eeq   
In the following, we will suppress the sterile index $i$ if not relevant (keeping in mind that $(g_s)_{ij}$ and $(m_s)_{ij}$ are matrices). Later, in section \ref{sec:microscopic_model}, we will propose a more complete dark sector model that explains the generation of the scales $d$ and $n$, too. In any event, when $\psi$ transitions to the true minimum  $\psi_\mathrm{True}$, due to \eqref{vertex}, the sterile $\nu_s$ acquires a temperature dependent mass 
\begin{align}\label{sterile_m}
m_s(T_d) = 
	\begin{cases}
	0						 & \text{for} \quad T_d > T_d^*\\
	 g_s \psi_\mathrm{true}(T_d) 	 & \text{for} \quad T_d < T_d^*
	\end{cases}
\end{align}
From \eqref{psimin} and \eqref{psi_true_large_mass}, we obtain after using~\eqref{dimless_vars} and \eqref{eq:dict}
\begin{align}
\psi_\mathrm{true}(T_d) =
\begin{cases}
 \frac{1}{2} \frac{E T_d}{\lambda} \left[ 3+ \sqrt{9 - 4 \delta_\mathrm{eff}(T_d)} \right] & \text{for} \quad \gamma \gg 1 \,,\\
\frac{E \sqrt{\pi \gamma}  }{\lambda} T_\circ & \text{for} \quad \gamma \lesssim 1\,.
\end{cases}
\end{align}
If we are interested in its mass today, we can use that $T^2_\circ / T^2_d \big|_\mathrm{today} \sim 10^{6} \gg 1 $, which in turn allows us to substitute for $\delta_\mathrm{eff}$ with \eqref{eq:delta_late}. Further plugging in the values of $D$ and $E$ in \eqref{def:D_E_Abelian_Higgs} as they arise in the Abelian Higgs model, we obtain
\begin{align}\label{true_zero_temp}
\psi_\mathrm{true} \big|_\mathrm{today}  =\frac{\sqrt{\pi}}{g_\mathrm{NEDE}\sqrt{\gamma}} T_\circ \,\,\, (=\mu/\sqrt{\lambda} = v_\Psi)\,,
\end{align}
which as expected holds irrespective of the value of $\gamma$ (because both potentials have the same zero temperature limit).
Next, we have from \eqref{delta_eff} that $T_\circ = T_d^* \sqrt{1-\frac{\delta_\mathrm{eff^*}}{2} \frac{E^2}{\lambda D}}$, which together with \eqref{eq:T_d_star_large_mass} and \eqref{def:D_E_Abelian_Higgs}  yields the final expression
\begin{align}\label{mass_sterile}
m_s \simeq (1.0 \, \mathrm{eV}) \times \frac{1}{\gamma^{1/4}} \frac{g_s}{g_\mathrm{NEDE}} \left[1-\frac{\delta^*_\mathrm{eff}}{\pi \gamma} \right]^{1/2} \left[\frac{f_{\textrm{NEDE}} /(1-f_{\textrm{NEDE}} )}{0.1}\right]^{1/4} \left[ \frac{1+z_*}{5000}\right]\,,
\end{align}
which for $\gamma \lesssim 1$  holds for $T_d < T_d^*$.   
We see that for $\gamma \lesssim 1$ and $g_s > g_\mathrm{NEDE}$ the natural expectation is to get a sterile mass above the $\mathrm{eV}$ scale. This would be harder to achieve in the small-mass/high-temperature limit for which $\gamma \gg 1$. To be specific, the requirement of having $m_s \gtrsim \mathrm{eV}$ (along with $f_\mathrm{NEDE}  \simeq 10\%$, $z_* \simeq 5000$ and $\delta^*_\mathrm{eff} \sim \gamma $) translates to the parameter condition
\begin{align}\label{lambda_constraint}
\lambda \lesssim \frac{g_s^4}{4 \pi}\,,
\end{align}  
where we used $ (4 \pi \lambda)^{1/4} = \gamma^{1/4} g_\mathrm{NEDE}$.

For our phenomenological discussion, it will be useful to relate the sterile mass scale to the vacuum mass of the NEDE field through  
\begin{align}\label{m_psi}
m_\psi \big|_\mathrm{today} 
&= \sqrt{2} \mu  = \frac{g_\mathrm{NEDE}^2}{g_s} \frac{\sqrt{\gamma}}{\sqrt{2 \pi}}\, m_s\,,
\end{align}
where we used \eqref{true_zero_temp}.
We impose the mass ordering $m_\psi \lesssim \mathrm{eV} \lesssim m_s $, where the upper bound makes our model compatible with neutrino observations and the lower bound ensures that dark sector thermal corrections $\propto T_d^2 < \mathrm{eV}$ become sizeable relative to the vacuum mass above $T_\mathrm{vis}\sim \mathrm{eV}$. In fact, for small couplings and/or a cold dark sector with $\xi \ll 1$ this even requires $m_\psi \ll \mathrm{eV}$. To see this, we can use \eqref{mass_sterile} to eliminate $g_s$ in \eqref{m_psi}, which then implies that 
\begin{align}\label{bound_m_psi}
m_\psi/\mathrm{eV} \sim g_\mathrm{NEDE} \gamma^{1/4} \ll 1\,.
\end{align}
Another, independent combination of \eqref{mass_sterile} and \eqref{m_psi} fixes $g_s$ in terms of both mass scales as
\begin{align}\label{gs_constraint}
g_s \sim  \frac{m_s}{\mathrm{eV}} \frac{m_\psi}{\mathrm{eV}}\,.
\end{align}

\begin{figure}
     \centering
     \begin{subfigure}[b]{0.49\textwidth}
         \centering
         \includegraphics[width=\textwidth]{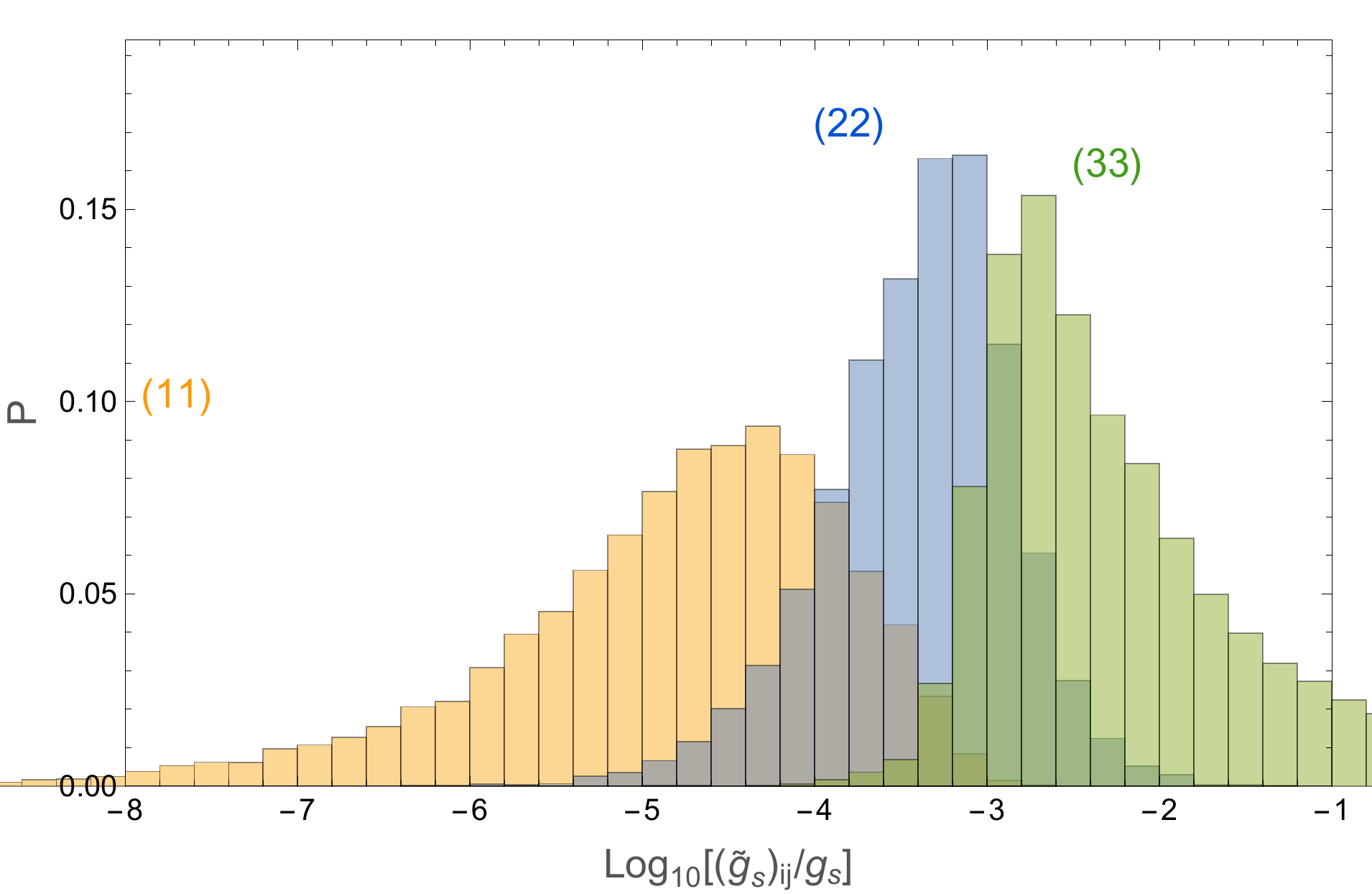}
         \caption{Diagonal elements $(11)$, $(22)$, $(33)$}
         \label{fig:hist1}
     \end{subfigure}
     \begin{subfigure}[b]{0.49\textwidth}
         \centering
         \includegraphics[width=\textwidth]{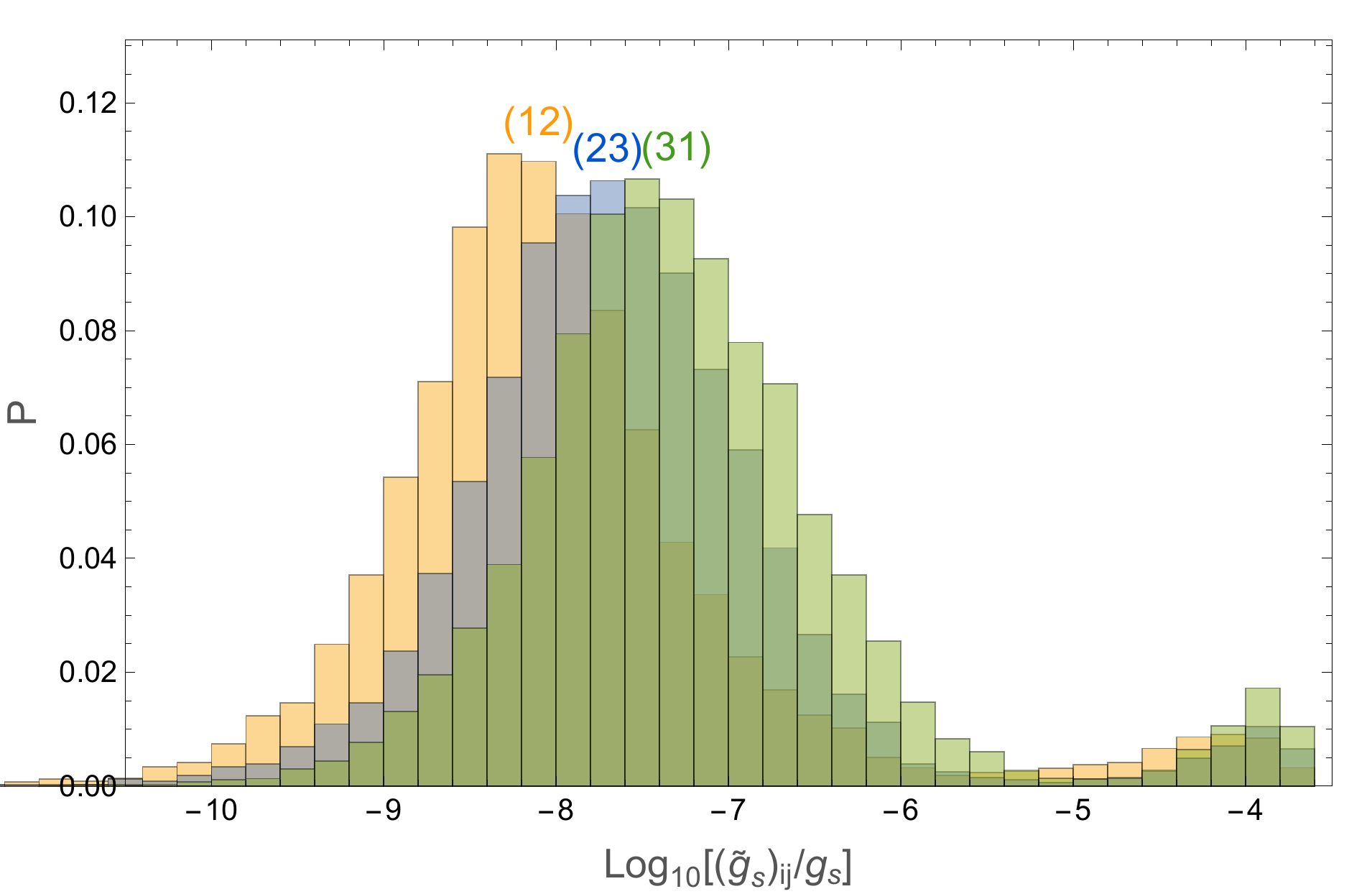}
          \caption{Off-diagonal elements $(12)$, $(23)$, $(31)$}
         \label{fig:hist2}
     \end{subfigure}
          \caption{Distribution of the coupling matrix $(\tilde g_s)_{ij} /g_s $ in the mass eigenbasis for $m_s = \mathcal{O}(100\,\mathrm{ eV})$, $d = \mathcal{O}(200\,\mathrm{GeV})$, $n = \mathcal{O}(10\, \mathrm{TeV})$ and $\kappa^2 \sim 10^{-3}$. For most cases, the off-diagonal terms are suppressed by more than four orders of magnitude, making them compatible with their stronger phenomenological constraints.  }
        \label{fig:hist}
\end{figure} 

The most stringent bound on $g_s$ comes from cosmology. It arises because the sterile neutrinos $\nu_s$ have a non-vanishing overlap with the active mass eigenstates $(\nu_1, \nu_2, \nu_3)$, which leads to late changes in their Boltzmann evolution~\cite{Archidiacono:2013dua}. To be precise, we can rewrite \eqref{vertex} in the mass eigenbasis as
\beq\label{vertex2}
-\frac{1}{\sqrt{2}}\sum_{ij}(g_s)_{ij} \Psi \overline{{(\nu_s)_i}^c} {(\nu_s)_j^{\phantom{c}}} \supset 
-\frac{1}{\sqrt{2}} \sum_{kl}(\tilde g_s)_{kl} \Psi \,\overline{{\nu_k}^c} {\nu_l^{\phantom{c}}} \,,
\eeq   
where $(\tilde{g}_{s})_{kl} =  \sum_{ij} U_{ki} U_{lj} (g_s)_{ij} $ is the rotated coupling matrix and $U_{ki}$ are the sterile-active components of the leptonic mixing matrix. Cosmological data then puts a limit on the diagonal elements, $(\tilde g_s)_{kk} \lesssim 10^{-7}$, and a significantly stronger one on the off-diagonal elements, $(\tilde g_s)_{kl} \lesssim 10^{-11}$ with $k\neq l${, although newer work indicates weaker bounds~\cite{Barenboim:2020vrr}}.  To estimate the size of $(\tilde g_s)_{kl}$,  we numerically diagonalized the neutrino mass matrix in \eqref{massmatrix} while requiring the sub-matrices $d$, $n$ and $m_s$ each to be set by a different scale subject to the hierarchy $\mathcal{O}(m_s)< \mathcal{O}(d)<\mathcal{O}(n)$. Individual entries within a submatrix were allowed to vary randomly within a range of the same order. Repeating this process for a large number of different mass matrices then admits a probabilistic statement about the mass spectrum and hence $(\tilde g_s)_{kl}$. We find that the diagonal entries are $(\tilde g_s)_{kk} = \mathcal{O}(g_s) \times \kappa^2$, where we assumed that all elements of $(g_s)_{ij}$ are set by a single scale $g_s$.\footnote{There is an additional mild hierarchy between different diagonal elements, $(\tilde g_s)_{11}<(\tilde g_s)_{22}<(\tilde g_s)_{33} =  \mathcal{O}(g_s) \times \kappa^2$, which is not relevant for this work though. A similar statement holds for the light neutrino masses.} The off-diagonal entries, on the other hand, do not only depend on $\kappa$ but also the ratio between $m_s$ and $d$. Provided $ 10^{-2} \lesssim \kappa < 1 $ and $\mathcal{O}(m_s)/\mathcal{O}(d) \gtrsim 10^{-13}/\kappa^2$, they are in the majority of cases suppressed with respect to their diagonal counterparts by at least four orders of magnitude, naturally accommodating their stronger bounds. We depict the distributions for a relevant example with $m_s = \mathcal{O}(100\,\mathrm{eV})$ in Fig.~\ref{fig:hist}. As a result, the cosmological constraints can be recast {conservatively} as 
\begin{align}\label{bound_gs}
g_s < 10^{-7} / \kappa^2\,,
\end{align}
which holds for both off-diagonal and diagonal elements. This, in turn, translates into a bound on $m_\psi$,
\begin{align}\label{bound_m_psi_2}
\frac{m_\psi}{\mathrm{eV}} \lesssim 10^{-7} \left( \frac{\mathrm{eV}}{m_i} \right)\sim 10^{-5}\,,
\end{align}
where we used \eqref{active_masses} and \eqref{gs_constraint}. It is compatible with \eqref{bound_m_psi} for a sufficiently small gauge coupling. We note that the same numerical analysis allowed us to reproduce the result from the literature cited in~\eqref{active_masses}.

Depending on the number of right-handed neutrinos and the scale of $m_s$, we discuss three complementary phenomenologies related to the sterile sector:

\begin{enumerate}

\item \textbf{No light sterile:}
In its conventional form the inverse seesaw mechanism introduces the same number of right-handed and sterile neutrinos. In particular, there is no additional light eigenstate $\nu_4$ in the spectrum (despite the occurrence of the low energy mass scale $m_s$ in the interaction basis). Instead, it leads to three pairs of heavy pseudo-Dirac neutrinos with masses  above the TeV scale that do not contribute to the low-energy dynamics. From a phenomenological perspective, this is the simplest scenario as the steriles are only important for generating the light neutrino masses.

\item  \textbf{eV mass sterile:}\footnote{If a distinction is not relevant, we follow the standard convention in the literature and use the name ``sterile'' also for the fourth mass eigenstate $\nu_4$.}

If the model only contains two right-handed neutrinos, the spectrum  contains an additional light mass eigenstate $\nu_4$ with mass  $m_4=\mathcal{O}(m_s)$. In this case, we quantify the small mixing between $\nu_4$ and the active neutrinos $\nu_L$ through the angle $\sin^2 2 \theta_4 = 4 \left( |U_{\mathrm{e}4}|^2 +|U_{\mu 4}|^2+|U_{\tau 4}|^2\right) \ll 1$. Using the same analysis as before, we find, in agreement with the literature, that its distribution is peaked around $\mathcal{O}(1) \kappa^2 \ll 1$. The smallness of  the mixing implies that $\nu_4$ is almost completely sterile with $\nu_4 \simeq \sum_i c_i (\nu_s)_i$. For a moderately small overlap of the order of $\sin^2 2 \theta_4 \sim 0.05$, this scenario was claimed to resolve oscillation anomalies in reactor~\cite{Mention:2011rk}, accelerator~\cite{Gariazzo:2015rra,Gonzalez-Garcia:2015qrr} and gallium experiments~\cite{Acero:2007su,Giunti:2010zu} (see also~\cite{Kopp:2013vaa,Boser:2019rta,Dasgupta:2021ies} for reviews), although a debate around that possibility has recently emerged~\cite{MicroBooNE:2021rmx,MicroBooNE:2021sne,Arguelles:2021meu}.

A priori, such an $\mathrm{eV}$ mass sterile neutrino is in tension with cosmology because the thermalization through the Dodelson-Widrow mechanism~\cite{Dodelson:1993je} would lead to an unacceptable increase in $N_\mathrm{eff}$. This conclusion can, however, be avoided in the presence of a ``secret interaction'' mediated by a light scalar, pseudo-scalar or vector field which prevent it from thermalizing with the active neutrinos before their decoupling~\cite{Hannestad:2013ana,Dasgupta:2013zpn,Archidiacono:2014nda,Archidiacono:2015oma,Archidiacono:2016kkh}. 
It is intriguing to speculate that the NEDE field $\Psi$ can both give mass to the  $\mathrm{eV}$ sterile when it acquires a  $\mathrm{vev}$ in the phase transition and, at the same time, provide the annihilation channel to avoid the cosmology bounds on $N_\mathrm{eff}$. To prevent full thermalization $g_s> 10^{-6}$ is required, which leads to a viable range $10^{-6} <g_s <10^{-5} $.
While the ``secret interaction'' in \eqref{vertex} prevents a full thermalization of the sterile with the visible sector \textit{before} neutrino decoupling, a late thermalization between $\nu_4$ and the active neutrinos will still take place \textit{after} decoupling.  Moreover, sterile neutrinos couple at late times to the dark sector plasma via $\nu_s \nu_s \leftrightarrow \psi \psi$ provided $g_s \gtrsim 10^{-6}$~\cite{Archidiacono:2015oma}. As a result, the dark sector equilibrates with the neutrino sector. This process will increase $\xi$ and change the fraction of free-streaming radiation, but most importantly it will not affect the value of $\Delta N_\mathrm{eff}$ as the energy is only reshuffled between both sectors. While the reduced fraction of free-streaming radiation is an interesting signature of this scenario, the $\mathrm{eV}$ mass threatens to violate mass bounds from LSS data. This, however, can be avoided if $\nu_4$ annihilates into $\psi$ at late times (after the NEDE phase transition). In fact, from \eqref{bound_m_psi_2}, we have that $m_\psi  \ll \mathrm{eV} $, demonstrating that this process is indeed kinematically allowed (as argued in~\cite{Archidiacono:2015oma}, it is also efficient for the values of $g_s$ considered here).

For our previous numerical example with $\gamma=0.024$ (needed to obtain $f_\mathrm{NEDE}= 10\%$ when $\xi_*=0.34$), we find that  an  $\mathrm{eV}$ mass sterile requires $g_\mathrm{NEDE} \simeq 2.5 g_s$, which can be achieved for the full viable range of Yukawa couplings $g_s$. To be specific, the bound on $g_s$ translates to $ g_\mathrm{NEDE}  < 2.5 \times 10^{-5}$, which, as illustrated in the first plot in Fig.~\eqref{fig:params_large_mass}, is deep inside the thick-wall regime where $\tilde \delta_\mathrm{eff}^* \ll 1$ and the decay of NEDE is dominated by oscillations around the true minimum (rather than the colliding bubble wall condensate). In principle, this conclusion can be  avoided if we assume the dark sector to be much colder $\xi_* \ll 1 $, which due to \eqref{xi_large_mass} is compatible with $\gamma \ll 10^{-2}$, relaxing the upper bound on $g_\mathrm{NEDE}$ and making it compatible with the thin-wall limit where $g_\mathrm{NEDE} \gtrsim 0.2$ (light gray region).  In summary, this scenario includes an $\mathrm{eV}$ mass sterile neutrino while providing enough early dark energy to resolve the $H_0$ tension. 

\item  \textbf{Super-eV mass sterile:} 

We again assume the presence of only two right-handed neutrinos. Moreover, we have $\lambda \ll g_s^4/(4 \pi) $, which leads to $m_s \simeq m_4 \gg \mathrm{eV}$. Since the active neutrinos have masses $ \kappa^2 \mathcal{O}(m_s)$, this situation requires  a smaller value of $\kappa = \mathcal{O}(d)/\mathcal{O}(n)$ and thus a weaker mixing between the active and the sterile sector where  $\sin^2 2 \theta_4 \ll 10^{-2}$. As a consequence, the sterile has a negligible effect on neutrino oscillations.  If we assume that $\sin^2 2 \theta_4 $ is small enough to prevent an early thermalization of $\nu_s$ through the  Dodelson-Widrow mechanism (which predicts a production rate proportional to $ \sin^2 2 \theta_4$), this provides another viable scenario.\footnote{Instead, we could again introduce a lower bound on $g_s$ to prevent a full thermalization like for the eV mass sterile. Also note that in a different context keV steriles are considered as a warm DM candidate~\cite{Abada:2014zra}. This is not possible in our scenario where the sterile becomes massive only very late in the expansion history shortly before matter-radiation equality. It might still lead to a late DM increase though.} As before, we can have a late thermalization between $\nu_s$ and the neutrino sector, which is decoupled from the visible sector. Once the sterile acquires its mass in the NEDE phase transition, it annihilates into the lighter $\psi$.

\end{enumerate}

\subsection{Towards a complete dark sector model}\label{sec:microscopic_model}
\begin{figure}[t]
    \centering
    \includegraphics[width=0.55\textwidth]{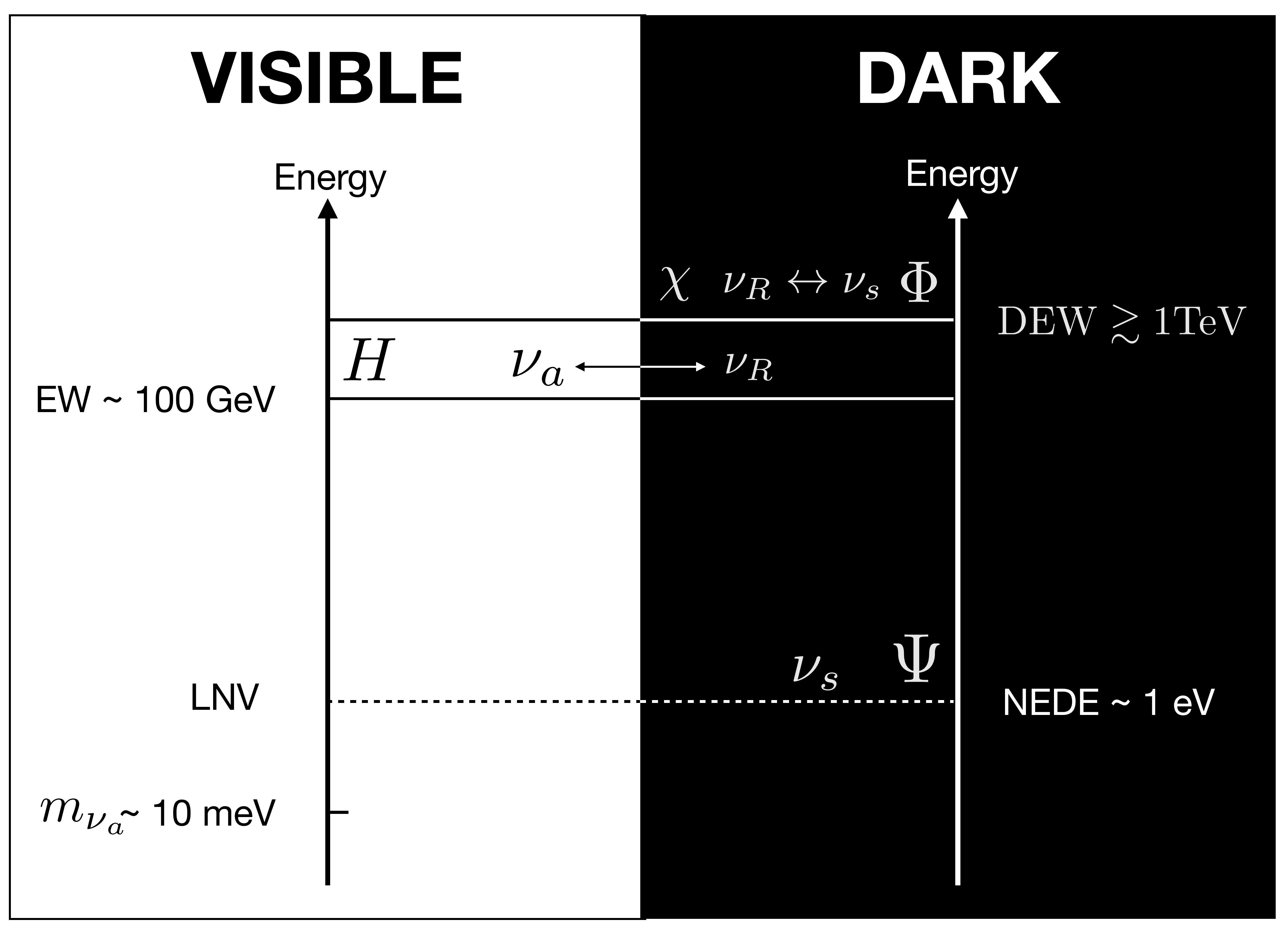}
    \caption{Schematic of the dark electroweak model (DEW). Besides the usual electroweak (EW) transition in the visible sector, we introduce two phase transitions in the dark sector at the $\mathrm{TeV}$ and $\mathrm{eV}$ scale. The first one corresponds to the DEW breaking $\mathrm{SU(2)}_\mathrm{D} \times \mathrm{U(1)_\mathrm{Y_D}} \to \mathrm{U(1)}_\mathrm{DEM}$ and the second one to a lepton number violating NEDE transition. The scalar fields that induce the transitions are the standard Higgs doublet $H$, the $\mathrm{SU(2)}_\mathrm{D}$ doublet $\Phi$ and the $\mathrm{SU(2)}_\mathrm{D}$ triplet $\Psi$. They are also responsible for creating the inverse seesaw mass matrix \protect\eqref{massmatrix}.}
    \label{fig:dark_visible}
\end{figure}

Building up on the previous section, we discuss the possibility that the NEDE phase transition is not only responsible for generating an eV-scale Majorana mass as a vital ingredient of the inverse seesaw mechanism, but also leads to a spontaneous breaking of a global $\mathrm{U(1)}_\mathrm{L}$  lepton symmetry. At the same time, we propose a dynamical mechanism to create the high-mass entries in the neutrino mass matrix $M$. To that end, we embed hot NEDE in a dark sector theory, dubbed dark electroweak model (DEW), that relies on two dark sector phase transitions, one at the $\mathrm{eV}$ and another one above the $\mathrm{TeV}$ scale, to generate \textit{all} the mass mixings needed to explain the active neutrino masses along with the mass of a super-TeV DM candidate $\chi$. {In short, the DEW model is a concrete example of a more complete dark sector that gives rise to the hot NEDE phase transition while explaining neutrino masses and featuring a DM candidate $\chi$.} A schematic overview is provided in Fig.~\ref{fig:dark_visible}. 

\subsubsection{Dark electroweak model}

The DEW model embeds NEDE into a bigger symmetry group in order to generate the Dirac mixing between three sterile neutrinos  $\nu_{s} = \left( \nu_{s,1},\, \nu_{s,2},\, \nu_{s,3}\right)^T$ and three right-handed neutrinos $\nu_{R} = \left( \nu_{R,1},\, \nu_{R,2},\, \nu_{R,3}\right)^T$. One way to do this is to assume that the steriles and the NEDE field transform in a dark $\mathrm{SU}(2)_\mathrm{D}\times \mathrm{U}(1)_\mathrm{Y_D}$ under which the SM fields, including the active and right-handed neutrinos, are singlets. Here, the index $\mathrm{Y_D}$ refers to some dark hypercharge. This symmetry group is then broken above the TeV scale from $\mathrm{SU}(2)_\mathrm{D} \times\mathrm{U}(1)_\mathrm{Y_D}$ down to a dark electromagnetism $\mathrm{U}(1)_\mathrm{DEM}$. The model  also features an approximate global $\mathrm{U(1)}_\mathrm{L}$ lepton symmetry, which is broken much later at the $\mathrm{eV}$ scale  during the NEDE phase transition when the sterile neutrinos acquire their Majorana mass. 

The first breaking can be achieved by adding  a dark Higgs doublet ${ \Phi}=(\Phi_+,\Phi_0)^T$ coupled to $(\nu_R)_\alpha$ and a doublet $S_i= (\nu_s, S_-)_i^T$, which contains the sterile $(\nu_s)_i$ as its neutral component (for notational simplicity, we will suppress the generation indices $i$ and $\alpha$ for the sterile and right-handed neutrinos, respectively). The second breaking is achieved in terms of the NEDE triplet  (similar to the triplet introduced in the Gelmini and Roncadelli model~\cite{Gelmini:1980re}),\footnote{Adopting the notation in \cite{Grimus:1999fz}.}
\begin{align}
{\Psi} = 
\left(\begin{matrix}
 \frac{1}{\sqrt{2}} \left( \Psi_0 + \Psi_{++} \right) \\
  - \frac{\mathrm{i}}{\sqrt{2}}  \left(\Psi_0 - \Psi_{++}\right)  \\
\Psi_{+} 
\end{matrix}\right)~.
\end{align}
It transforms in the adjoint representation of $\mathrm{SU}(2)_\mathrm{D}$. Our parametrization anticipates the charges that different components carry under $\mathrm{U}(1)_\mathrm{DEM}$. Accordingly, $\Psi_0$ and $\Phi_0$ are the neutral, $\Psi_+$ and $\Phi_+$ the single-charged and $\Psi_{++}$ the double-charged components.  For $U \in \mathrm{SU}(2)$, the fields transform as 
\begin{align}
\Delta \to U \Delta U^\dagger \, && \Phi \to U \Phi  \,&& \nu_s \to U \nu_s\,, && \nu_R \to  \nu_R\,,
\end{align} 
where we introduced $\Delta = \Psi \cdot \tau $  with $\tau = (\tau_1,\tau_2 ,\tau_3)$ denoting the Pauli matrices.
\begin{table}[t]
\begin{tabular}{c|c|c| c |c |c|c|c} 
\hline
\hline
 &$S$ & $\nu_R$ & $\Phi$ & $\Psi$ &H&$ \chi $& L\\
 \hline
 $\mathrm{SU(2)}_\mathrm{D}$ &$\bf 2$&$\bf 1$&$\bf 2$&$\bf 3$&$\bf 1$& $\bf 2$&$\bf 1$\\
$\mathrm{U(1)}_\mathrm{Y_D}$ & -1& 0 &1 & 2 & 0 & $Y_{\mathrm{D},\chi}$ &0\\
$\mathrm{U(1)}_\mathrm{L} $ & 1 & 1 & 0 & -2  & 0& 1 & 1\\
 \hline
 \hline
\end{tabular}
\caption{Quantum numbers in DEW model. The DEM charge is $Q_\mathrm{D}=T_{\mathrm{D},3}+\frac{1}{2} Y_\mathrm{D}$. The NEDE field $\psi$ arises as the neutral component of $\Psi$. The doublet $\Phi$ and singlet $H$ are the dark and visible sector Higgs fields, respectively. The Sm neutrino sector, represented by the SM lepton doublet $L$, is supplemented with the sterile doublet $S$ and the right-handed singlet $\nu_R$. Beyond that, $\chi$ denotes our DM candidate. }
\label{tab:quantumnumbers}
\end{table}
The full $\mathrm{SU}(2)_\mathrm{D} \times \mathrm{U(1)}_\mathrm{Y_D}$ invariant dark sector action then reads\footnote{We note that this theory might require an extended particle content to make it anomaly-free. We leave a corresponding investigation to future work.}
\begin{align}
\mathcal{L}_\mathrm{DEW} =  \mathcal{L}_\mathrm{kin}+ \mathcal{L}_\mathrm{Y} + \mathcal{L}_\mathrm{gauge}  - V(\Psi,\Phi) 
\end{align} 
where 
\begin{subequations} \label{DEW}
\begin{multline}\label{fullpotential}
V(\Psi,\Phi) = a \Phi^\dagger \Phi + c \left( \Phi^\dagger \Phi \right)^2  - \frac{\mu^2}{2}\, \mathrm{Tr} \left( \Delta^\dagger \Delta\right) + \frac{\lambda}{4} \left[ \mathrm{Tr} \left( \Delta^{\dagger}\Delta \right) \right]^2\\
 + \frac{e-h}{2} \Phi^\dagger \Phi  \mathrm{Tr} \left( \Delta^{\dagger}\Delta \right) + h \Phi^\dagger \Delta^\dagger \Delta \Phi +\frac{f}{4} \mathrm{Tr}\left( \Delta^\dagger \Delta^\dagger\right)   \mathrm{Tr}\left( \Delta \Delta\right)  - \bar{\epsilon} \left(\Phi^\dagger \Delta \epsilon \Phi^* + \mathrm{h.c.} \right)
\end{multline}
is the general, renormalizable dark sector Higgs potential and  $\epsilon = \mathrm{i} \tau_2$. The kinetic and Yukawa terms are
\begin{align}
 \mathcal{L}_\mathrm{kin} &= -\frac{\mathrm{i}}{2} \overline{S^c} \slashed{D} S- \mathrm{i} \overline{\nu_R} \slashed{D} \nu_R  - \mathrm{i} \overline{\chi}  \slashed{D}  \chi  -  \left(D_\mu \Phi\right)^{\dagger} D^{\mu} \Phi - \left(D_\mu \Psi \right)^{\dagger} D^{\mu} \Psi \,,\\
 \mathcal{L}_\mathrm{Y} &=-g_{\Phi} \overline{\nu_R} S^T \epsilon \Phi  - \frac{ g_s}{2} \overline{S^c} \epsilon \Delta S + g_H \overline{\nu_R} L^T \epsilon H + \mathrm{h.c.}~, \label{Yukawas}
\end{align}
\end{subequations}
where $L^T = (\nu_L, e_L)$ and $H$ are the SM lepton and Higgs doublet, respectively.  The gauge kinetic terms are collected in $\mathcal{L}_\mathrm{gauge}$.  The gauge couplings corresponding to $\mathrm{U(1)}_\mathrm{Y_D}$ and $\mathrm{SU}(2)_\mathrm{D}$ are $g'_d$ and $g_d \simeq g'_d$, respectively. 
For reasons that will become clear later, we will not identify $g_d$ with the coupling parameter $g_\mathrm{NEDE}$ that was responsible for the  hot NEDE thermal corrections  (instead, that role will be played by $f$). 
We further assign the dark hypercharge and lepton numbers according to Tab.~\ref{tab:quantumnumbers}. In particular, this choice implies that only the last term in \eqref{fullpotential} proportional to $\bar{\epsilon}$ breaks lepton number explicitly (while preserving dark isospin and hypercharge symmetry). The dark hypercharges $Y_\mathrm{D}$ are chosen such that $\Psi$ and $\Phi$ both contain a neutral component that is not charged under the DEM.
Imposing charge conservation during both transitions then fixes the breaking directions to
\begin{align}\label{vacua_directions}
\langle\Phi\rangle_0 = \frac{1}{\sqrt{2}}
\left(\begin{matrix} 0\\
v_\Phi
 \end{matrix}\right)
 &&
 \text{and}
 &&
 \langle\Delta\rangle_0 =
 \left(\begin{matrix} 0 & 0\\
v_\Psi& 0
 \end{matrix}\right)\,,
\end{align}
where $v_\Phi > \mathrm{TeV}$ and $v_{\Psi}$ are the high and low energy vev, respectively.   We therefore have $v_\Psi \ll v_\Phi$. For this configuration to be a stationary point of $V(\Psi,\Phi)$, two parameter conditions have to hold:
\begin{subequations}
\begin{align}
a + c v_\Phi^2  + \frac{1}{2} \left( e-h \right) v_\Psi^2 - 2 \bar{\epsilon} v_\Psi &= 0 \label{eq:vev_phi}\\
-\mu^2 + \lambda v_\Psi^2  + \frac{1}{2} \left( e-h \right) v_\Phi^2 -  \bar{\epsilon} \frac{v_\Phi^2}{v_\Psi} &= 0 \label{eq:vev_eq}
\end{align}
\end{subequations}
As argued in \cite{Lusignoli:1990yk,Grimus:1999fz}, this system admits stable solutions while being compatible with both $v_\Psi$ and  $v_\Phi$ real and positive. In order to avoid a parameter tuning, we need to effectively decouple both equations, which requires
\begin{align}\label{unmixing}
e,h \lesssim \lambda \frac{v_\Psi^2}{v_\Phi^2} \ll 1&& \text{and} && \frac{\bar{\epsilon}}{\lambda} \lesssim \frac{v_\Psi^3}{v_\Phi^2} \ll 1 \;.
\end{align}
From Eq.~\ref{eq:vev_eq}, we then obtain $v_\Psi^2 \simeq  \mu^2/ \lambda$ in agreement with \eqref{true_zero_temp}.  Substituting back into~\eqref{unmixing} yields $e, h \lesssim \mu^2/v_\Phi^2 \ll 1$.  
\begin{figure}
     \centering
     \begin{subfigure}[b]{0.25\textwidth}
         \centering
         \includegraphics[width=\textwidth]{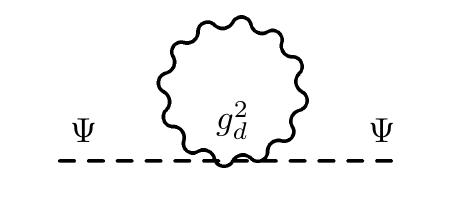}
         \caption{}
         \label{fig:mu_corr}
     \end{subfigure}
     \quad
     \begin{subfigure}[b]{0.25\textwidth}
         \centering
         \includegraphics[width=\textwidth]{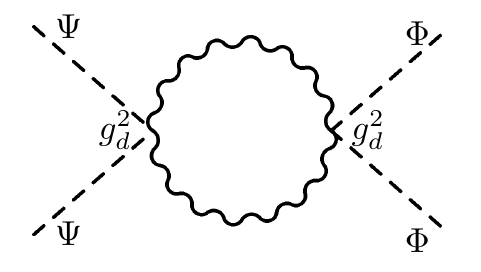}
          \caption{}
         \label{fig:h_e_corr}
     \end{subfigure}
      \quad
     \begin{subfigure}[b]{0.25\textwidth}
         \centering
         \includegraphics[width=\textwidth]{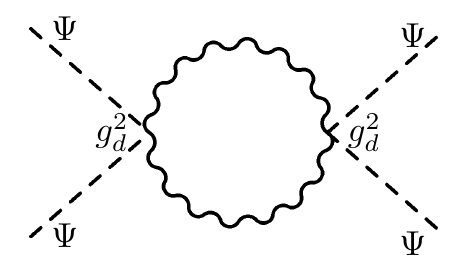}
          \caption{}
         \label{fig:lambda_corr}
     \end{subfigure}
          \caption{Loop corrections arising from the massive gauge bosons. Diagram (a), (b) and (c) correct  $ \mu^2$, $\{e,\, h\}$ and $\lambda$, respectively. The vacuum decoupling conditions in \protect\eqref{unmixing} are radiatively stable provided that the gauge coupling $g_d \lesssim  \sqrt{\mu/v_\Phi} < 10^{-6}$. }
        \label{fig:three graphs}
\end{figure}
These couplings, quadratic in $\Phi$ and $\Psi$, receive radiative corrections arising from massive gauge boson loops of order $g_d^4$ (using dimensional regularization and suppressing logarithmic factors; see also Fig.~\ref{fig:h_e_corr}). Radiative stability thus imposes 
\begin{subequations}
\begin{align}\label{bound_g_d}
g_d^2 \lesssim  \frac{\mu}{v_\Phi} (\ll 1)\, .
\end{align}
This bound also ensures that  the mass correction depicted in Fig.~\ref{fig:mu_corr}, which is of order of $g_d^4 v_\Phi^2$, is not significantly larger than the classical value $ \mu^2$. Similarly, for the quartic self-coupling to be radiatively stable, it has to obey 
\begin{align}\label{bound_lambda}
 g_d^4 \lesssim \lambda\, .
\end{align} 
\end{subequations}
Since the NEDE phase transition occurs close to matter-radiation  equality, we further have $\mu < \mathrm{eV}$ as, otherwise, the symmetry cannot be restored without heating the dark sector above the visible sector. In fact, as shown in \eqref{bound_m_psi_2},  $\mu = m_\psi/\sqrt{2} \ll \mathrm{eV}$ is more realistic as the thermal corrections are further suppressed by a small coupling $f \sim g^2_\mathrm{NEDE}$.  This, as a consequence of \eqref{bound_g_d}, implies a rather tight bound on the gauge coupling, which amounts to $g_d \ll 10^{-6}$.  The bound on $\bar{\epsilon}$ in \eqref{unmixing}, on the other hand, is radiatively stable as lepton symmetry is restored in the limit $\bar{\epsilon} \to 0$.

The second term in \eqref{Yukawas}, $\propto g_s \overline{S^c} \epsilon \Delta S$, is crucial for understanding the model. It promotes the interaction \eqref{vertex} to a $\mathrm{U(1)}_\mathrm{L}$ symmetric term, provides a portal between the neutrino and the dark sector, and creates the sterile mass through a spontaneous breaking of $\mathrm{U(1)}_\mathrm{L}$. To be precise,
once the triplet $\Psi$ picks up its vev, it generates the low-mass entry in the neutrino mass matrix M,
 \begin{align}\label{ms}
 m_s =  g_s v_\Psi \,,
 \end{align} 
 which we obtained by substituting \eqref{vacua_directions}, and which reproduces \eqref{sterile_m} for small temperatures. The corresponding breaking of lepton number by two units then gives rise to the \textit{majoron} as the corresponding (pseudo) Nambu-Goldstone boson denoted as $\eta$. In the absence of any explicit breaking ($\bar \epsilon = 0$) it is massless. We can identify it as $\eta \propto \eta_\Psi - 2 v_\Psi \eta_\Phi /v_\Phi $ by splitting off the complex phases,
\begin{align}\label{phase_split}
\Psi_0 \equiv \frac{\psi}{\sqrt{2}} \mathrm{e}^{\mathrm{i} \eta_\Psi/v_\Psi} && \text{and} && \Phi_0 \equiv \frac{\phi}{\sqrt{2}} \mathrm{e}^{\mathrm{i} \eta_\Phi/v_\Phi}\,,
\end{align}
where $\phi \equiv v_\Phi + \delta \phi $ and $\psi \equiv  v_\Psi + \delta \psi $.
 The orthogonal direction $\propto   \eta_\Phi + 2 v_\Psi \eta_\Psi /v_\Phi $ is also massless and corresponds to the Goldstone eaten by one of the neutral gauge bosons.
For $\bar{\epsilon} > 0$, $\eta$ acquires a small mass~\cite{Lusignoli:1990yk}
\begin{align}\label{majoran_mass}
m_\eta^2 =  \frac{\bar{\epsilon}}{v_\Psi} \left(v_\Phi^2 + 4 v_\Psi^2 \right) \simeq \bar{\epsilon} \frac{v_\Phi^2}{v_\Psi}   \lesssim \lambda v_\Psi^2\,,
\end{align}
where we used \eqref{unmixing} in deriving the upper bound. The mass mixing of the moduli fields $\delta \psi$ and $\delta \phi$ is then described in terms of the matrix
\begin{align}
\frac{1}{2}\left( \delta \phi, \delta \psi \right) \left(\begin{matrix}{}
  2 c v_\Phi^2 & v_\Phi \left[ (e-h) v_\Psi -  2\bar{\epsilon}\right] \\
   v_\Phi \left[ (e-h) v_\Psi - 2 \bar{\epsilon}\right]  & 2 \lambda v_\Psi^2 +  \bar{\epsilon} v_\Phi^2/v_\Psi
\end{matrix}\right) 
\left(\begin{matrix}
 \delta \phi\\
\delta \psi
\end{matrix}\right)\,.
\end{align}
It decouples at leading order in $v_\Phi^2/v_\Psi^2 \ll 1$, resulting in
\begin{align}\label{m_psi_2}
m_\psi^2 \simeq    2 \lambda v_\Psi^2 +  m_\eta^2 \,,
\end{align}
which recovers \eqref{m_psi} for sufficiently small majoron mass [compatible with  the bound in \eqref{majoran_mass}] and $v_\Psi \simeq \mu / \sqrt{\lambda}$. There is also a heavy mode in the spectrum with mass $m_\phi^2 = 2 c v_\Phi^2 \gtrsim \mathrm{TeV}$.
The double-charged triplet component $\Psi_{++}$ has a mass 
\begin{align}
m^2_{\Psi_{++}} \simeq h v_\Phi^2 +2 f v_\Psi^2 +m_\eta^2\,
\end{align}
which is above $m_\psi^2$ for $f \gg \lambda$. Finally, there is the single-charged sector with mass matrix
\begin{align}
\left[ \bar{\epsilon} + \frac{h}{2} v_\Psi \right]\left( \Phi_+, \Psi_+ \right) \left(\begin{matrix}{}
  2 v_\Psi & -\sqrt{2} v_\Phi \\
  -\sqrt{2} v_\Phi &  v_\Phi^2/v_\Psi
\end{matrix}\right) 
\left(\begin{matrix}
 \Phi_+^*\\
\Psi_+^*
\end{matrix}\right)\,.
\end{align}
I admits a massless mode $\propto \left(\Phi_+ + \sqrt{2} v_\Psi \Psi_+/v_\Phi   \right) \simeq \Phi_+  $, which is the Goldstone eaten by the charged gauge boson, along with its orthogonal counterpart  $\propto \left(\Psi_+ - \sqrt{2} v_\Psi \Phi_+/v_\Phi   \right) \simeq \Psi_+  $ with mass
\begin{align}
m^2_{\Psi_+} = \left[ \bar{\epsilon} + \frac{h}{2} v_\Psi \right] \frac{v_\Phi^2 + 2 v_\Psi^2}{v_\Psi} \lesssim m_\psi^2\,.
\end{align}

\subsubsection{DEW phenomenology}

To  be compatible with a thermally triggered phase transition, we will assume that the dark sector has a non-vanishing temperature $\xi $. As discussed in the context of the (N)ADM model, this gives rise to a change in the effective number of neutrino species,
\begin{align}
\Delta N_\mathrm{eff} = \frac{4}{7}  \left(\frac{11}{4}\right)^{4/3} g_\mathrm{rel,d} \, \xi^4\, ,
\end{align}
where we generalized the expression in \eqref{Delta_Neff} by introducing the effective number of dark, relativistic degrees of freedom $g_\mathrm{rel,d}$. We will demand that $\Delta N_\mathrm{eff} \lesssim 0.1$ during big bang nucleosynthesis (BBN) to stay clear of cosmological bounds related to DR. If we further use that at least $\delta \phi$, $S_-$ and the heavy neutrino mass eigenstates have become non-relativistic by the time of BBN (the gauge bosons might still be relativistic), we have   $g_\mathrm{rel,d} (t_\mathrm{BBN}) <  17   $ (corresponding to three massive and one massless gauge bosons, $\delta \psi$, $\eta$, $\Psi_{++}$ and $\Psi_+$). It then follows that 
\begin{align}\label{xi_bound}
\xi < 0.2\,,
\end{align}
which has to be understood as a conservative bound that can be relaxed in scenarios with heavier and/or decoupled dark sector fields. There is a potential increase of $\xi$ and hence $\Delta N_\mathrm{eff}$ of a few tens percent when the gauge bosons (and $\psi_{++}$ for $f \gg \lambda$) become massive, which is too small to be of any phenomenological concern. Beyond that, as mentioned in the previous section, there can be a late equilibration through the neutrino portal provided $g_s$ is large enough. As this is happening after neutrino decoupling it will not change the value of $N_\mathrm{eff}$. We also reiterate the point that having a supercooled phase transition is compatible with having a rather cold dark sector fully compatible with \eqref{xi_bound}.
In any event, with regard to the DEW model, we will remain agnostic about how this dark sector temperature is achieved. One possibility is that it arises from a freeze-in  through (non-renormalizable) gravitational interactions at high temperatures similar to the Planckian interacting dark matter (PIDM) scenario, which also offers a way of producing $\chi$ to account for the right DM abundance~\cite{Garny:2015sjg,Garny:2017kha}. If the gauge coupling is not too small, $g_d \gtrsim 10^{-6.5}$~\cite{Buen-Abad:2015ova}, we can also rely on the (N)ADM mechanism to heat the dark sector. This is compatible with our radiative stability bound for $\mu \simeq \mathrm{eV}$ and $v_\Phi \lesssim 10\, \mathrm{TeV}$. However, in order to be compatible with the bound in \eqref{xi_bound}, a freeze-in version of the (N)ADM model should be considered to only equilibrate the dark sector partially. A third possibility would be to have full equilibration between the dark and visible sector up until before the TeV scale where $g_\mathrm{rel,vis}  \gtrsim 110 $. Subsequently, $T_\mathrm{vis}$ (and hence $\xi$) would decrease as more and more particles deposit their entropy into the SM plasma.

We now discuss the three phase transitions in turn:

\begin{enumerate}

\item[(a)]

\textbf{DEW transition:} The first breaking  $\mathrm{SU}(2)_\mathrm{D}\times \mathrm{U}(1)_\mathrm{Y_D} \to  \mathrm{U}(1)_\mathrm{DEM}$  is induced when $\Phi$ acquires its vev $v_\phi \gtrsim \mathrm{TeV}$. At this stage, we do not need to make any assumption about the character of the transition. In particular, it could, like the NEDE transition, be triggered by subsiding thermal corrections or another scalar field. What is important, however, is that this generates the Dirac mass which couples $\nu_R$ and $\nu_s$ through the first term in \eqref{Yukawas}.  Specifically, we obtain $n= g_\Phi v_\Phi / \sqrt{2} \gtrsim \mathrm{TeV}$, where we assumed an Yukawa coupling of order unity.  It also makes three of the four dark gauge bosons massive with masses of order $~g_d^2 v_\phi^2 $. We identify the NEDE scalar with neutral triplet component $ \Psi_0 $ along with $\psi = \sqrt{2} |\Psi_0|$ as in \eqref{phase_split}. Its zero-temperature, tree-level potential then follows from \eqref{fullpotential} as (in unitary gauge)
\begin{multline}\label{phi_pot_after_breaking}
V_\mathrm{cl}(\psi) = \left[\frac{- \mu^2}{2} + \frac{e-h}{4}v_\Phi^2  \right]\psi^2 + \frac{1}{4}\lambda \psi^4 +  \left[ \lambda |\Psi_+|^2 + (\lambda+ 2 f ) |\Psi_{++}|^2  \right] \psi^2 \\
+ \sqrt{2} f \left[ \Psi_+^2 \Psi_{++}^*  \left\{ 1 +\frac{\mathrm{i}}{2} \frac{\eta}{v_\Psi} \right\} + \mathrm{h.c.}\right] \psi  -  \bar{\epsilon} v_\Phi^2\left[ 1 - \frac{\eta^2}{2 v_\Psi^2}\right]  \psi+ \ldots
\end{multline}
where the ellipsis stands for terms dependent on the heavy mode $\delta \phi$, and we neglected  terms of order of $\eta^3$. 
Comparing with the Abelian Higgs model in \eqref{pot_vac}, the main difference is the occurrence of additional interaction terms between $\psi$ and the independent triplet components $\Psi_+$, $\Psi_{++}$ and $\eta$. In general, these triplet interactions are an additional source of thermal corrections. A detailed derivation of the finite-temperature effective potential within the DEW model is beyond the scope of the present work. 
However, we can still make contact with our previous analysis in Sec.~\ref{sec:hotNEDE} in a regime where  $g_d^2 \ll f$.  This corresponds to a situation where the thermal corrections arise mainly from the coupling of $\psi$ to $\Psi_{++}$ and gauge sector contributions can be neglected.
This motivates to formally identify $g^2_\mathrm{NEDE} \sim f$. The reason is that $f$ controls the strength of the $|\Psi_{++}|^2\psi^2$ vertex and thus leads to the same power counting as the gauge boson coupling $g_\mathrm{NEDE}^2 A_\mu A^\mu \psi^2$ in \eqref{classical} when computing gauge boson loops (an analogous identification was employed in Sec.~III in \cite{Arnold:1992rz}). The same identification holds for the mass $m^2_{\Psi_{++}} \simeq 2 f v_\Psi^2$, which correctly mimics the gauge boson mass $\propto g_\mathrm{NEDE}^2 v_\Psi^2$. As a result, we have $\gamma \sim \lambda/f^2 \lesssim 1$ as the condition for a strong first-order phase transition where the false vacuum energy dominates over the thermal plasma. This regime also corresponds to the hierarchy $ \lambda \lesssim f^2 \ll f \ll 1$, which in turn tells us that $m_{\Psi_{++}} \gg m_\psi$.\footnote{We point out that the disjunct case where $f \lesssim g_d^4$ could also be described within our formalism. However, in that case the radiative stability condition in \eqref{bound_lambda} arising from the diagram in Fig.~\ref{fig:lambda_corr} cannot be satisfied in the supercooled regime where $\gamma = 4 \pi \lambda / g_d^4 \lesssim 1$.}
Another subtlety involves the last term in \eqref{phi_pot_after_breaking}. To be precise, it gives rise to a linear correction of the potential $\propto \bar{\epsilon} v_\Phi^2 \psi $, which, in principle, can interfere with the thermal corrections. To avoid this we have to slightly strengthen the bound in \eqref{unmixing} to $\bar{\epsilon} \lesssim \sqrt{\gamma} \lambda v_\Psi^3/v_\Phi^2 $. This makes sure that the linear correction is subdominant around the maximum of the potential where $\psi \sim \mu/g_d^2$. As a consequence, we always have $m_\eta^2 < m_\psi^2$ in the supercooled regime where $\gamma \lesssim 1$.

\item[(b)]

\textbf{SM transition:}  When the SM Higgs doublet $H$ acquires its vev, $v_H =  246 \,\mathrm{GeV}$, it will introduce the mixing between $\nu_R$ and $\nu_L$ inducing the off-diagonal mass term  $d= g_H v_H / \sqrt{2} \sim 10^2\, \mathrm{GeV} $. In particular, we find $d < n$ as required for the inverse seesaw to give rise to the active neutrino masses. To be precise, for their masses to be of order of $10^{-2} \mathrm{eV}$, we need $v_\Phi \sim \sqrt{2} g_\Phi^{-1} \sqrt{m_s/\mathrm{eV}} \,\mathrm{TeV} \gtrsim \mathrm{TeV}$, where we used \eqref{active_masses}. Beyond that, the electroweak transition remains decoupled from the dark sector.

\item[(c)]

\textbf{NEDE transition:} Finally, at much lower energies, when the dark sector temperature falls below $T_d^*$ in \eqref{eq:T_d_star_large_mass}, corresponding to 
 $T_\mathrm{vis}^* \simeq \mathrm{eV}$ in the visible sector, the NEDE phase transition takes place. It occurs when the thermal corrections subside. As discussed in the previous section, it also gives rise to the Majorana mass entry $m_s$ as defined in~\eqref{ms}. We obtain $v_\Psi/v_\Phi \sim (g_\Phi/g_s) 10^{-12} \sqrt{m_s/\mathrm{eV}}$, which shows that the assumption $v_\Psi \ll v_\Phi$ was justified for a wide range of sterile masses and Yukawa couplings. Beyond that, the hot NEDE phenomenology can still be described in terms of $f_\mathrm{NEDE}$, $H_*\bar{\beta}^{-1}$ and $T_d^*$ (or $z_*$ equivalently) defined in \eqref{f_NEDE_large_mass}, \eqref{beta_large_mass} and \eqref{eq:T_d_star_large_mass}, respectively. In particular, the transition leads to a small-scale condensate of colliding bubble walls consisting of $\psi$ quanta.  
 
 \end{enumerate}

\begin{figure}
     \centering
     \begin{subfigure}[b]{0.2\textwidth}
         \centering
         \includegraphics[width=\textwidth]{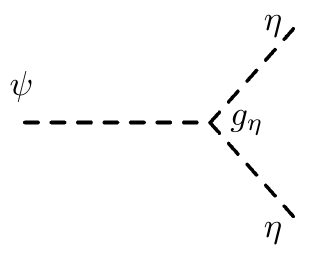}
         \caption{}
         \label{fig:decay_eta}
     \end{subfigure}
     \quad
      \quad
      \quad
      \quad
            \quad
      \quad
     \begin{subfigure}[b]{0.2\textwidth}
         \centering
         \includegraphics[width=\textwidth]{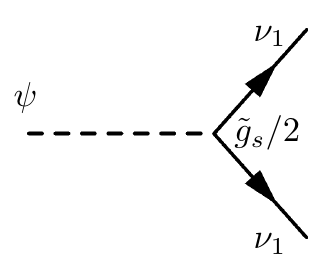}
          \caption{}
         \label{fig:decay_nu}
     \end{subfigure}
          \caption{The $\psi$ condensate can decay into (a) the majoran or (b) the lightest active neutrino. The respective couplings can be inferred from \protect\eqref{vertex_eta} and \protect\eqref{vertex2} as $g_\eta = m_\eta^2 / (\sqrt{2} v_\Psi) $ and $\tilde g_s /2 $. If $g_s \ll \kappa^2$ then the fermionic decay $\Gamma\left( \delta \psi \to \nu_j\nu_j\right)  = \tilde g_s ^2 m_\psi / (32 \pi)$ dominates over the bosonic decay  $\Gamma\left(\delta \psi \to \eta \eta \right) = g_\eta^2  / (8 \pi m_\psi)$.  }
        \label{fig:decays}
\end{figure}

We now turn to the decay of NEDE within the DEW model.
The last term in \eqref{phi_pot_after_breaking} describes the decay into the majoron $\eta$ as in Fig.~\ref{fig:decay_eta}; more explicitly, the interaction vertex is
\begin{align}\label{vertex_eta}
-\frac{m_\eta^2}{\sqrt{2} v_\Psi} \eta^2 \delta \psi\;.
\end{align}
The corresponding perturbative decay rate becomes (neglecting Bose enhancement effects)
\begin{align}\label{gamma_eta}
\Gamma\left(\delta \psi \to \eta \eta \right) &= \frac{1}{16 \pi } \frac{m_{\eta}^4}{v_\Psi^2 m_\psi} \ll \frac{g_s^4}{32 \pi^2}m_\psi\,,
\end{align}
where we substituted $v_\Psi= m_\psi / \sqrt{2\lambda}$ and used $m_\psi > m_\eta$ along with \eqref{lambda_constraint} to derive the upper bound. There is a second decay channel that arises from the coupling between $\delta \psi$ and $\nu_s$ (see Fig.~\ref{fig:decay_nu}). Since $\delta \psi$ is much lighter than the heavy pseudo-Dirac fermions, the decay can only produce the three lightest neutrino mass states $(\nu_1,\nu_2,\nu_3)$ (where we assume normal mass ordering $m_1 < m_2 < m_3$). In fact, using the bound in \eqref{bound_m_psi_2} alongside constraints from solar and atmospheric neutrino observations~\cite{Esteban:2020cvm} on the square-mass differences, $\psi$ can only decay into $\nu_1$.    The interaction vertex can be inferred from~\eqref{vertex2} as $\propto (\tilde g_s)_{11} \psi \,\overline{{\nu_1}^c} \nu_1 /2$, where the coupling in the mass eigenbasis is $(\tilde g_s)_{11} = \mathcal{O}(g_s) \times \kappa^2$.
This gives rise to the decay rate
 \begin{align}
\Gamma\left( \delta \psi \to \nu_1\nu_1\right) 
= \frac{(\tilde g_s)_{11}^2 m_\psi}{ 32\, \pi} \sim  \frac{\kappa^4 g_s^2 m_\psi}{ 32\, \pi} \,.
\end{align}
Whether the decay into $\eta$ or $\nu_1$ dominates mainly depends on the majoron mass $m_\eta$ (when the upper bound in \eqref{gamma_eta} is saturated, it also depends on the ratio $ g_s / \kappa^2 < 10^{-7} \kappa^4$).
For concreteness, we will assume $m_\eta$ to be sufficiently small (which is natural within our low-energy theory) such that the~$\psi$ condensate decays into $\nu_1$, although the opposite case can provide a viable scenario too. Using $\kappa^2 \sim 10^{-2} \left(\mathrm{eV}/m_s\right) $, we find that $\Gamma\left( \psi \to \nu_1\nu_1\right) \gg H_* $ provided $g_s$ satisfies the relatively weak bound $g_s \gg 10^{-11} (m_s/ \mathrm{eV}) \sqrt{\mathrm{eV}/m_\psi}$.
In fact, having at least one of the two decay channels is key to realizing scenario B in Sec.~\ref{sec:decay} where the NEDE condensate decays into radiation that subsequently becomes non-relativistic. As a numerical example, if we want the decay product (with typical energy $m_\psi/2$) to become non-relativistic before matter-radiation equality, we need a relative mass difference of order of  $ (m_\psi - m_1) / m_\psi \simeq   (z_*-z_\mathrm{eq} )/ z_* \sim 0.4$. This example assumes the produced radiation to be free-streaming (which is justified for small enough mixing angle).  We also note that a similar discussion would apply to the decay into the majoron $\eta$. In particular, we could obtain the right mass scale by appropriately choosing the breaking parameter $\bar{\epsilon}$. However, we would need to make sure that $\eta$ is not tightly coupled with the DR plasma as this would lead to a late reheating in the dark sector in conflict with our mixed DM scenario.  We leave a more quantitative investigation of the condensate's decay to future work.

We also included $\chi$ in \eqref{DEW} as our DM candidate. The idea is to dynamically create its mass through a Yukawa coupling with $\Phi$ during the DEW phase transition. Correspondingly, we expect a mass of order of $ v_\Phi  \sim \sqrt{2} g_\Phi^{-1} \sqrt{m_s/\mathrm{eV}} \,\mathrm{TeV} > \mathrm{TeV}$. It transforms as a doublet under the dark $\mathrm{SU(2)}_\mathrm{D}$ and carries dark hypercharge $Y_{\mathrm{D},\chi}$. {For a large enough gauge coupling parameter $g_d> 10^{-6.5}$, it also serves as the agent responsible for heating the dark sector as discussed in Sec.~\ref{sec:NADM} in the context of the (N)ADM model}. {Otherwise, we assume it to be produced via the freeze-in mechanism through gravitational interactions as in the PIDM model~\cite{Garny:2015sjg,Garny:2017kha}. This minimal scenario can account for the right DM abundance for a super-TeV mass, which is indeed achieved in the (preferred) heavy sterile case with $m_s > \mathrm{eV}$.}  As discussed before, interactions  of $\chi$ with the DR fluid  have the potential to relieve the $S_8$ tension.  
{The coupling parameter between $\chi$ and the massless gauge boson (after the DEW transition) is given by 
\begin{align}
 \mathcal{O}(1) \times g'_d Y_{\mathrm{D},\chi} \lesssim \mathcal{O}(1)  \times 10^{-8} \, \left( \frac{m_\psi}{\mathrm{meV}}\right)^{1/2}  \left( \frac{m_s}{\mathrm{100 \, eV}}\right)^{1/4}  \sqrt{g_\Phi} \,Y_{\mathrm{D},\chi}\,,
\end{align}
where we have used the radiative stability bound inferred from \eqref{bound_g_d} and substituted $v_\Phi$.} For the interaction to have an effect on structure formation, it has to be greater than $10^{-5}$~\cite{Buen-Abad:2015ova}, which requires  a sufficiently large hypercharge\footnote{For a string theory setup which gives rise to large charges see for example~\cite{Raghuram:2018hjn}.} $Y_{\mathrm{D},\chi} \gtrsim 10^3$.

Finally, we provide an explicit parameter example that has the potential to resolve the $H_0$ and $S_8$ tension while explaining the neutrino masses. To be specific, we consider a gauge coupling parameter $g'_d \simeq g_d =  2 \times 10^{-10}$ along with a sterile coupling $ g_s \simeq 10^{-4}$. The mass of $\psi$ (around the true vacuum) is $m_\psi= \sqrt{2}  \mu = 1.4 \times 10^{-6}\, \mathrm{eV}$. We fix the couplings $\lambda = 6 \times 10^{-25} (\gg g_d^4)$, $ g_\mathrm{NEDE}^2 \sim f=2 \times 10^{-10}$ and impose the bounds $e, h < 10^{-39}$. This then fixes $v_\psi = 1.3\times 10^{6} \, \mathrm{eV}$. The parameters $a \sim 10\,\mathrm{TeV}$ and $c \sim 1$ in \eqref{eq:vev_phi} are chosen such that $v_\Phi = 20 \mathrm{TeV}$. With these choices it can be checked that the loop corrections in  Fig.~\ref{fig:three graphs} are under control. The explicit lepton number breaking can either vanish completely or  taken to be $\bar{\epsilon} \ll 10^{-34} \mathrm{eV}$, which is small enough to avoid any interference with the thermal corrections and gives rise to a majoron mass $m_\eta \ll m_\psi$.\footnote{Strictly speaking, global symmetries are not allowed by quantum gravity but non-perturbative instanton corrections can lead to an exponential suppression of $\bar{\epsilon}$~\cite{Kallosh:1995hi}.} The sterile mass parameter, on the other hand, becomes $m_s = g_s v_\Psi / \sqrt{2} \simeq 100 \,\mathrm{eV}$, which via the inverse seesaw mechanism explains the active neutrino masses $m_j \lesssim \kappa^2 m_s \sim 10^{-2}\, \mathrm{eV} $, where we used that $\kappa = \mathcal{O}(d)/\mathcal{O}(n) \sim 10^{-2}$ for  $g_\Phi$ and $g_H$ of order unity. 
Within this scenario, the $\psi$ condensate decays into the lightest neutrino $\nu_1$ with mass $m_1 \sim \mathcal{O}(0.1) m_\psi$, thereby realizing the mixed DM scenario for the decay of NEDE. Moreover, the DM field $\chi$ acquires a mass $M_\chi \lesssim v_\Phi \sim 10\,\mathrm{TeV}$, which can lead to the right DM abundance via the minimal PIDM scenario. With regards to the hot NEDE phenomenology, we have $\gamma \simeq 4 \pi \lambda / f^2 \simeq 1.8\times 10^{-4}$, corresponding to a supercooled transition where the released vacuum energy dominates over the DR fluid [with strength parameter $\alpha \simeq 200$ as defined in \eqref{alpha_large_mass}]. Due to \eqref{xi_large_mass}, an NEDE fraction of $f_\mathrm{NEDE} = 10 \%$, as required for resolving the Hubble tension, then implies a relative dark sector temperature $\xi_*=T_d^*/T_\mathrm{vis}^* \simeq 0.1$ (which is compatible with the bound in \eqref{xi_bound} needed for suppressing $\Delta N_\mathrm{eff}$). In absolute terms, we obtain from  \eqref{eq:T_d_star_large_mass} that for a transition taking place at $z_*=5000$, we have $T_d^* \simeq 0.08\, \mathrm{eV}$ and $T_\mathrm{vis}^* \simeq 0.81\, \mathrm{eV}$. The bubble percolation is extremely quick with a duration $H_* \bar{\beta}^{-1} \sim 10^{-12} $ inferred from \eqref{beta_large_mass}, which prevents bubbles from growing to cosmological sizes. Of course, this only serves as a numerical example, and, ultimately, a Boltzmann code implementation of hot NEDE is needed to constrain these parameters further.

\section{Discussion}

In the first part of this work, we proposed a new trigger mechanism for NEDE~\cite{Niedermann:2019olb,Niedermann:2020dwg}. It relies on finite-temperature corrections to its effective potential that restore the symmetry of a dark gauge group at early times.  The NEDE transition then happens when the dark sector temperature falls below a critical temperature of order of $\mathrm{eV}$. As our working example, we used the Abelian Higgs model with gauge coupling parameter $g_\mathrm{NEDE}$ and self-interaction strength $\lambda$, although we expect our general parameterization of the finite-temperature effective potential to also apply to weakly-coupled non-Abelian gauge theories. We derived the condition under which this transition is a strong first-order phase transition with $\alpha \gg 1$ and capable of accommodating a sizable fraction of NEDE required for resolving the Hubble tension. To be specific, after computing the effective potential valid for low temperatures and a large gauge boson mass, we found that $\gamma = \lambda / (4 \pi g_\mathrm{NEDE}^4) \lesssim 1$ is the preferred region in parameter space. Here, the false vacuum energy dominates over the DR fluid during an extended supercooled period before it decays as a consequence of the phase transition. The supercooling is important for two reasons: First, it provides a short energy injection in the cosmic fluid that builds up quickly relative to the dominant radiation component in the visible sector; and second, it avoids cosmological bounds on the effective number of relativistic degrees of freedom. At the same time, we showed that the phase transition is sufficiently short to avoid problems with large anisotropies that arise when bubbles of true vacuum grow too large. We then proposed different decay scenarios for hot NEDE: One building on previous work where the fluid dynamics after the phase transition is controlled by the bubble wall condensate and its small-scale anisotropic stress; and another one, which considers the microscopic decay of the condensate into particles that, after a short relativistic phase, turn non-relativistic before recombination and provide a fraction of DM.  
By comparing the background evolution with that of the $\Lambda$CDM model, we argued that both cases are compatible with a short energy injection around matter-radiation equality as required by phenomenology. 

We stress that up to this point hot NEDE is a rather general low-energy framework compatible with different microphysics. Therefore, in the second part of this work, we used our simple implementation of hot NEDE as an anchor for finding  a more complete description of the dark sector. We started by proposing a mechanism for heating up the dark sector through a fermionic mediator field that plays the role of DM and has been proposed in the context of the (N)ADM model~\cite{Buen-Abad:2015ova,Lesgourgues:2015wza,Buen-Abad:2017gxg}. This leads to a preferred value of the dark sector temperature that is slightly colder than the visible sector, and, at the same time, offers a mechanism to address the S8 tension through DM-DR interactions. After that, we argued that the NEDE phase transition is responsible for generating the active neutrino masses. This was done in two steps: First, the NEDE phase transition gives rise to the sub-GeV sector of the inverse seesaw mechanism by generating a (Majorana) mass term for a set of sterile neutrinos $\nu_s$. This is achieved in terms of a Yukawa interaction between $\nu_s$ and the NEDE scalar $\Psi$ of strength $g_s$. As a consequence, the active neutrinos acquire their sub-eV masses during the NEDE phase transition when $\psi$ picks up its vev. The virtue of this mechanism is that it ties the neutrino masses to the NEDE energy scale, making both of them less arbitrary. In particular, we found that we can accommodate the neutrino mass spectrum and oscillation patterns without violating phenomenological bounds on the coupling $g_s$. Second, we embedded NEDE in a  larger symmetry group within the dark sector. This local symmetry is partially broken above the TeV scale, giving rise to the the high-energy sector of the inverse seesaw mechanism and, simultaneously, generating the mass of a fermionic DM field $\chi$.  This model also introduces an approximate global lepton symmetry, which is broken spontaneously during the NEDE phase transition and {stabilizes the active neutrino masses against loop corrections}. It is intriguing to speculate that such a breaking  could explain the recent $R_K$ and $R_{K^*}$ anomalies observed at CERN and the $(g-2)_{\mu}$ anomalous magnetic moment of the muon observed at Fermilab~\cite{DelleRose:2019ukt,DelleRose:2020oaa}.

Hot NEDE has the potential to resolve both the Hubble and S8 tension  while shedding light on the origin of neutrino masses {and dark matter} through a more complete modeling of the dark sector. To further scrutinize this proposal, different theoretical and phenomenological aspects need to be investigated. This includes finding a link between the parameters of the effective NEDE fluid and the fundamental model parameters. In particular, this raises complicated questions about the evolution and microscopic decay of the colliding bubble wall condensate after the phase transition. Beyond that, a full cosmological parameter extraction should be performed using a cosmological model for hot NEDE that incorporates both decay scenarios. This can be done by extending the present code {\tt TriggerCLASS}\footnote{\url{https://github.com/flo1984/TriggerCLASS}} describing the cold NEDE case. On a more theoretical level, the consistency of the microscopic model outlined in Sec.~\ref{sec:microscopic_model} needs to be further investigated,  including a detailed derivation of its quantum and thermal corrections. Beyond these most pressing issues related to the viability of hot NEDE, a search for unique signatures that set  hot (and cold) NEDE apart from other early time proposals such as old EDE should be undertaken.  For example, signatures unique to cold NEDE were predicted to arise on small angular scales in the CMB power spectrum, in LSS data and the low-frequency gravitational wave spectrum~\cite{Niedermann:2020dwg}.  Hot NEDE, and in particular the DEW model with its richer microscopic underpinning, is further adding to this list by predicting neutrinos to interact with the dark sector plasma and to become massless above the hot NEDE breaking scale.

\begin{acknowledgments}
We would like to thank Edmund Copeland and Steen Hannestad for useful comments on the draft.
This work is supported by Villum Fonden grant 13384 and Independent Research Fund Denmark grant 0135-00378B.
\end{acknowledgments}

\appendix

\section{Thermal corrections in Abelian Higgs model} \label{appendix_potential}

We consider the simple case of a $\mathrm{U(1)}$ gauge theory with gauge field $A_\mu$ and coupling $g_\mathrm{NEDE}<1$ to establish the connection between the fundamental model parameters and the finite temperature effective potential in \eqref{eq:effective_T_pot}. To that end, we will review (and slightly extend) the discussion in~\cite{Dolan:1973qd,Arnold:1992rz}. 
Our starting point is the zero-temperature tree-level potential of a complex scalar field $\Psi$, 

\begin{align}\label{pot_vac}
V_\mathrm{cl}(|\Psi|^2) = -  \mu^2 |\Psi|^2 + \lambda |\Psi|^4
\end{align}
which we assume to be charged under the $\mathrm{U(1)}$. The action is
\begin{align}\label{classical}
\mathcal{L} = - \frac{1}{4} F^2 - |D\Psi|^2-V_\mathrm{cl}(|\Psi|^2) + \mathrm{g.f.}\,,
\end{align}
with the gauge-covariant derivative $D_\mu \Psi =\left(\partial_\mu - \frac{\mathrm{i}}{2} g_\mathrm{NEDE} A_\mu\right) \Psi$ and a gauge-fixing term $\mathrm{g.f.\,}$. The corrections to $V_\mathrm{cl}$ at $1$-loop order can then be calculated using the standard procedure for the calculation of the effective potential at finite temperature (see Sec.~5 in~\cite{Dolan:1973qd} for the Abelian Higgs model).  They contain the thermal corrections $\Delta V_\mathrm{thermal}^{(1)}$ alongside the standard vacuum corrections (which are absorbed in renormalized couplings). Shifting \eqref{classical} around the background field $\psi= \sqrt{2}|\Psi|$, the result in Landau gauge is\footnote{In the notation of~\cite{Dolan:1973qd}, Landau gauge corresponds to setting $\alpha=0$.}
\begin{align}\label{eq:thermal_corr}
\Delta V_\mathrm{thermal}^{(1)} = \sum_i n_i T_d \int \frac{\mathrm{d}^3 k}{\left( 2 \pi \right)^3} \ln \left\{ 1 - \exp{\left[-\frac{1}{T_d} \sqrt{k^2 + m_i^2(\psi)}\right] }\right\}
\end{align}
where the index $i$ runs over all modes (physical and unphysical) with different masses $m_i(\psi)$, and $n_i$ is the corresponding number of degrees of freedom; explicitly
\begin{align}\label{eq:masses_standard}
m_1^2(\psi)=- \mu^2 + 3 \lambda \psi^2 \,,&&m_2^2(\psi)=- \mu^2 +  \lambda \psi^2 \,,&&m_3^2(\psi)=g_\mathrm{NEDE}^2 \psi^2 \,,&&m_4^2=0\;.
\end{align}
where $m_1$ is the mass of the scalar field fluctuations ($n_1=1$), $m_2$ the mass of the unphysical scalar gauge mode ($n_2=1$), $m_3$ the mass of the transverse vector modes ($n_3=3$), and $m_4=0$  ($n_4=1$) reflects the presence of the unphysical $k_\mu$ vector mode. The momentum integral can be performed in the high-temperature/small-mass limit where $m_i^2/T_d^2 \ll 1$, yielding
\begin{align}\label{eq:thermal_high_T}
\Delta V_\mathrm{thermal}^{(1)} = \sum_i n_i \left[ -\frac{\pi^2}{90}T_d^4+\frac{1}{24} m_i^2(\psi) T_d^2- \frac{1}{12 \pi} m_i^3(\psi) T_d  + \mathcal{O}(m_i^4) \right]\;.
\end{align}
Away from this limit, we find that the integral is well approximated by

\begin{align}\label{eq:thermal_corr_gen}
\Delta V_\mathrm{thermal}^{(1)} = \sum_i n_i T_d^4 K\left(\frac{m_i(\psi)}{T_d} \right) \mathrm{e}^{-m_i(\psi)/T_d}
\end{align}
where 
\begin{align}\label{eq:K}
 K(a)= -0.1134 \, (1+a) - 0.0113 \, a^2 +4.32 \times 10^{-6} \ln{(a)}\,a^{3.58}+0.0038\,\mathrm{e}^{-a(a-1)} 
\end{align}
is a fitting function valid in the range $0<a<30 $ with a relative error $<4 \%$ , which we determined by numerically solving the momentum integral. It matches on smoothly to \eqref{eq:thermal_high_T} for high temperatures. As argued in \cite{Arnold:1992rz}, the above result can be further improved by including thermal corrections to the masses.  To be precise, the authors suggest to replace in \eqref{eq:thermal_corr} 
\begin{subequations}
\label{eq:masses}
\begin{align}
m_i^2 \to m_{\mathrm{eff},i}^2=m_i^2 + \left(\frac{\lambda}{3}+ \frac{g_\mathrm{NEDE}^2}{4}\right)  T_d^2 \quad \mathrm{for} \quad i=1,2 
\end{align}
as well as
\begin{align}
 m_3 \to M_{\mathrm{eff},\perp}^2= m_3^2, && m_3 \to M_{\mathrm{eff},L}^2 = m_3^2+ \frac{1}{3} g_\mathrm{NEDE}^2 T_d^2  \,,
\end{align}
\end{subequations}
for the two transverse ($\perp$) and the longitudinal ($L$) gauge field polarizations, respectively.  This substitution corresponds to a resummation of one-loop ring (or daisy) diagrams. Somewhat heuristically, $m_{\mathrm{eff},i}$ can be obtained by first deriving the potential using \eqref{eq:masses_standard} and then repeating the algorithm for calculating the effective potential. In any case, plugging this back into \eqref{eq:thermal_high_T}, we obtain the one-loop, ring-improved potential in the high-temperature/small-mass limit
\begin{align} \label{eq:effective_T_pot_gen}
V(\psi;T_d) 
&=  V_\mathrm{cl}(\psi)+ \Delta V^{(1,\mathrm{ring})}_\mathrm{thermal}(\psi) + \ldots\nonumber \\
&= \frac{1}{2} \left[ - \mu^2 + \left(\frac{\lambda}{3} +  \frac{g_\mathrm{NEDE}^2}{4}\right) T^2_d \right] \psi^2 - \frac{1}{12 \pi} \left[ 2 + \left(1 + \frac{T_d^2}{3\psi^2}  \right)^{3/2}+ \left(\frac{m^2_{\mathrm{eff},1}+m^2_{\mathrm{eff},2}}{g_\mathrm{NEDE}^2 \psi^2}\right)^{3/2}\right] g_\mathrm{NEDE}^3 T_d \psi^3 \nonumber \\
&\quad \quad \quad \quad \quad \quad + \frac{\lambda}{4}\psi^4 + V_0(T_d)  + \ldots \,,
\end{align}
where the ellipsis stands for terms of order of $m_i^4$ and zero-temperature one-loop corrections. Field-independent contributions are collected in $V_0(T_d)$.
%\begin{align}
%V_0(T_d) = \left( - \frac{4}{3} \frac{\pi^2}{30} + \frac{\lambda}{4} + \frac{g_\mathrm{NEDE}^2}{3}\right) T_d^4 - \frac{\mu^2}{24} T_d^2\,,
%\end{align}
%where the first term always comes with a negative sign.  

\begin{figure}[t]
    \centering
    \includegraphics[width=0.75\textwidth]{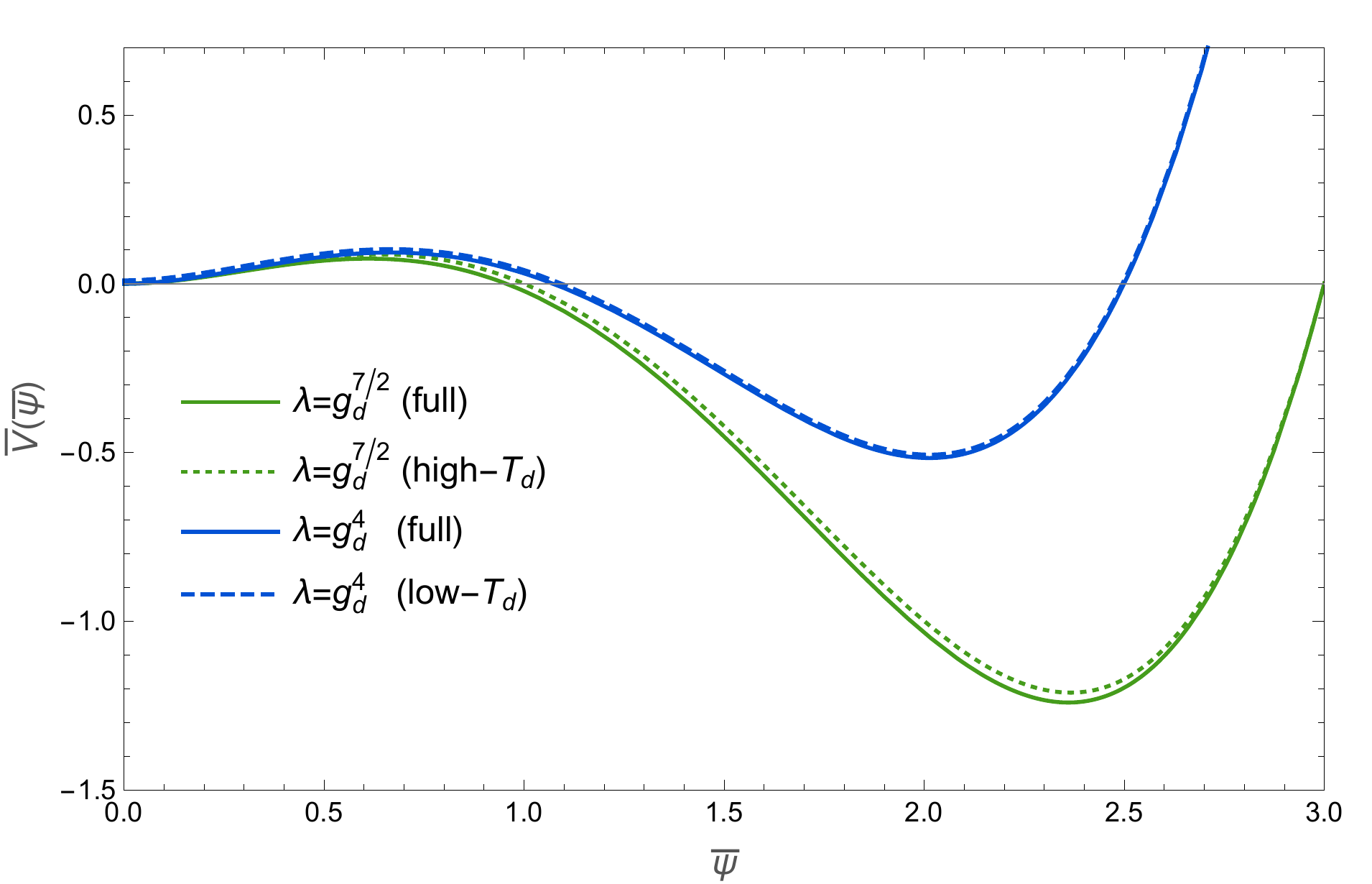}
    \caption{Dimensionless potential $\bar{V}= \lambda^3 V(\psi; T_d)/(T_d^4 E^3)$ as a function of $\bar{\psi} = \lambda \psi / (T_d E)$ for $\delta_\mathrm{eff} = \frac{2 \lambda D}{E^2} (1-T_\circ^2/T_d^2) = 3/2 $ and $g_\mathrm{NEDE}=10^{-3}$. The green solid line corresponds to $\lambda = g_\mathrm{NEDE}^{7/2}$ and the blue one to  $\lambda = g_\mathrm{NEDE}^{4}$, making the small-mass (dotted line) and large-mass (dashed line) approximations applicable, respectively.}
    \label{fig:Potential_comparison}
\end{figure}

For this  expression to be valid, a set of necessary conditions applies. The small-mass expansion demands
\begin{subequations}
\begin{align}\label{eq:high_T_cond}
m^2_\mathrm{eff,i} (\psi) \ll T_d^2 \,, && M_{\mathrm{eff},\perp}^2(\psi), M_{\mathrm{eff},L}^2(\psi)  \ll T_d^2\,,
\end{align}
and, at the same time, the underlying perturbative treatment requires the scalar and vector loop expansion parameters to fulfill~\cite{Arnold:1992rz} (for a more recent discussion on the applicability of perturbation theory see \cite{Croon:2020cgk,Gould:2021oba})\footnote{We note that the second condition is violated in the case of the electroweak phase transition for realistic values of the Higgs mass, making a non-perturbative treatment necessary~\cite{Dine:1992wr,Kajantie:1996mn,Laine:1998jb,Arnold:1992rz}. There is another subtlety involved in this case: For $\psi \to 0 $, i.e.\ when approaching the symmetric vacuum, the perturbative expansion breaks down as the transverse gauge boson mass vanishes. It has been argued that this leads to an uncertainty of order~\cite{Arnold:1992rz} $g_\mathrm{NEDE}^6 T_d^4$. While this can affect the determination of $f_\mathrm{NEDE}$ in principle, we can neglect such a contribution for sufficiently small gauge couplings or a sufficiently cold dark sector. In the same limit $m_\mathrm{eff}^2$ in \eqref{eq:masses} can turn negative (provided $T_d$ is small enough), leading to an imaginary effective potential. This signals the quantum instability of the corresponding state~\cite{Weinberg:1987vp}. }
\begin{align}
\frac{\lambda T_d}{m_\mathrm{eff,i} (\psi) } \ll 1\,, && \frac{g_\mathrm{NEDE}^2 T_d}{M_{\mathrm{eff},\perp}} \ll 1\;.
\end{align}
If we further assume that
\begin{align}\label{eq:cond_eff_pot_3}
T_d/\psi \lesssim 1\,, && m_{\mathrm{eff,i}}/ (g_\mathrm{NEDE} \psi) \lesssim 1 \,,
\end{align}
\end{subequations}
the coefficient of the cubic term is approximately constant, and we can identify
\begin{align}\label{D_E_abelian}
D=\frac{1}{2}  \left(\frac{\lambda}{3} + \frac{g_\mathrm{NEDE}^2}{4}\right)\,, && E \simeq \frac{g_\mathrm{NEDE}^3}{4 \pi}\,, && T_\circ^2 = \frac{\mu^2}{2D} \,,
\end{align} 
which then reproduces the potential \eqref{eq:effective_T_pot}. Now, as a minimal requirement, we want the potential to be valid around the true minimum $\psi \sim \psi_\mathrm{True}$. In this regime, the cubic and quartic term balance each other, implying $ \psi_\mathrm{True} \sim g_\mathrm{NEDE}^3 T_d /\lambda$. With this identification, we can show that all the above conditions are satisfied if
\begin{align}\label{lambda_gd_cond}
g_\mathrm{NEDE}^4 \ll \lambda \lesssim g_\mathrm{NEDE}^3\,,
\end{align}
where we also implied that $\mu^2 < g_\mathrm{NEDE}^2 T_d^2$ for a potential barrier to exist.
For example, the upper bound  follows directly from the first condition in  \eqref{eq:cond_eff_pot_3} and the lower bound from the last condition in \eqref{eq:high_T_cond} (small-mass limit). The others can then be shown to be fulfilled when using the definitions in \eqref{eq:masses}.
Our result slightly tightens the upper bound derived by the authors in~\cite{Arnold:1992rz}. The reason is that their analysis did not impose the conditions \eqref{eq:cond_eff_pot_3}, which, in our case, were necessary to make contact with the simple form in  \eqref{eq:effective_T_pot}. Effectively, this is a regime where the effect of the ring resummation, which manifests itself through a field dependence of the cubic coefficient, becomes negligible, making contact with the earlier analysis in~\cite{Dine:1992wr}. The validity of the approximation is demonstrated by the solid and dotted green line in Fig.~\ref{fig:Potential_comparison}, depicting the full ring-resummed, one-loop result~\eqref{eq:thermal_corr_gen} alongside its small-mass approximation~\eqref{eq:effective_T_pot}, respectively.

In this work, we contrast the previous regime where $\lambda \gg g_\mathrm{NEDE}^4$ with the complementary situation where
\begin{align}\label{low_T_cond}
\lambda \lesssim g_\mathrm{NEDE}^4\,.
\end{align}
This regime is needed for realizing a strong first-order phase transition along with supercooling as preferred by the hot NEDE phenomenology. It requires us to drop the small-mass assumption in \eqref{eq:high_T_cond} and resort to the more general formula \eqref{eq:thermal_corr_gen}. The main contribution in this case arises from the three gauge boson modes, for which we find $M_{\mathrm{eff},L} \simeq M_{\mathrm{eff},\perp} = g_\mathrm{NEDE} \psi$ due to \eqref{low_T_cond}. The potential therefore becomes
\begin{align}
V(\psi; T_d) =  V_\mathrm{cl}(\psi) + 3 T_d^4 K\left(\frac{m_3(\psi)}{T_d} \right) \mathrm{e}^{-m_3(\psi)/T_d}\,,
\end{align}
which translates to \eqref{eq:effective_T_pot_low_T} when using $8 D \simeq (4 \pi E)^{2/3} =g_\mathrm{NEDE}^2 $. We compare this approximation (dashed blue) to the full result (blue solid) in Fig.~\ref{fig:Potential_comparison} [not neglecting the scalar field modes in \eqref{eq:thermal_corr_gen}] and find excellent agreement. This approximation is relevant for values $\lambda \lesssim g_\mathrm{NEDE}^4 $ for which $m_3(\psi) / T_d$ is still order unity. 
%For even smaller values $\lambda \ll g_\mathrm{NEDE}^4 $ the exponential suppresses the thermal corrections and we are left with the vacuum potential \eqref{pot_vac}, which cannot give rise to a first-order transition. 

\bibliography{Hot_NEDE}

\begin{thebibliography}{128}%
\makeatletter
\providecommand \@ifxundefined [1]{%
 \@ifx{#1\undefined}
}%
\providecommand \@ifnum [1]{%
 \ifnum #1\expandafter \@firstoftwo
 \else \expandafter \@secondoftwo
 \fi
}%
\providecommand \@ifx [1]{%
 \ifx #1\expandafter \@firstoftwo
 \else \expandafter \@secondoftwo
 \fi
}%
\providecommand \natexlab [1]{#1}%
\providecommand \enquote  [1]{``#1''}%
\providecommand \bibnamefont  [1]{#1}%
\providecommand \bibfnamefont [1]{#1}%
\providecommand \citenamefont [1]{#1}%
\providecommand \href@noop [0]{\@secondoftwo}%
\providecommand \href [0]{\begingroup \@sanitize@url \@href}%
\providecommand \@href[1]{\@@startlink{#1}\@@href}%
\providecommand \@@href[1]{\endgroup#1\@@endlink}%
\providecommand \@sanitize@url [0]{\catcode `\\12\catcode `\$12\catcode
  `\&12\catcode `\#12\catcode `\^12\catcode `\_12\catcode `\%12\relax}%
\providecommand \@@startlink[1]{}%
\providecommand \@@endlink[0]{}%
\providecommand \url  [0]{\begingroup\@sanitize@url \@url }%
\providecommand \@url [1]{\endgroup\@href {#1}{\urlprefix }}%
\providecommand \urlprefix  [0]{URL }%
\providecommand \Eprint [0]{\href }%
\providecommand \doibase [0]{https://doi.org/}%
\providecommand \selectlanguage [0]{\@gobble}%
\providecommand \bibinfo  [0]{\@secondoftwo}%
\providecommand \bibfield  [0]{\@secondoftwo}%
\providecommand \translation [1]{[#1]}%
\providecommand \BibitemOpen [0]{}%
\providecommand \bibitemStop [0]{}%
\providecommand \bibitemNoStop [0]{.\EOS\space}%
\providecommand \EOS [0]{\spacefactor3000\relax}%
\providecommand \BibitemShut  [1]{\csname bibitem#1\endcsname}%
\let\auto@bib@innerbib\@empty
%</preamble>
\bibitem [{\citenamefont {Aghanim}\ \emph {et~al.}(2018)\citenamefont {Aghanim}
  \emph {et~al.}}]{Aghanim:2018eyx}%
  \BibitemOpen
  \bibfield  {author} {\bibinfo {author} {\bibfnamefont {N.}~\bibnamefont
  {Aghanim}} \emph {et~al.} (\bibinfo {collaboration} {Planck}),\ }\bibfield
  {title} {\bibinfo {title} {{Planck 2018 results. VI. Cosmological
  parameters}},\ }\href@noop {} {\  (\bibinfo {year} {2018})},\ \Eprint
  {https://arxiv.org/abs/1807.06209} {arXiv:1807.06209 [astro-ph.CO]}
  \BibitemShut {NoStop}%
%%CITATION = ARXIV:1807.06209;%%
\bibitem [{\citenamefont {Alam}\ \emph {et~al.}(2017)\citenamefont {Alam} \emph
  {et~al.}}]{Alam:2016hwk}%
  \BibitemOpen
  \bibfield  {author} {\bibinfo {author} {\bibfnamefont {S.}~\bibnamefont
  {Alam}} \emph {et~al.} (\bibinfo {collaboration} {BOSS}),\ }\bibfield
  {title} {\bibinfo {title} {{The clustering of galaxies in the completed
  SDSS-III Baryon Oscillation Spectroscopic Survey: cosmological analysis of
  the DR12 galaxy sample}},\ }\href {https://doi.org/10.1093/mnras/stx721}
  {\bibfield  {journal} {\bibinfo  {journal} {Mon. Not. Roy. Astron. Soc.}\
  }\textbf {\bibinfo {volume} {470}},\ \bibinfo {pages} {2617} (\bibinfo {year}
  {2017})},\ \Eprint {https://arxiv.org/abs/1607.03155} {arXiv:1607.03155
  [astro-ph.CO]} \BibitemShut {NoStop}%
%%CITATION = ARXIV:1607.03155;%%
\bibitem [{\citenamefont {Freedman}(2017)}]{Freedman:2017yms}%
  \BibitemOpen
  \bibfield  {author} {\bibinfo {author} {\bibfnamefont {W.~L.}\ \bibnamefont
  {Freedman}},\ }\bibfield  {title} {\bibinfo {title} {{Cosmology at a
  Crossroads}},\ }\href {https://doi.org/10.1038/s41550-017-0121} {\bibfield
  {journal} {\bibinfo  {journal} {Nature Astron.}\ }\textbf {\bibinfo {volume}
  {1}},\ \bibinfo {pages} {0121} (\bibinfo {year} {2017})},\ \Eprint
  {https://arxiv.org/abs/1706.02739} {arXiv:1706.02739 [astro-ph.CO]}
  \BibitemShut {NoStop}%
\bibitem [{\citenamefont {Verde}\ \emph {et~al.}(2019)\citenamefont {Verde},
  \citenamefont {Treu},\ and\ \citenamefont {Riess}}]{Verde:2019ivm}%
  \BibitemOpen
  \bibfield  {author} {\bibinfo {author} {\bibfnamefont {L.}~\bibnamefont
  {Verde}}, \bibinfo {author} {\bibfnamefont {T.}~\bibnamefont {Treu}},\ and\
  \bibinfo {author} {\bibfnamefont {A.}~\bibnamefont {Riess}},\ }\bibfield
  {title} {\bibinfo {title} {{Tensions between the Early and the Late
  Universe}}\ }(\bibinfo {year} {2019})\ \Eprint
  {https://arxiv.org/abs/1907.10625} {arXiv:1907.10625 [astro-ph.CO]}
  \BibitemShut {NoStop}%
\bibitem [{\citenamefont {Di~Valentino}\ \emph
  {et~al.}(2021{\natexlab{a}})\citenamefont {Di~Valentino} \emph
  {et~al.}}]{DiValentino:2020zio}%
  \BibitemOpen
  \bibfield  {author} {\bibinfo {author} {\bibfnamefont {E.}~\bibnamefont
  {Di~Valentino}} \emph {et~al.},\ }\bibfield  {title} {\bibinfo {title}
  {{Snowmass2021 - Letter of interest cosmology intertwined II: The hubble
  constant tension}},\ }\href
  {https://doi.org/10.1016/j.astropartphys.2021.102605} {\bibfield  {journal}
  {\bibinfo  {journal} {Astropart. Phys.}\ }\textbf {\bibinfo {volume} {131}},\
  \bibinfo {pages} {102605} (\bibinfo {year} {2021}{\natexlab{a}})},\ \Eprint
  {https://arxiv.org/abs/2008.11284} {arXiv:2008.11284 [astro-ph.CO]}
  \BibitemShut {NoStop}%
\bibitem [{\citenamefont {Riess}\ \emph {et~al.}(2021)\citenamefont {Riess},
  \citenamefont {Casertano}, \citenamefont {Yuan}, \citenamefont {Bowers},
  \citenamefont {Macri}, \citenamefont {Zinn},\ and\ \citenamefont
  {Scolnic}}]{Riess:2020fzl}%
  \BibitemOpen
  \bibfield  {author} {\bibinfo {author} {\bibfnamefont {A.~G.}\ \bibnamefont
  {Riess}}, \bibinfo {author} {\bibfnamefont {S.}~\bibnamefont {Casertano}},
  \bibinfo {author} {\bibfnamefont {W.}~\bibnamefont {Yuan}}, \bibinfo {author}
  {\bibfnamefont {J.~B.}\ \bibnamefont {Bowers}}, \bibinfo {author}
  {\bibfnamefont {L.}~\bibnamefont {Macri}}, \bibinfo {author} {\bibfnamefont
  {J.~C.}\ \bibnamefont {Zinn}},\ and\ \bibinfo {author} {\bibfnamefont
  {D.}~\bibnamefont {Scolnic}},\ }\bibfield  {title} {\bibinfo {title} {{Cosmic
  Distances Calibrated to 1\% Precision with Gaia EDR3 Parallaxes and Hubble
  Space Telescope Photometry of 75 Milky Way Cepheids Confirm Tension with
  $\Lambda$CDM}},\ }\href {https://doi.org/10.3847/2041-8213/abdbaf} {\bibfield
   {journal} {\bibinfo  {journal} {Astrophys. J. Lett.}\ }\textbf {\bibinfo
  {volume} {908}},\ \bibinfo {pages} {L6} (\bibinfo {year} {2021})},\ \Eprint
  {https://arxiv.org/abs/2012.08534} {arXiv:2012.08534 [astro-ph.CO]}
  \BibitemShut {NoStop}%
\bibitem [{\citenamefont {Freedman}\ \emph {et~al.}(2020)\citenamefont
  {Freedman}, \citenamefont {Madore}, \citenamefont {Hoyt}, \citenamefont
  {Jang}, \citenamefont {Beaton}, \citenamefont {Lee}, \citenamefont {Monson},
  \citenamefont {Neeley},\ and\ \citenamefont {Rich}}]{Freedman:2020dne}%
  \BibitemOpen
  \bibfield  {author} {\bibinfo {author} {\bibfnamefont {W.~L.}\ \bibnamefont
  {Freedman}}, \bibinfo {author} {\bibfnamefont {B.~F.}\ \bibnamefont
  {Madore}}, \bibinfo {author} {\bibfnamefont {T.}~\bibnamefont {Hoyt}},
  \bibinfo {author} {\bibfnamefont {I.~S.}\ \bibnamefont {Jang}}, \bibinfo
  {author} {\bibfnamefont {R.}~\bibnamefont {Beaton}}, \bibinfo {author}
  {\bibfnamefont {M.~G.}\ \bibnamefont {Lee}}, \bibinfo {author} {\bibfnamefont
  {A.}~\bibnamefont {Monson}}, \bibinfo {author} {\bibfnamefont
  {J.}~\bibnamefont {Neeley}},\ and\ \bibinfo {author} {\bibfnamefont
  {J.}~\bibnamefont {Rich}},\ }\bibfield  {title} {\bibinfo {title}
  {{Calibration of the Tip of the Red Giant Branch (TRGB)}}\ }\href
  {https://doi.org/10.3847/1538-4357/ab7339} {10.3847/1538-4357/ab7339}
  (\bibinfo {year} {2020}),\ \Eprint {https://arxiv.org/abs/2002.01550}
  {arXiv:2002.01550 [astro-ph.GA]} \BibitemShut {NoStop}%
\bibitem [{\citenamefont {Freedman}(2021)}]{Freedman:2021ahq}%
  \BibitemOpen
  \bibfield  {author} {\bibinfo {author} {\bibfnamefont {W.~L.}\ \bibnamefont
  {Freedman}},\ }\bibfield  {title} {\bibinfo {title} {{Measurements of the
  Hubble Constant: Tensions in Perspective}},\ }\href@noop {} {\  (\bibinfo
  {year} {2021})},\ \Eprint {https://arxiv.org/abs/2106.15656}
  {arXiv:2106.15656 [astro-ph.CO]} \BibitemShut {NoStop}%
\bibitem [{\citenamefont {Yuan}\ \emph {et~al.}(2019)\citenamefont {Yuan},
  \citenamefont {Riess}, \citenamefont {Macri}, \citenamefont {Casertano},\
  and\ \citenamefont {Scolnic}}]{Yuan:2019npk}%
  \BibitemOpen
  \bibfield  {author} {\bibinfo {author} {\bibfnamefont {W.}~\bibnamefont
  {Yuan}}, \bibinfo {author} {\bibfnamefont {A.~G.}\ \bibnamefont {Riess}},
  \bibinfo {author} {\bibfnamefont {L.~M.}\ \bibnamefont {Macri}}, \bibinfo
  {author} {\bibfnamefont {S.}~\bibnamefont {Casertano}},\ and\ \bibinfo
  {author} {\bibfnamefont {D.}~\bibnamefont {Scolnic}},\ }\bibfield  {title}
  {\bibinfo {title} {{Consistent Calibration of the Tip of the Red Giant Branch
  in the Large Magellanic Cloud on the Hubble Space Telescope Photometric
  System and a Re-determination of the Hubble Constant}},\ }\href
  {https://doi.org/10.3847/1538-4357/ab4bc9} {\bibfield  {journal} {\bibinfo
  {journal} {Astrophys. J.}\ }\textbf {\bibinfo {volume} {886}},\ \bibinfo
  {pages} {61} (\bibinfo {year} {2019})},\ \Eprint
  {https://arxiv.org/abs/1908.00993} {arXiv:1908.00993 [astro-ph.GA]}
  \BibitemShut {NoStop}%
\bibitem [{\citenamefont {Soltis}\ \emph {et~al.}(2021)\citenamefont {Soltis},
  \citenamefont {Casertano},\ and\ \citenamefont {Riess}}]{Soltis:2020gpl}%
  \BibitemOpen
  \bibfield  {author} {\bibinfo {author} {\bibfnamefont {J.}~\bibnamefont
  {Soltis}}, \bibinfo {author} {\bibfnamefont {S.}~\bibnamefont {Casertano}},\
  and\ \bibinfo {author} {\bibfnamefont {A.~G.}\ \bibnamefont {Riess}},\
  }\bibfield  {title} {\bibinfo {title} {{The Parallax of $\omega$ Centauri
  Measured from Gaia EDR3 and a Direct, Geometric Calibration of the Tip of the
  Red Giant Branch and the Hubble Constant}},\ }\href
  {https://doi.org/10.3847/2041-8213/abdbad} {\bibfield  {journal} {\bibinfo
  {journal} {Astrophys. J. Lett.}\ }\textbf {\bibinfo {volume} {908}},\
  \bibinfo {pages} {L5} (\bibinfo {year} {2021})},\ \Eprint
  {https://arxiv.org/abs/2012.09196} {arXiv:2012.09196 [astro-ph.GA]}
  \BibitemShut {NoStop}%
\bibitem [{\citenamefont {Anand}\ \emph {et~al.}(2021)\citenamefont {Anand},
  \citenamefont {Tully}, \citenamefont {Rizzi}, \citenamefont {Riess},\ and\
  \citenamefont {Yuan}}]{Anand:2021sum}%
  \BibitemOpen
  \bibfield  {author} {\bibinfo {author} {\bibfnamefont {G.~S.}\ \bibnamefont
  {Anand}}, \bibinfo {author} {\bibfnamefont {R.~B.}\ \bibnamefont {Tully}},
  \bibinfo {author} {\bibfnamefont {L.}~\bibnamefont {Rizzi}}, \bibinfo
  {author} {\bibfnamefont {A.~G.}\ \bibnamefont {Riess}},\ and\ \bibinfo
  {author} {\bibfnamefont {W.}~\bibnamefont {Yuan}},\ }\bibfield  {title}
  {\bibinfo {title} {{Comparing Tip of the Red Giant Branch Distance Scales: An
  Independent Reduction of the Carnegie-Chicago Hubble Program and the Value of
  the Hubble Constant}},\ }\href@noop {} {\  (\bibinfo {year} {2021})},\
  \Eprint {https://arxiv.org/abs/2108.00007} {arXiv:2108.00007 [astro-ph.CO]}
  \BibitemShut {NoStop}%
\bibitem [{\citenamefont {Abbott}\ \emph {et~al.}(2019)\citenamefont {Abbott}
  \emph {et~al.}}]{Abbott:2019yzh}%
  \BibitemOpen
  \bibfield  {author} {\bibinfo {author} {\bibfnamefont {B.}~\bibnamefont
  {Abbott}} \emph {et~al.} (\bibinfo {collaboration} {LIGO Scientific,
  Virgo}),\ }\bibfield  {title} {\bibinfo {title} {{A gravitational-wave
  measurement of the Hubble constant following the second observing run of
  Advanced LIGO and Virgo}},\ }\href@noop {} {\  (\bibinfo {year} {2019})},\
  \Eprint {https://arxiv.org/abs/1908.06060} {arXiv:1908.06060 [astro-ph.CO]}
  \BibitemShut {NoStop}%
\bibitem [{\citenamefont {Wong}\ \emph {et~al.}(2019)\citenamefont {Wong} \emph
  {et~al.}}]{Wong:2019kwg}%
  \BibitemOpen
  \bibfield  {author} {\bibinfo {author} {\bibfnamefont {K.~C.}\ \bibnamefont
  {Wong}} \emph {et~al.},\ }\bibfield  {title} {\bibinfo {title} {{H0LiCOW
  XIII. A 2.4\% measurement of $H_{0}$ from lensed quasars: $5.3\sigma$ tension
  between early and late-Universe probes}},\ }\href@noop {} {\  (\bibinfo
  {year} {2019})},\ \Eprint {https://arxiv.org/abs/1907.04869}
  {arXiv:1907.04869 [astro-ph.CO]} \BibitemShut {NoStop}%
%%CITATION = ARXIV:1907.04869;%%
\bibitem [{\citenamefont {Birrer}\ \emph {et~al.}(2020)\citenamefont {Birrer}
  \emph {et~al.}}]{Birrer:2020tax}%
  \BibitemOpen
  \bibfield  {author} {\bibinfo {author} {\bibfnamefont {S.}~\bibnamefont
  {Birrer}} \emph {et~al.},\ }\bibfield  {title} {\bibinfo {title} {{TDCOSMO
  IV: Hierarchical time-delay cosmography -- joint inference of the Hubble
  constant and galaxy density profiles}},\ }\href
  {https://doi.org/10.1051/0004-6361/202038861} {\bibfield  {journal} {\bibinfo
   {journal} {Astron. Astrophys.}\ }\textbf {\bibinfo {volume} {643}},\
  \bibinfo {pages} {A165} (\bibinfo {year} {2020})},\ \Eprint
  {https://arxiv.org/abs/2007.02941} {arXiv:2007.02941 [astro-ph.CO]}
  \BibitemShut {NoStop}%
\bibitem [{\citenamefont {Di~Valentino}\ \emph
  {et~al.}(2021{\natexlab{b}})\citenamefont {Di~Valentino}, \citenamefont
  {Mena}, \citenamefont {Pan}, \citenamefont {Visinelli}, \citenamefont {Yang},
  \citenamefont {Melchiorri}, \citenamefont {Mota}, \citenamefont {Riess},\
  and\ \citenamefont {Silk}}]{DiValentino:2021izs}%
  \BibitemOpen
  \bibfield  {author} {\bibinfo {author} {\bibfnamefont {E.}~\bibnamefont
  {Di~Valentino}}, \bibinfo {author} {\bibfnamefont {O.}~\bibnamefont {Mena}},
  \bibinfo {author} {\bibfnamefont {S.}~\bibnamefont {Pan}}, \bibinfo {author}
  {\bibfnamefont {L.}~\bibnamefont {Visinelli}}, \bibinfo {author}
  {\bibfnamefont {W.}~\bibnamefont {Yang}}, \bibinfo {author} {\bibfnamefont
  {A.}~\bibnamefont {Melchiorri}}, \bibinfo {author} {\bibfnamefont {D.~F.}\
  \bibnamefont {Mota}}, \bibinfo {author} {\bibfnamefont {A.~G.}\ \bibnamefont
  {Riess}},\ and\ \bibinfo {author} {\bibfnamefont {J.}~\bibnamefont {Silk}},\
  }\bibfield  {title} {\bibinfo {title} {{In the realm of the Hubble
  tension\textemdash{}a review of solutions}},\ }\href
  {https://doi.org/10.1088/1361-6382/ac086d} {\bibfield  {journal} {\bibinfo
  {journal} {Class. Quant. Grav.}\ }\textbf {\bibinfo {volume} {38}},\ \bibinfo
  {pages} {153001} (\bibinfo {year} {2021}{\natexlab{b}})},\ \Eprint
  {https://arxiv.org/abs/2103.01183} {arXiv:2103.01183 [astro-ph.CO]}
  \BibitemShut {NoStop}%
\bibitem [{\citenamefont {Verde}\ \emph {et~al.}(2017)\citenamefont {Verde},
  \citenamefont {Bernal}, \citenamefont {Heavens},\ and\ \citenamefont
  {Jimenez}}]{Verde:2016ccp}%
  \BibitemOpen
  \bibfield  {author} {\bibinfo {author} {\bibfnamefont {L.}~\bibnamefont
  {Verde}}, \bibinfo {author} {\bibfnamefont {J.~L.}\ \bibnamefont {Bernal}},
  \bibinfo {author} {\bibfnamefont {A.~F.}\ \bibnamefont {Heavens}},\ and\
  \bibinfo {author} {\bibfnamefont {R.}~\bibnamefont {Jimenez}},\ }\bibfield
  {title} {\bibinfo {title} {{The length of the low-redshift standard ruler}},\
  }\href {https://doi.org/10.1093/mnras/stx116} {\bibfield  {journal} {\bibinfo
   {journal} {Mon. Not. Roy. Astron. Soc.}\ }\textbf {\bibinfo {volume}
  {467}},\ \bibinfo {pages} {731} (\bibinfo {year} {2017})},\ \Eprint
  {https://arxiv.org/abs/1607.05297} {arXiv:1607.05297 [astro-ph.CO]}
  \BibitemShut {NoStop}%
\bibitem [{\citenamefont {Bernal}\ \emph {et~al.}(2016)\citenamefont {Bernal},
  \citenamefont {Verde},\ and\ \citenamefont {Riess}}]{Bernal:2016gxb}%
  \BibitemOpen
  \bibfield  {author} {\bibinfo {author} {\bibfnamefont {J.~L.}\ \bibnamefont
  {Bernal}}, \bibinfo {author} {\bibfnamefont {L.}~\bibnamefont {Verde}},\ and\
  \bibinfo {author} {\bibfnamefont {A.~G.}\ \bibnamefont {Riess}},\ }\bibfield
  {title} {\bibinfo {title} {{The trouble with $H_0$}},\ }\href
  {https://doi.org/10.1088/1475-7516/2016/10/019} {\bibfield  {journal}
  {\bibinfo  {journal} {JCAP}\ }\textbf {\bibinfo {volume} {1610}}\bibfield
  {number} {\bibinfo  {number} { (10)},\ \bibinfo {pages} {019}},\ }\Eprint
  {https://arxiv.org/abs/1607.05617} {arXiv:1607.05617 [astro-ph.CO]}
  \BibitemShut {NoStop}%
%%CITATION = ARXIV:1607.05617;%%
\bibitem [{\citenamefont {Knox}\ and\ \citenamefont
  {Millea}(2020)}]{Knox:2019rjx}%
  \BibitemOpen
  \bibfield  {author} {\bibinfo {author} {\bibfnamefont {L.}~\bibnamefont
  {Knox}}\ and\ \bibinfo {author} {\bibfnamefont {M.}~\bibnamefont {Millea}},\
  }\bibfield  {title} {\bibinfo {title} {{Hubble constant hunter's guide}},\
  }\href {https://doi.org/10.1103/PhysRevD.101.043533} {\bibfield  {journal}
  {\bibinfo  {journal} {Phys. Rev. D}\ }\textbf {\bibinfo {volume} {101}},\
  \bibinfo {pages} {043533} (\bibinfo {year} {2020})},\ \Eprint
  {https://arxiv.org/abs/1908.03663} {arXiv:1908.03663 [astro-ph.CO]}
  \BibitemShut {NoStop}%
\bibitem [{\citenamefont {Aylor}\ \emph {et~al.}(2019)\citenamefont {Aylor},
  \citenamefont {Joy}, \citenamefont {Knox}, \citenamefont {Millea},
  \citenamefont {Raghunathan},\ and\ \citenamefont {Wu}}]{Aylor_2019}%
  \BibitemOpen
  \bibfield  {author} {\bibinfo {author} {\bibfnamefont {K.}~\bibnamefont
  {Aylor}}, \bibinfo {author} {\bibfnamefont {M.}~\bibnamefont {Joy}}, \bibinfo
  {author} {\bibfnamefont {L.}~\bibnamefont {Knox}}, \bibinfo {author}
  {\bibfnamefont {M.}~\bibnamefont {Millea}}, \bibinfo {author} {\bibfnamefont
  {S.}~\bibnamefont {Raghunathan}},\ and\ \bibinfo {author} {\bibfnamefont
  {W.~L.~K.}\ \bibnamefont {Wu}},\ }\bibfield  {title} {\bibinfo {title}
  {Sounds discordant: Classical distance ladder and $\lambda$cdm-based
  determinations of the cosmological sound horizon},\ }\href
  {https://doi.org/10.3847/1538-4357/ab0898} {\bibfield  {journal} {\bibinfo
  {journal} {The Astrophysical Journal}\ }\textbf {\bibinfo {volume} {874}},\
  \bibinfo {pages} {4} (\bibinfo {year} {2019})}\BibitemShut {NoStop}%
\bibitem [{\citenamefont {Arendse}\ \emph {et~al.}(2019)\citenamefont
  {Arendse}, \citenamefont {Agnello},\ and\ \citenamefont
  {Wojtak}}]{Arendse:2019itb}%
  \BibitemOpen
  \bibfield  {author} {\bibinfo {author} {\bibfnamefont {N.}~\bibnamefont
  {Arendse}}, \bibinfo {author} {\bibfnamefont {A.}~\bibnamefont {Agnello}},\
  and\ \bibinfo {author} {\bibfnamefont {R.}~\bibnamefont {Wojtak}},\
  }\bibfield  {title} {\bibinfo {title} {{Low-redshift measurement of the sound
  horizon through gravitational time-delays}},\ }\href
  {https://doi.org/10.1051/0004-6361/201935972} {\bibfield  {journal} {\bibinfo
   {journal} {Astron. Astrophys.}\ }\textbf {\bibinfo {volume} {632}},\
  \bibinfo {pages} {A91} (\bibinfo {year} {2019})},\ \Eprint
  {https://arxiv.org/abs/1905.12000} {arXiv:1905.12000 [astro-ph.CO]}
  \BibitemShut {NoStop}%
%%CITATION = ARXIV:1905.12000;%%
\bibitem [{\citenamefont {Efstathiou}(2021)}]{Efstathiou:2021ocp}%
  \BibitemOpen
  \bibfield  {author} {\bibinfo {author} {\bibfnamefont {G.}~\bibnamefont
  {Efstathiou}},\ }\bibfield  {title} {\bibinfo {title} {{To H0 or not to
  H0?}},\ }\href {https://doi.org/10.1093/mnras/stab1588} {\bibfield  {journal}
  {\bibinfo  {journal} {Mon. Not. Roy. Astron. Soc.}\ }\textbf {\bibinfo
  {volume} {505}},\ \bibinfo {pages} {3866} (\bibinfo {year} {2021})},\ \Eprint
  {https://arxiv.org/abs/2103.08723} {arXiv:2103.08723 [astro-ph.CO]}
  \BibitemShut {NoStop}%
\bibitem [{\citenamefont {Poulin}\ \emph {et~al.}(2018)\citenamefont {Poulin},
  \citenamefont {Smith}, \citenamefont {Grin}, \citenamefont {Karwal},\ and\
  \citenamefont {Kamionkowski}}]{Poulin:2018dzj}%
  \BibitemOpen
  \bibfield  {author} {\bibinfo {author} {\bibfnamefont {V.}~\bibnamefont
  {Poulin}}, \bibinfo {author} {\bibfnamefont {T.~L.}\ \bibnamefont {Smith}},
  \bibinfo {author} {\bibfnamefont {D.}~\bibnamefont {Grin}}, \bibinfo {author}
  {\bibfnamefont {T.}~\bibnamefont {Karwal}},\ and\ \bibinfo {author}
  {\bibfnamefont {M.}~\bibnamefont {Kamionkowski}},\ }\bibfield  {title}
  {\bibinfo {title} {{Cosmological implications of ultralight axionlike
  fields}},\ }\href {https://doi.org/10.1103/PhysRevD.98.083525} {\bibfield
  {journal} {\bibinfo  {journal} {Phys. Rev.}\ }\textbf {\bibinfo {volume}
  {D98}},\ \bibinfo {pages} {083525} (\bibinfo {year} {2018})},\ \Eprint
  {https://arxiv.org/abs/1806.10608} {arXiv:1806.10608 [astro-ph.CO]}
  \BibitemShut {NoStop}%
%%CITATION = ARXIV:1806.10608;%%
\bibitem [{\citenamefont {Poulin}\ \emph {et~al.}(2019)\citenamefont {Poulin},
  \citenamefont {Smith}, \citenamefont {Karwal},\ and\ \citenamefont
  {Kamionkowski}}]{Poulin:2018cxd}%
  \BibitemOpen
  \bibfield  {author} {\bibinfo {author} {\bibfnamefont {V.}~\bibnamefont
  {Poulin}}, \bibinfo {author} {\bibfnamefont {T.~L.}\ \bibnamefont {Smith}},
  \bibinfo {author} {\bibfnamefont {T.}~\bibnamefont {Karwal}},\ and\ \bibinfo
  {author} {\bibfnamefont {M.}~\bibnamefont {Kamionkowski}},\ }\bibfield
  {title} {\bibinfo {title} {{Early Dark Energy Can Resolve The Hubble
  Tension}},\ }\href {https://doi.org/10.1103/PhysRevLett.122.221301}
  {\bibfield  {journal} {\bibinfo  {journal} {Phys. Rev. Lett.}\ }\textbf
  {\bibinfo {volume} {122}},\ \bibinfo {pages} {221301} (\bibinfo {year}
  {2019})},\ \Eprint {https://arxiv.org/abs/1811.04083} {arXiv:1811.04083
  [astro-ph.CO]} \BibitemShut {NoStop}%
%%CITATION = ARXIV:1811.04083;%%
\bibitem [{\citenamefont {Smith}\ \emph {et~al.}(2020)\citenamefont {Smith},
  \citenamefont {Poulin},\ and\ \citenamefont {Amin}}]{Smith:2019ihp}%
  \BibitemOpen
  \bibfield  {author} {\bibinfo {author} {\bibfnamefont {T.~L.}\ \bibnamefont
  {Smith}}, \bibinfo {author} {\bibfnamefont {V.}~\bibnamefont {Poulin}},\ and\
  \bibinfo {author} {\bibfnamefont {M.~A.}\ \bibnamefont {Amin}},\ }\bibfield
  {title} {\bibinfo {title} {{Oscillating scalar fields and the Hubble tension:
  a resolution with novel signatures}},\ }\href
  {https://doi.org/10.1103/PhysRevD.101.063523} {\bibfield  {journal} {\bibinfo
   {journal} {Phys. Rev. D}\ }\textbf {\bibinfo {volume} {101}},\ \bibinfo
  {pages} {063523} (\bibinfo {year} {2020})},\ \Eprint
  {https://arxiv.org/abs/1908.06995} {arXiv:1908.06995 [astro-ph.CO]}
  \BibitemShut {NoStop}%
\bibitem [{\citenamefont {Smith}\ \emph {et~al.}(2021)\citenamefont {Smith},
  \citenamefont {Poulin}, \citenamefont {Bernal}, \citenamefont {Boddy},
  \citenamefont {Kamionkowski},\ and\ \citenamefont {Murgia}}]{Smith:2020rxx}%
  \BibitemOpen
  \bibfield  {author} {\bibinfo {author} {\bibfnamefont {T.~L.}\ \bibnamefont
  {Smith}}, \bibinfo {author} {\bibfnamefont {V.}~\bibnamefont {Poulin}},
  \bibinfo {author} {\bibfnamefont {J.~L.}\ \bibnamefont {Bernal}}, \bibinfo
  {author} {\bibfnamefont {K.~K.}\ \bibnamefont {Boddy}}, \bibinfo {author}
  {\bibfnamefont {M.}~\bibnamefont {Kamionkowski}},\ and\ \bibinfo {author}
  {\bibfnamefont {R.}~\bibnamefont {Murgia}},\ }\bibfield  {title} {\bibinfo
  {title} {{Early dark energy is not excluded by current large-scale structure
  data}},\ }\href {https://doi.org/10.1103/PhysRevD.103.123542} {\bibfield
  {journal} {\bibinfo  {journal} {Phys. Rev. D}\ }\textbf {\bibinfo {volume}
  {103}},\ \bibinfo {pages} {123542} (\bibinfo {year} {2021})},\ \Eprint
  {https://arxiv.org/abs/2009.10740} {arXiv:2009.10740 [astro-ph.CO]}
  \BibitemShut {NoStop}%
\bibitem [{\citenamefont {Murgia}\ \emph {et~al.}(2021)\citenamefont {Murgia},
  \citenamefont {Abell\'an},\ and\ \citenamefont {Poulin}}]{Murgia:2020ryi}%
  \BibitemOpen
  \bibfield  {author} {\bibinfo {author} {\bibfnamefont {R.}~\bibnamefont
  {Murgia}}, \bibinfo {author} {\bibfnamefont {G.~F.}\ \bibnamefont
  {Abell\'an}},\ and\ \bibinfo {author} {\bibfnamefont {V.}~\bibnamefont
  {Poulin}},\ }\bibfield  {title} {\bibinfo {title} {{Early dark energy
  resolution to the Hubble tension in light of weak lensing surveys and lensing
  anomalies}},\ }\href {https://doi.org/10.1103/PhysRevD.103.063502} {\bibfield
   {journal} {\bibinfo  {journal} {Phys. Rev. D}\ }\textbf {\bibinfo {volume}
  {103}},\ \bibinfo {pages} {063502} (\bibinfo {year} {2021})},\ \Eprint
  {https://arxiv.org/abs/2009.10733} {arXiv:2009.10733 [astro-ph.CO]}
  \BibitemShut {NoStop}%
\bibitem [{\citenamefont {Poulin}\ \emph {et~al.}(2021)\citenamefont {Poulin},
  \citenamefont {Smith},\ and\ \citenamefont {Bartlett}}]{Poulin:2021bjr}%
  \BibitemOpen
  \bibfield  {author} {\bibinfo {author} {\bibfnamefont {V.}~\bibnamefont
  {Poulin}}, \bibinfo {author} {\bibfnamefont {T.~L.}\ \bibnamefont {Smith}},\
  and\ \bibinfo {author} {\bibfnamefont {A.}~\bibnamefont {Bartlett}},\
  }\bibfield  {title} {\bibinfo {title} {{Dark Energy at early times and ACT: a
  larger Hubble constant without late-time priors}},\ }\href@noop {} {\
  (\bibinfo {year} {2021})},\ \Eprint {https://arxiv.org/abs/2109.06229}
  {arXiv:2109.06229 [astro-ph.CO]} \BibitemShut {NoStop}%
\bibitem [{\citenamefont {Lin}\ \emph {et~al.}(2019)\citenamefont {Lin},
  \citenamefont {Benevento}, \citenamefont {Hu},\ and\ \citenamefont
  {Raveri}}]{Lin:2019qug}%
  \BibitemOpen
  \bibfield  {author} {\bibinfo {author} {\bibfnamefont {M.-X.}\ \bibnamefont
  {Lin}}, \bibinfo {author} {\bibfnamefont {G.}~\bibnamefont {Benevento}},
  \bibinfo {author} {\bibfnamefont {W.}~\bibnamefont {Hu}},\ and\ \bibinfo
  {author} {\bibfnamefont {M.}~\bibnamefont {Raveri}},\ }\bibfield  {title}
  {\bibinfo {title} {{Acoustic Dark Energy: Potential Conversion of the Hubble
  Tension}},\ }\href {https://doi.org/10.1103/PhysRevD.100.063542} {\bibfield
  {journal} {\bibinfo  {journal} {Phys. Rev.}\ }\textbf {\bibinfo {volume}
  {D100}},\ \bibinfo {pages} {063542} (\bibinfo {year} {2019})},\ \Eprint
  {https://arxiv.org/abs/1905.12618} {arXiv:1905.12618 [astro-ph.CO]}
  \BibitemShut {NoStop}%
%%CITATION = ARXIV:1905.12618;%%
\bibitem [{\citenamefont {Kaloper}(2019)}]{Kaloper:2019lpl}%
  \BibitemOpen
  \bibfield  {author} {\bibinfo {author} {\bibfnamefont {N.}~\bibnamefont
  {Kaloper}},\ }\bibfield  {title} {\bibinfo {title} {{Dark energy, $H_0$ and
  weak gravity conjecture}},\ }\href
  {https://doi.org/10.1142/S0218271819440176} {\bibfield  {journal} {\bibinfo
  {journal} {Int. J. Mod. Phys. D}\ }\textbf {\bibinfo {volume} {28}},\
  \bibinfo {pages} {1944017} (\bibinfo {year} {2019})},\ \Eprint
  {https://arxiv.org/abs/1903.11676} {arXiv:1903.11676 [hep-th]} \BibitemShut
  {NoStop}%
\bibitem [{\citenamefont {Alexander}\ and\ \citenamefont
  {McDonough}(2019)}]{Alexander:2019rsc}%
  \BibitemOpen
  \bibfield  {author} {\bibinfo {author} {\bibfnamefont {S.}~\bibnamefont
  {Alexander}}\ and\ \bibinfo {author} {\bibfnamefont {E.}~\bibnamefont
  {McDonough}},\ }\bibfield  {title} {\bibinfo {title} {{Axion-Dilaton
  Destabilization and the Hubble Tension}},\ }\href
  {https://doi.org/10.1016/j.physletb.2019.134830} {\bibfield  {journal}
  {\bibinfo  {journal} {Phys. Lett.}\ }\textbf {\bibinfo {volume} {B797}},\
  \bibinfo {pages} {134830} (\bibinfo {year} {2019})},\ \Eprint
  {https://arxiv.org/abs/1904.08912} {arXiv:1904.08912 [astro-ph.CO]}
  \BibitemShut {NoStop}%
%%CITATION = ARXIV:1904.08912;%%
\bibitem [{\citenamefont {Hardy}\ and\ \citenamefont
  {Parameswaran}(2020)}]{Hardy:2019apu}%
  \BibitemOpen
  \bibfield  {author} {\bibinfo {author} {\bibfnamefont {E.}~\bibnamefont
  {Hardy}}\ and\ \bibinfo {author} {\bibfnamefont {S.}~\bibnamefont
  {Parameswaran}},\ }\bibfield  {title} {\bibinfo {title} {{Thermal Dark
  Energy}},\ }\href {https://doi.org/10.1103/PhysRevD.101.023503} {\bibfield
  {journal} {\bibinfo  {journal} {Phys. Rev.}\ }\textbf {\bibinfo {volume}
  {D101}},\ \bibinfo {pages} {023503} (\bibinfo {year} {2020})},\ \Eprint
  {https://arxiv.org/abs/1907.10141} {arXiv:1907.10141 [hep-th]} \BibitemShut
  {NoStop}%
%%CITATION = ARXIV:1907.10141;%%
\bibitem [{\citenamefont {Sakstein}\ and\ \citenamefont
  {Trodden}(2020)}]{Sakstein:2019fmf}%
  \BibitemOpen
  \bibfield  {author} {\bibinfo {author} {\bibfnamefont {J.}~\bibnamefont
  {Sakstein}}\ and\ \bibinfo {author} {\bibfnamefont {M.}~\bibnamefont
  {Trodden}},\ }\bibfield  {title} {\bibinfo {title} {{Early dark energy from
  massive neutrinos -- a natural resolution of the Hubble tension}},\ }\href
  {https://doi.org/10.1103/PhysRevLett.124.161301} {\bibfield  {journal}
  {\bibinfo  {journal} {Phys. Rev. Lett.}\ }\textbf {\bibinfo {volume} {124}},\
  \bibinfo {pages} {161301} (\bibinfo {year} {2020})},\ \Eprint
  {https://arxiv.org/abs/1911.11760} {arXiv:1911.11760 [astro-ph.CO]}
  \BibitemShut {NoStop}%
\bibitem [{\citenamefont {Berghaus}\ and\ \citenamefont
  {Karwal}(2020)}]{Berghaus:2019cls}%
  \BibitemOpen
  \bibfield  {author} {\bibinfo {author} {\bibfnamefont {K.~V.}\ \bibnamefont
  {Berghaus}}\ and\ \bibinfo {author} {\bibfnamefont {T.}~\bibnamefont
  {Karwal}},\ }\bibfield  {title} {\bibinfo {title} {{Thermal Friction as a
  Solution to the Hubble Tension}},\ }\href
  {https://doi.org/10.1103/PhysRevD.101.083537} {\bibfield  {journal} {\bibinfo
   {journal} {Phys. Rev. D}\ }\textbf {\bibinfo {volume} {101}},\ \bibinfo
  {pages} {083537} (\bibinfo {year} {2020})},\ \Eprint
  {https://arxiv.org/abs/1911.06281} {arXiv:1911.06281 [astro-ph.CO]}
  \BibitemShut {NoStop}%
\bibitem [{\citenamefont {Braglia}\ \emph {et~al.}(2020)\citenamefont
  {Braglia}, \citenamefont {Emond}, \citenamefont {Finelli}, \citenamefont
  {Gumrukcuoglu},\ and\ \citenamefont {Koyama}}]{1798362}%
  \BibitemOpen
  \bibfield  {author} {\bibinfo {author} {\bibfnamefont {M.}~\bibnamefont
  {Braglia}}, \bibinfo {author} {\bibfnamefont {W.~T.}\ \bibnamefont {Emond}},
  \bibinfo {author} {\bibfnamefont {F.}~\bibnamefont {Finelli}}, \bibinfo
  {author} {\bibfnamefont {A.~E.}\ \bibnamefont {Gumrukcuoglu}},\ and\ \bibinfo
  {author} {\bibfnamefont {K.}~\bibnamefont {Koyama}},\ }\bibfield  {title}
  {\bibinfo {title} {{Unified framework for Early Dark Energy from
  $\alpha$-attractors}},\ }\href@noop {} {\  (\bibinfo {year} {2020})},\
  \Eprint {https://arxiv.org/abs/2005.14053} {arXiv:2005.14053 [astro-ph.CO]}
  \BibitemShut {NoStop}%
\bibitem [{\citenamefont {Lin}\ \emph {et~al.}(2020)\citenamefont {Lin},
  \citenamefont {Hu},\ and\ \citenamefont {Raveri}}]{Lin:2020jcb}%
  \BibitemOpen
  \bibfield  {author} {\bibinfo {author} {\bibfnamefont {M.-X.}\ \bibnamefont
  {Lin}}, \bibinfo {author} {\bibfnamefont {W.}~\bibnamefont {Hu}},\ and\
  \bibinfo {author} {\bibfnamefont {M.}~\bibnamefont {Raveri}},\ }\bibfield
  {title} {\bibinfo {title} {{Testing $H_0$ in Acoustic Dark Energy with Planck
  and ACT Polarization}},\ }\href {https://doi.org/10.1103/PhysRevD.102.123523}
  {\bibfield  {journal} {\bibinfo  {journal} {Phys. Rev. D}\ }\textbf {\bibinfo
  {volume} {102}},\ \bibinfo {pages} {123523} (\bibinfo {year} {2020})},\
  \Eprint {https://arxiv.org/abs/2009.08974} {arXiv:2009.08974 [astro-ph.CO]}
  \BibitemShut {NoStop}%
\bibitem [{\citenamefont {Carrillo~Gonz\'alez}\ \emph
  {et~al.}(2021)\citenamefont {Carrillo~Gonz\'alez}, \citenamefont {Liang},
  \citenamefont {Sakstein},\ and\ \citenamefont
  {Trodden}}]{CarrilloGonzalez:2020oac}%
  \BibitemOpen
  \bibfield  {author} {\bibinfo {author} {\bibfnamefont {M.}~\bibnamefont
  {Carrillo~Gonz\'alez}}, \bibinfo {author} {\bibfnamefont {Q.}~\bibnamefont
  {Liang}}, \bibinfo {author} {\bibfnamefont {J.}~\bibnamefont {Sakstein}},\
  and\ \bibinfo {author} {\bibfnamefont {M.}~\bibnamefont {Trodden}},\
  }\bibfield  {title} {\bibinfo {title} {{Neutrino-Assisted Early Dark Energy:
  Theory and Cosmology}},\ }\href
  {https://doi.org/10.1088/1475-7516/2021/04/063} {\bibfield  {journal}
  {\bibinfo  {journal} {JCAP}\ }\textbf {\bibinfo {volume} {04}},\ \bibinfo
  {pages} {063}},\ \Eprint {https://arxiv.org/abs/2011.09895} {arXiv:2011.09895
  [astro-ph.CO]} \BibitemShut {NoStop}%
\bibitem [{\citenamefont {Freese}\ and\ \citenamefont
  {Winkler}(2021)}]{Freese:2021rjq}%
  \BibitemOpen
  \bibfield  {author} {\bibinfo {author} {\bibfnamefont {K.}~\bibnamefont
  {Freese}}\ and\ \bibinfo {author} {\bibfnamefont {M.~W.}\ \bibnamefont
  {Winkler}},\ }\bibfield  {title} {\bibinfo {title} {{Chain Early Dark Energy:
  Solving the Hubble Tension and Explaining Today's Dark Energy}},\ }\href@noop
  {} {\  (\bibinfo {year} {2021})},\ \Eprint {https://arxiv.org/abs/2102.13655}
  {arXiv:2102.13655 [astro-ph.CO]} \BibitemShut {NoStop}%
\bibitem [{\citenamefont {Allali}\ \emph {et~al.}(2021)\citenamefont {Allali},
  \citenamefont {Hertzberg},\ and\ \citenamefont {Rompineve}}]{Allali:2021azp}%
  \BibitemOpen
  \bibfield  {author} {\bibinfo {author} {\bibfnamefont {I.~J.}\ \bibnamefont
  {Allali}}, \bibinfo {author} {\bibfnamefont {M.~P.}\ \bibnamefont
  {Hertzberg}},\ and\ \bibinfo {author} {\bibfnamefont {F.}~\bibnamefont
  {Rompineve}},\ }\bibfield  {title} {\bibinfo {title} {{A Dark Sector to
  Restore Cosmological Concordance}},\ }\href@noop {} {\  (\bibinfo {year}
  {2021})},\ \Eprint {https://arxiv.org/abs/2104.12798} {arXiv:2104.12798
  [astro-ph.CO]} \BibitemShut {NoStop}%
\bibitem [{\citenamefont {Sabla}\ and\ \citenamefont
  {Caldwell}(2021)}]{Sabla:2021nfy}%
  \BibitemOpen
  \bibfield  {author} {\bibinfo {author} {\bibfnamefont {V.~I.}\ \bibnamefont
  {Sabla}}\ and\ \bibinfo {author} {\bibfnamefont {R.~R.}\ \bibnamefont
  {Caldwell}},\ }\bibfield  {title} {\bibinfo {title} {{No $H_0$ assistance
  from assisted quintessence}},\ }\href
  {https://doi.org/10.1103/PhysRevD.103.103506} {\bibfield  {journal} {\bibinfo
   {journal} {Phys. Rev. D}\ }\textbf {\bibinfo {volume} {103}},\ \bibinfo
  {pages} {103506} (\bibinfo {year} {2021})},\ \Eprint
  {https://arxiv.org/abs/2103.04999} {arXiv:2103.04999 [astro-ph.CO]}
  \BibitemShut {NoStop}%
\bibitem [{\citenamefont {Karwal}\ \emph {et~al.}(2021)\citenamefont {Karwal},
  \citenamefont {Raveri}, \citenamefont {Jain}, \citenamefont {Khoury},\ and\
  \citenamefont {Trodden}}]{Karwal:2021vpk}%
  \BibitemOpen
  \bibfield  {author} {\bibinfo {author} {\bibfnamefont {T.}~\bibnamefont
  {Karwal}}, \bibinfo {author} {\bibfnamefont {M.}~\bibnamefont {Raveri}},
  \bibinfo {author} {\bibfnamefont {B.}~\bibnamefont {Jain}}, \bibinfo {author}
  {\bibfnamefont {J.}~\bibnamefont {Khoury}},\ and\ \bibinfo {author}
  {\bibfnamefont {M.}~\bibnamefont {Trodden}},\ }\bibfield  {title} {\bibinfo
  {title} {{Chameleon Early Dark Energy and the Hubble Tension}},\ }\href@noop
  {} {\  (\bibinfo {year} {2021})},\ \Eprint {https://arxiv.org/abs/2106.13290}
  {arXiv:2106.13290 [astro-ph.CO]} \BibitemShut {NoStop}%
\bibitem [{\citenamefont {Vagnozzi}(2021)}]{Vagnozzi:2021gjh}%
  \BibitemOpen
  \bibfield  {author} {\bibinfo {author} {\bibfnamefont {S.}~\bibnamefont
  {Vagnozzi}},\ }\bibfield  {title} {\bibinfo {title} {{Consistency tests of
  $\Lambda$CDM from the early integrated Sachs-Wolfe effect: Implications for
  early-time new physics and the Hubble tension}},\ }\href@noop {} {\
  (\bibinfo {year} {2021})},\ \Eprint {https://arxiv.org/abs/2105.10425}
  {arXiv:2105.10425 [astro-ph.CO]} \BibitemShut {NoStop}%
\bibitem [{\citenamefont {G\'omez-Valent}\ \emph {et~al.}(2021)\citenamefont
  {G\'omez-Valent}, \citenamefont {Zheng}, \citenamefont {Amendola},
  \citenamefont {Pettorino},\ and\ \citenamefont
  {Wetterich}}]{Gomez-Valent:2021cbe}%
  \BibitemOpen
  \bibfield  {author} {\bibinfo {author} {\bibfnamefont {A.}~\bibnamefont
  {G\'omez-Valent}}, \bibinfo {author} {\bibfnamefont {Z.}~\bibnamefont
  {Zheng}}, \bibinfo {author} {\bibfnamefont {L.}~\bibnamefont {Amendola}},
  \bibinfo {author} {\bibfnamefont {V.}~\bibnamefont {Pettorino}},\ and\
  \bibinfo {author} {\bibfnamefont {C.}~\bibnamefont {Wetterich}},\ }\bibfield
  {title} {\bibinfo {title} {{Early dark energy in the pre- and
  post-recombination epochs}},\ }\href@noop {} {\  (\bibinfo {year} {2021})},\
  \Eprint {https://arxiv.org/abs/2107.11065} {arXiv:2107.11065 [astro-ph.CO]}
  \BibitemShut {NoStop}%
\bibitem [{\citenamefont {Moss}\ \emph {et~al.}(2021)\citenamefont {Moss},
  \citenamefont {Copeland}, \citenamefont {Bamford},\ and\ \citenamefont
  {Clarke}}]{Moss:2021obd}%
  \BibitemOpen
  \bibfield  {author} {\bibinfo {author} {\bibfnamefont {A.}~\bibnamefont
  {Moss}}, \bibinfo {author} {\bibfnamefont {E.}~\bibnamefont {Copeland}},
  \bibinfo {author} {\bibfnamefont {S.}~\bibnamefont {Bamford}},\ and\ \bibinfo
  {author} {\bibfnamefont {T.}~\bibnamefont {Clarke}},\ }\bibfield  {title}
  {\bibinfo {title} {{A model-independent reconstruction of dark energy to very
  high redshift}},\ }\href@noop {} {\  (\bibinfo {year} {2021})},\ \Eprint
  {https://arxiv.org/abs/2109.14848} {arXiv:2109.14848 [astro-ph.CO]}
  \BibitemShut {NoStop}%
\bibitem [{\citenamefont {Clark}\ \emph {et~al.}(2021)\citenamefont {Clark},
  \citenamefont {Vattis}, \citenamefont {Fan},\ and\ \citenamefont
  {Koushiappas}}]{Clark:2021hlo}%
  \BibitemOpen
  \bibfield  {author} {\bibinfo {author} {\bibfnamefont {S.~J.}\ \bibnamefont
  {Clark}}, \bibinfo {author} {\bibfnamefont {K.}~\bibnamefont {Vattis}},
  \bibinfo {author} {\bibfnamefont {J.}~\bibnamefont {Fan}},\ and\ \bibinfo
  {author} {\bibfnamefont {S.~M.}\ \bibnamefont {Koushiappas}},\ }\bibfield
  {title} {\bibinfo {title} {{The $H_0$ and $S_8$ tensions necessitate early
  and late time changes to $\Lambda$CDM}},\ }\href@noop {} {\  (\bibinfo {year}
  {2021})},\ \Eprint {https://arxiv.org/abs/2110.09562} {arXiv:2110.09562
  [astro-ph.CO]} \BibitemShut {NoStop}%
\bibitem [{\citenamefont {Escudero}\ and\ \citenamefont
  {Witte}(2020)}]{Escudero:2019gvw}%
  \BibitemOpen
  \bibfield  {author} {\bibinfo {author} {\bibfnamefont {M.}~\bibnamefont
  {Escudero}}\ and\ \bibinfo {author} {\bibfnamefont {S.~J.}\ \bibnamefont
  {Witte}},\ }\bibfield  {title} {\bibinfo {title} {{A CMB search for the
  neutrino mass mechanism and its relation to the Hubble tension}},\ }\href
  {https://doi.org/10.1140/epjc/s10052-020-7854-5} {\bibfield  {journal}
  {\bibinfo  {journal} {Eur. Phys. J. C}\ }\textbf {\bibinfo {volume} {80}},\
  \bibinfo {pages} {294} (\bibinfo {year} {2020})},\ \Eprint
  {https://arxiv.org/abs/1909.04044} {arXiv:1909.04044 [astro-ph.CO]}
  \BibitemShut {NoStop}%
\bibitem [{\citenamefont {Kreisch}\ \emph {et~al.}(2020)\citenamefont
  {Kreisch}, \citenamefont {Cyr-Racine},\ and\ \citenamefont
  {Dore}}]{Kreisch:2019yzn}%
  \BibitemOpen
  \bibfield  {author} {\bibinfo {author} {\bibfnamefont {C.~D.}\ \bibnamefont
  {Kreisch}}, \bibinfo {author} {\bibfnamefont {F.-Y.}\ \bibnamefont
  {Cyr-Racine}},\ and\ \bibinfo {author} {\bibfnamefont {O.}~\bibnamefont
  {Dore}},\ }\bibfield  {title} {\bibinfo {title} {{The Neutrino Puzzle:
  Anomalies, Interactions, and Cosmological Tensions}},\ }\href
  {https://doi.org/10.1103/PhysRevD.101.123505} {\bibfield  {journal} {\bibinfo
   {journal} {Phys. Rev. D}\ }\textbf {\bibinfo {volume} {101}},\ \bibinfo
  {pages} {123505} (\bibinfo {year} {2020})},\ \Eprint
  {https://arxiv.org/abs/1902.00534} {arXiv:1902.00534 [astro-ph.CO]}
  \BibitemShut {NoStop}%
\bibitem [{\citenamefont {Park}\ \emph {et~al.}(2019)\citenamefont {Park},
  \citenamefont {Kreisch}, \citenamefont {Dunkley}, \citenamefont
  {Hadzhiyska},\ and\ \citenamefont {Cyr-Racine}}]{Park:2019ibn}%
  \BibitemOpen
  \bibfield  {author} {\bibinfo {author} {\bibfnamefont {M.}~\bibnamefont
  {Park}}, \bibinfo {author} {\bibfnamefont {C.~D.}\ \bibnamefont {Kreisch}},
  \bibinfo {author} {\bibfnamefont {J.}~\bibnamefont {Dunkley}}, \bibinfo
  {author} {\bibfnamefont {B.}~\bibnamefont {Hadzhiyska}},\ and\ \bibinfo
  {author} {\bibfnamefont {F.-Y.}\ \bibnamefont {Cyr-Racine}},\ }\bibfield
  {title} {\bibinfo {title} {{$\Lambda$CDM or self-interacting neutrinos: How
  CMB data can tell the two models apart}},\ }\href
  {https://doi.org/10.1103/PhysRevD.100.063524} {\bibfield  {journal} {\bibinfo
   {journal} {Phys. Rev. D}\ }\textbf {\bibinfo {volume} {100}},\ \bibinfo
  {pages} {063524} (\bibinfo {year} {2019})},\ \Eprint
  {https://arxiv.org/abs/1904.02625} {arXiv:1904.02625 [astro-ph.CO]}
  \BibitemShut {NoStop}%
\bibitem [{\citenamefont {Pandey}\ \emph {et~al.}(2020)\citenamefont {Pandey},
  \citenamefont {Karwal},\ and\ \citenamefont {Das}}]{Pandey:2019plg}%
  \BibitemOpen
  \bibfield  {author} {\bibinfo {author} {\bibfnamefont {K.~L.}\ \bibnamefont
  {Pandey}}, \bibinfo {author} {\bibfnamefont {T.}~\bibnamefont {Karwal}},\
  and\ \bibinfo {author} {\bibfnamefont {S.}~\bibnamefont {Das}},\ }\bibfield
  {title} {\bibinfo {title} {{Alleviating the $H_0$ and $\sigma_8$ anomalies
  with a decaying dark matter model}},\ }\href
  {https://doi.org/10.1088/1475-7516/2020/07/026} {\bibfield  {journal}
  {\bibinfo  {journal} {JCAP}\ }\textbf {\bibinfo {volume} {07}},\ \bibinfo
  {pages} {026}},\ \Eprint {https://arxiv.org/abs/1902.10636} {arXiv:1902.10636
  [astro-ph.CO]} \BibitemShut {NoStop}%
\bibitem [{\citenamefont {Sekiguchi}\ and\ \citenamefont
  {Takahashi}(2021)}]{Sekiguchi:2020teg}%
  \BibitemOpen
  \bibfield  {author} {\bibinfo {author} {\bibfnamefont {T.}~\bibnamefont
  {Sekiguchi}}\ and\ \bibinfo {author} {\bibfnamefont {T.}~\bibnamefont
  {Takahashi}},\ }\bibfield  {title} {\bibinfo {title} {{Early recombination as
  a solution to the $H_0$ tension}},\ }\href
  {https://doi.org/10.1103/PhysRevD.103.083507} {\bibfield  {journal} {\bibinfo
   {journal} {Phys. Rev. D}\ }\textbf {\bibinfo {volume} {103}},\ \bibinfo
  {pages} {083507} (\bibinfo {year} {2021})},\ \Eprint
  {https://arxiv.org/abs/2007.03381} {arXiv:2007.03381 [astro-ph.CO]}
  \BibitemShut {NoStop}%
\bibitem [{\citenamefont {Jedamzik}\ and\ \citenamefont
  {Pogosian}(2020)}]{Jedamzik:2020krr}%
  \BibitemOpen
  \bibfield  {author} {\bibinfo {author} {\bibfnamefont {K.}~\bibnamefont
  {Jedamzik}}\ and\ \bibinfo {author} {\bibfnamefont {L.}~\bibnamefont
  {Pogosian}},\ }\bibfield  {title} {\bibinfo {title} {{Relieving the Hubble
  tension with primordial magnetic fields}},\ }\href
  {https://doi.org/10.1103/PhysRevLett.125.181302} {\bibfield  {journal}
  {\bibinfo  {journal} {Phys. Rev. Lett.}\ }\textbf {\bibinfo {volume} {125}},\
  \bibinfo {pages} {181302} (\bibinfo {year} {2020})},\ \Eprint
  {https://arxiv.org/abs/2004.09487} {arXiv:2004.09487 [astro-ph.CO]}
  \BibitemShut {NoStop}%
\bibitem [{\citenamefont {Escudero}\ and\ \citenamefont
  {Witte}(2021)}]{Escudero:2021rfi}%
  \BibitemOpen
  \bibfield  {author} {\bibinfo {author} {\bibfnamefont {M.}~\bibnamefont
  {Escudero}}\ and\ \bibinfo {author} {\bibfnamefont {S.~J.}\ \bibnamefont
  {Witte}},\ }\bibfield  {title} {\bibinfo {title} {{The hubble tension as a
  hint of leptogenesis and neutrino mass generation}},\ }\href
  {https://doi.org/10.1140/epjc/s10052-021-09276-5} {\bibfield  {journal}
  {\bibinfo  {journal} {Eur. Phys. J. C}\ }\textbf {\bibinfo {volume} {81}},\
  \bibinfo {pages} {515} (\bibinfo {year} {2021})},\ \Eprint
  {https://arxiv.org/abs/2103.03249} {arXiv:2103.03249 [hep-ph]} \BibitemShut
  {NoStop}%
\bibitem [{\citenamefont {Bansal}\ \emph {et~al.}(2021)\citenamefont {Bansal},
  \citenamefont {Kim}, \citenamefont {Kolda}, \citenamefont {Low},\ and\
  \citenamefont {Tsai}}]{Bansal:2021dfh}%
  \BibitemOpen
  \bibfield  {author} {\bibinfo {author} {\bibfnamefont {S.}~\bibnamefont
  {Bansal}}, \bibinfo {author} {\bibfnamefont {J.~H.}\ \bibnamefont {Kim}},
  \bibinfo {author} {\bibfnamefont {C.}~\bibnamefont {Kolda}}, \bibinfo
  {author} {\bibfnamefont {M.}~\bibnamefont {Low}},\ and\ \bibinfo {author}
  {\bibfnamefont {Y.}~\bibnamefont {Tsai}},\ }\bibfield  {title} {\bibinfo
  {title} {{Mirror Twin Higgs Cosmology: Constraints and a Possible Resolution
  to the $H_0$ and $S_8$ Tensions}},\ }\href@noop {} {\  (\bibinfo {year}
  {2021})},\ \Eprint {https://arxiv.org/abs/2110.04317} {arXiv:2110.04317
  [hep-ph]} \BibitemShut {NoStop}%
\bibitem [{\citenamefont {Aloni}\ \emph {et~al.}(2021)\citenamefont {Aloni},
  \citenamefont {Berlin}, \citenamefont {Joseph}, \citenamefont {Schmaltz},\
  and\ \citenamefont {Weiner}}]{Aloni:2021eaq}%
  \BibitemOpen
  \bibfield  {author} {\bibinfo {author} {\bibfnamefont {D.}~\bibnamefont
  {Aloni}}, \bibinfo {author} {\bibfnamefont {A.}~\bibnamefont {Berlin}},
  \bibinfo {author} {\bibfnamefont {M.}~\bibnamefont {Joseph}}, \bibinfo
  {author} {\bibfnamefont {M.}~\bibnamefont {Schmaltz}},\ and\ \bibinfo
  {author} {\bibfnamefont {N.}~\bibnamefont {Weiner}},\ }\bibfield  {title}
  {\bibinfo {title} {{A Step in Understanding the Hubble Tension}},\
  }\href@noop {} {\  (\bibinfo {year} {2021})},\ \Eprint
  {https://arxiv.org/abs/2111.00014} {arXiv:2111.00014 [astro-ph.CO]}
  \BibitemShut {NoStop}%
\bibitem [{\citenamefont {Agrawal}\ \emph {et~al.}(2019)\citenamefont
  {Agrawal}, \citenamefont {Cyr-Racine}, \citenamefont {Pinner},\ and\
  \citenamefont {Randall}}]{Agrawal:2019lmo}%
  \BibitemOpen
  \bibfield  {author} {\bibinfo {author} {\bibfnamefont {P.}~\bibnamefont
  {Agrawal}}, \bibinfo {author} {\bibfnamefont {F.-Y.}\ \bibnamefont
  {Cyr-Racine}}, \bibinfo {author} {\bibfnamefont {D.}~\bibnamefont {Pinner}},\
  and\ \bibinfo {author} {\bibfnamefont {L.}~\bibnamefont {Randall}},\
  }\bibfield  {title} {\bibinfo {title} {{Rock 'n' Roll Solutions to the Hubble
  Tension}},\ }\href@noop {} {\  (\bibinfo {year} {2019})},\ \Eprint
  {https://arxiv.org/abs/1904.01016} {arXiv:1904.01016 [astro-ph.CO]}
  \BibitemShut {NoStop}%
%%CITATION = ARXIV:1904.01016;%%
\bibitem [{\citenamefont {Ivanov}\ \emph {et~al.}(2020)\citenamefont {Ivanov},
  \citenamefont {McDonough}, \citenamefont {Hill}, \citenamefont {Simonovi\'c},
  \citenamefont {Toomey}, \citenamefont {Alexander},\ and\ \citenamefont
  {Zaldarriaga}}]{Ivanov:2020ril}%
  \BibitemOpen
  \bibfield  {author} {\bibinfo {author} {\bibfnamefont {M.~M.}\ \bibnamefont
  {Ivanov}}, \bibinfo {author} {\bibfnamefont {E.}~\bibnamefont {McDonough}},
  \bibinfo {author} {\bibfnamefont {J.~C.}\ \bibnamefont {Hill}}, \bibinfo
  {author} {\bibfnamefont {M.}~\bibnamefont {Simonovi\'c}}, \bibinfo {author}
  {\bibfnamefont {M.~W.}\ \bibnamefont {Toomey}}, \bibinfo {author}
  {\bibfnamefont {S.}~\bibnamefont {Alexander}},\ and\ \bibinfo {author}
  {\bibfnamefont {M.}~\bibnamefont {Zaldarriaga}},\ }\bibfield  {title}
  {\bibinfo {title} {{Constraining Early Dark Energy with Large-Scale
  Structure}},\ }\href@noop {} {\  (\bibinfo {year} {2020})},\ \Eprint
  {https://arxiv.org/abs/2006.11235} {arXiv:2006.11235 [astro-ph.CO]}
  \BibitemShut {NoStop}%
\bibitem [{\citenamefont {Hill}\ \emph {et~al.}(2020)\citenamefont {Hill},
  \citenamefont {McDonough}, \citenamefont {Toomey},\ and\ \citenamefont
  {Alexander}}]{Hill:2020osr}%
  \BibitemOpen
  \bibfield  {author} {\bibinfo {author} {\bibfnamefont {J.~C.}\ \bibnamefont
  {Hill}}, \bibinfo {author} {\bibfnamefont {E.}~\bibnamefont {McDonough}},
  \bibinfo {author} {\bibfnamefont {M.~W.}\ \bibnamefont {Toomey}},\ and\
  \bibinfo {author} {\bibfnamefont {S.}~\bibnamefont {Alexander}},\ }\bibfield
  {title} {\bibinfo {title} {{Early Dark Energy Does Not Restore Cosmological
  Concordance}},\ }\href@noop {} {\  (\bibinfo {year} {2020})},\ \Eprint
  {https://arxiv.org/abs/2003.07355} {arXiv:2003.07355 [astro-ph.CO]}
  \BibitemShut {NoStop}%
\bibitem [{\citenamefont {Di~Valentino}\ \emph
  {et~al.}(2021{\natexlab{c}})\citenamefont {Di~Valentino} \emph
  {et~al.}}]{DiValentino:2020vvd}%
  \BibitemOpen
  \bibfield  {author} {\bibinfo {author} {\bibfnamefont {E.}~\bibnamefont
  {Di~Valentino}} \emph {et~al.},\ }\bibfield  {title} {\bibinfo {title}
  {{Cosmology intertwined III: $f\sigma_8$ and $S_8$}},\ }\href
  {https://doi.org/10.1016/j.astropartphys.2021.102604} {\bibfield  {journal}
  {\bibinfo  {journal} {Astropart. Phys.}\ }\textbf {\bibinfo {volume} {131}},\
  \bibinfo {pages} {102604} (\bibinfo {year} {2021}{\natexlab{c}})},\ \Eprint
  {https://arxiv.org/abs/2008.11285} {arXiv:2008.11285 [astro-ph.CO]}
  \BibitemShut {NoStop}%
\bibitem [{\citenamefont {Niedermann}\ and\ \citenamefont
  {Sloth}(2021{\natexlab{a}})}]{Niedermann:2019olb}%
  \BibitemOpen
  \bibfield  {author} {\bibinfo {author} {\bibfnamefont {F.}~\bibnamefont
  {Niedermann}}\ and\ \bibinfo {author} {\bibfnamefont {M.~S.}\ \bibnamefont
  {Sloth}},\ }\bibfield  {title} {\bibinfo {title} {{New Early Dark Energy}},\
  }\href {https://doi.org/10.1103/PhysRevD.103.L041303} {\bibfield  {journal}
  {\bibinfo  {journal} {Phys. Rev. D}\ }\textbf {\bibinfo {volume} {103}},\
  \bibinfo {pages} {L041303} (\bibinfo {year} {2021}{\natexlab{a}})},\ \Eprint
  {https://arxiv.org/abs/1910.10739} {arXiv:1910.10739 [astro-ph.CO]}
  \BibitemShut {NoStop}%
\bibitem [{\citenamefont {Niedermann}\ and\ \citenamefont
  {Sloth}(2020{\natexlab{a}})}]{Niedermann:2020dwg}%
  \BibitemOpen
  \bibfield  {author} {\bibinfo {author} {\bibfnamefont {F.}~\bibnamefont
  {Niedermann}}\ and\ \bibinfo {author} {\bibfnamefont {M.~S.}\ \bibnamefont
  {Sloth}},\ }\bibfield  {title} {\bibinfo {title} {{Resolving the Hubble
  tension with new early dark energy}},\ }\href
  {https://doi.org/10.1103/PhysRevD.102.063527} {\bibfield  {journal} {\bibinfo
   {journal} {Phys. Rev. D}\ }\textbf {\bibinfo {volume} {102}},\ \bibinfo
  {pages} {063527} (\bibinfo {year} {2020}{\natexlab{a}})},\ \Eprint
  {https://arxiv.org/abs/2006.06686} {arXiv:2006.06686 [astro-ph.CO]}
  \BibitemShut {NoStop}%
\bibitem [{\citenamefont {Niedermann}\ and\ \citenamefont
  {Sloth}(2020{\natexlab{b}})}]{Niedermann:2020qbw}%
  \BibitemOpen
  \bibfield  {author} {\bibinfo {author} {\bibfnamefont {F.}~\bibnamefont
  {Niedermann}}\ and\ \bibinfo {author} {\bibfnamefont {M.~S.}\ \bibnamefont
  {Sloth}},\ }\bibfield  {title} {\bibinfo {title} {{New Early Dark Energy is
  compatible with current LSS data}},\ }\href@noop {} {\  (\bibinfo {year}
  {2020}{\natexlab{b}})},\ \Eprint {https://arxiv.org/abs/2009.00006}
  {arXiv:2009.00006 [astro-ph.CO]} \BibitemShut {NoStop}%
\bibitem [{\citenamefont {Sch\"oneberg}\ \emph {et~al.}(2021)\citenamefont
  {Sch\"oneberg}, \citenamefont {Franco~Abell\'an}, \citenamefont
  {P\'erez~S\'anchez}, \citenamefont {Witte}, \citenamefont {Poulin},\ and\
  \citenamefont {Lesgourgues}}]{Schoneberg:2021qvd}%
  \BibitemOpen
  \bibfield  {author} {\bibinfo {author} {\bibfnamefont {N.}~\bibnamefont
  {Sch\"oneberg}}, \bibinfo {author} {\bibfnamefont {G.}~\bibnamefont
  {Franco~Abell\'an}}, \bibinfo {author} {\bibfnamefont {A.}~\bibnamefont
  {P\'erez~S\'anchez}}, \bibinfo {author} {\bibfnamefont {S.~J.}\ \bibnamefont
  {Witte}}, \bibinfo {author} {\bibfnamefont {V.}~\bibnamefont {Poulin}},\ and\
  \bibinfo {author} {\bibfnamefont {J.}~\bibnamefont {Lesgourgues}},\
  }\bibfield  {title} {\bibinfo {title} {{The $H_0$ Olympics: A fair ranking of
  proposed models}},\ }\href@noop {} {\  (\bibinfo {year} {2021})},\ \Eprint
  {https://arxiv.org/abs/2107.10291} {arXiv:2107.10291 [astro-ph.CO]}
  \BibitemShut {NoStop}%
\bibitem [{\citenamefont {Hill}\ \emph {et~al.}(2021)\citenamefont {Hill} \emph
  {et~al.}}]{Hill:2021yec}%
  \BibitemOpen
  \bibfield  {author} {\bibinfo {author} {\bibfnamefont {J.~C.}\ \bibnamefont
  {Hill}} \emph {et~al.},\ }\bibfield  {title} {\bibinfo {title} {{The Atacama
  Cosmology Telescope: Constraints on Pre-Recombination Early Dark Energy}},\
  }\href@noop {} {\  (\bibinfo {year} {2021})},\ \Eprint
  {https://arxiv.org/abs/2109.04451} {arXiv:2109.04451 [astro-ph.CO]}
  \BibitemShut {NoStop}%
\bibitem [{\citenamefont {D'Amico}\ \emph {et~al.}(2019)\citenamefont
  {D'Amico}, \citenamefont {Hamill},\ and\ \citenamefont
  {Kaloper}}]{DAmico:2018hgc}%
  \BibitemOpen
  \bibfield  {author} {\bibinfo {author} {\bibfnamefont {G.}~\bibnamefont
  {D'Amico}}, \bibinfo {author} {\bibfnamefont {T.}~\bibnamefont {Hamill}},\
  and\ \bibinfo {author} {\bibfnamefont {N.}~\bibnamefont {Kaloper}},\
  }\bibfield  {title} {\bibinfo {title} {{Neutrino Masses from Outer Space}},\
  }\href {https://doi.org/10.1016/j.physletb.2019.134846} {\bibfield  {journal}
  {\bibinfo  {journal} {Phys. Lett. B}\ }\textbf {\bibinfo {volume} {797}},\
  \bibinfo {pages} {134846} (\bibinfo {year} {2019})},\ \Eprint
  {https://arxiv.org/abs/1804.01542} {arXiv:1804.01542 [hep-ph]} \BibitemShut
  {NoStop}%
\bibitem [{\citenamefont {Fernandez-Martinez}\ \emph
  {et~al.}(2021)\citenamefont {Fernandez-Martinez}, \citenamefont {Pierre},
  \citenamefont {Pinsard},\ and\ \citenamefont
  {Rosauro-Alcaraz}}]{Fernandez-Martinez:2021ypo}%
  \BibitemOpen
  \bibfield  {author} {\bibinfo {author} {\bibfnamefont {E.}~\bibnamefont
  {Fernandez-Martinez}}, \bibinfo {author} {\bibfnamefont {M.}~\bibnamefont
  {Pierre}}, \bibinfo {author} {\bibfnamefont {E.}~\bibnamefont {Pinsard}},\
  and\ \bibinfo {author} {\bibfnamefont {S.}~\bibnamefont {Rosauro-Alcaraz}},\
  }\bibfield  {title} {\bibinfo {title} {{Inverse Seesaw, dark matter and the
  Hubble tension}},\ }\href@noop {} {\  (\bibinfo {year} {2021})},\ \Eprint
  {https://arxiv.org/abs/2106.05298} {arXiv:2106.05298 [hep-ph]} \BibitemShut
  {NoStop}%
\bibitem [{\citenamefont {Di~Bari}\ \emph {et~al.}(2021)\citenamefont
  {Di~Bari}, \citenamefont {Marfatia},\ and\ \citenamefont
  {Zhou}}]{DiBari:2021dri}%
  \BibitemOpen
  \bibfield  {author} {\bibinfo {author} {\bibfnamefont {P.}~\bibnamefont
  {Di~Bari}}, \bibinfo {author} {\bibfnamefont {D.}~\bibnamefont {Marfatia}},\
  and\ \bibinfo {author} {\bibfnamefont {Y.-L.}\ \bibnamefont {Zhou}},\
  }\bibfield  {title} {\bibinfo {title} {{Gravitational waves from first-order
  phase transitions in Majoron models of neutrino mass}},\ }\href@noop {} {\
  (\bibinfo {year} {2021})},\ \Eprint {https://arxiv.org/abs/2106.00025}
  {arXiv:2106.00025 [hep-ph]} \BibitemShut {NoStop}%
\bibitem [{\citenamefont {Barreiro}\ \emph {et~al.}(1996)\citenamefont
  {Barreiro}, \citenamefont {Copeland}, \citenamefont {Lyth},\ and\
  \citenamefont {Prokopec}}]{Barreiro:1996dx}%
  \BibitemOpen
  \bibfield  {author} {\bibinfo {author} {\bibfnamefont {T.}~\bibnamefont
  {Barreiro}}, \bibinfo {author} {\bibfnamefont {E.~J.}\ \bibnamefont
  {Copeland}}, \bibinfo {author} {\bibfnamefont {D.~H.}\ \bibnamefont {Lyth}},\
  and\ \bibinfo {author} {\bibfnamefont {T.}~\bibnamefont {Prokopec}},\
  }\bibfield  {title} {\bibinfo {title} {{Some aspects of thermal inflation:
  The Finite temperature potential and topological defects}},\ }\href
  {https://doi.org/10.1103/PhysRevD.54.1379} {\bibfield  {journal} {\bibinfo
  {journal} {Phys. Rev. D}\ }\textbf {\bibinfo {volume} {54}},\ \bibinfo
  {pages} {1379} (\bibinfo {year} {1996})},\ \Eprint
  {https://arxiv.org/abs/hep-ph/9602263} {arXiv:hep-ph/9602263} \BibitemShut
  {NoStop}%
\bibitem [{\citenamefont {Buen-Abad}\ \emph {et~al.}(2015)\citenamefont
  {Buen-Abad}, \citenamefont {Marques-Tavares},\ and\ \citenamefont
  {Schmaltz}}]{Buen-Abad:2015ova}%
  \BibitemOpen
  \bibfield  {author} {\bibinfo {author} {\bibfnamefont {M.~A.}\ \bibnamefont
  {Buen-Abad}}, \bibinfo {author} {\bibfnamefont {G.}~\bibnamefont
  {Marques-Tavares}},\ and\ \bibinfo {author} {\bibfnamefont {M.}~\bibnamefont
  {Schmaltz}},\ }\bibfield  {title} {\bibinfo {title} {{Non-Abelian dark matter
  and dark radiation}},\ }\href {https://doi.org/10.1103/PhysRevD.92.023531}
  {\bibfield  {journal} {\bibinfo  {journal} {Phys. Rev. D}\ }\textbf {\bibinfo
  {volume} {92}},\ \bibinfo {pages} {023531} (\bibinfo {year} {2015})},\
  \Eprint {https://arxiv.org/abs/1505.03542} {arXiv:1505.03542 [hep-ph]}
  \BibitemShut {NoStop}%
\bibitem [{\citenamefont {Lesgourgues}\ \emph {et~al.}(2016)\citenamefont
  {Lesgourgues}, \citenamefont {Marques-Tavares},\ and\ \citenamefont
  {Schmaltz}}]{Lesgourgues:2015wza}%
  \BibitemOpen
  \bibfield  {author} {\bibinfo {author} {\bibfnamefont {J.}~\bibnamefont
  {Lesgourgues}}, \bibinfo {author} {\bibfnamefont {G.}~\bibnamefont
  {Marques-Tavares}},\ and\ \bibinfo {author} {\bibfnamefont {M.}~\bibnamefont
  {Schmaltz}},\ }\bibfield  {title} {\bibinfo {title} {{Evidence for dark
  matter interactions in cosmological precision data?}},\ }\href
  {https://doi.org/10.1088/1475-7516/2016/02/037} {\bibfield  {journal}
  {\bibinfo  {journal} {JCAP}\ }\textbf {\bibinfo {volume} {02}},\ \bibinfo
  {pages} {037}},\ \Eprint {https://arxiv.org/abs/1507.04351} {arXiv:1507.04351
  [astro-ph.CO]} \BibitemShut {NoStop}%
\bibitem [{\citenamefont {Buen-Abad}\ \emph {et~al.}(2018)\citenamefont
  {Buen-Abad}, \citenamefont {Schmaltz}, \citenamefont {Lesgourgues},\ and\
  \citenamefont {Brinckmann}}]{Buen-Abad:2017gxg}%
  \BibitemOpen
  \bibfield  {author} {\bibinfo {author} {\bibfnamefont {M.~A.}\ \bibnamefont
  {Buen-Abad}}, \bibinfo {author} {\bibfnamefont {M.}~\bibnamefont {Schmaltz}},
  \bibinfo {author} {\bibfnamefont {J.}~\bibnamefont {Lesgourgues}},\ and\
  \bibinfo {author} {\bibfnamefont {T.}~\bibnamefont {Brinckmann}},\ }\bibfield
   {title} {\bibinfo {title} {{Interacting Dark Sector and Precision
  Cosmology}},\ }\href {https://doi.org/10.1088/1475-7516/2018/01/008}
  {\bibfield  {journal} {\bibinfo  {journal} {JCAP}\ }\textbf {\bibinfo
  {volume} {01}},\ \bibinfo {pages} {008}},\ \Eprint
  {https://arxiv.org/abs/1708.09406} {arXiv:1708.09406 [astro-ph.CO]}
  \BibitemShut {NoStop}%
\bibitem [{\citenamefont {Abada}\ and\ \citenamefont
  {Lucente}(2014)}]{Abada:2014vea}%
  \BibitemOpen
  \bibfield  {author} {\bibinfo {author} {\bibfnamefont {A.}~\bibnamefont
  {Abada}}\ and\ \bibinfo {author} {\bibfnamefont {M.}~\bibnamefont
  {Lucente}},\ }\bibfield  {title} {\bibinfo {title} {{Looking for the minimal
  inverse seesaw realisation}},\ }\href
  {https://doi.org/10.1016/j.nuclphysb.2014.06.003} {\bibfield  {journal}
  {\bibinfo  {journal} {Nucl. Phys. B}\ }\textbf {\bibinfo {volume} {885}},\
  \bibinfo {pages} {651} (\bibinfo {year} {2014})},\ \Eprint
  {https://arxiv.org/abs/1401.1507} {arXiv:1401.1507 [hep-ph]} \BibitemShut
  {NoStop}%
\bibitem [{\citenamefont {Kopp}\ \emph {et~al.}(2013)\citenamefont {Kopp},
  \citenamefont {Machado}, \citenamefont {Maltoni},\ and\ \citenamefont
  {Schwetz}}]{Kopp:2013vaa}%
  \BibitemOpen
  \bibfield  {author} {\bibinfo {author} {\bibfnamefont {J.}~\bibnamefont
  {Kopp}}, \bibinfo {author} {\bibfnamefont {P.~A.~N.}\ \bibnamefont
  {Machado}}, \bibinfo {author} {\bibfnamefont {M.}~\bibnamefont {Maltoni}},\
  and\ \bibinfo {author} {\bibfnamefont {T.}~\bibnamefont {Schwetz}},\
  }\bibfield  {title} {\bibinfo {title} {{Sterile Neutrino Oscillations: The
  Global Picture}},\ }\href {https://doi.org/10.1007/JHEP05(2013)050}
  {\bibfield  {journal} {\bibinfo  {journal} {JHEP}\ }\textbf {\bibinfo
  {volume} {05}},\ \bibinfo {pages} {050}},\ \Eprint
  {https://arxiv.org/abs/1303.3011} {arXiv:1303.3011 [hep-ph]} \BibitemShut
  {NoStop}%
\bibitem [{\citenamefont {B\"oser}\ \emph {et~al.}(2020)\citenamefont
  {B\"oser}, \citenamefont {Buck}, \citenamefont {Giunti}, \citenamefont
  {Lesgourgues}, \citenamefont {Ludhova}, \citenamefont {Mertens},
  \citenamefont {Schukraft},\ and\ \citenamefont {Wurm}}]{Boser:2019rta}%
  \BibitemOpen
  \bibfield  {author} {\bibinfo {author} {\bibfnamefont {S.}~\bibnamefont
  {B\"oser}}, \bibinfo {author} {\bibfnamefont {C.}~\bibnamefont {Buck}},
  \bibinfo {author} {\bibfnamefont {C.}~\bibnamefont {Giunti}}, \bibinfo
  {author} {\bibfnamefont {J.}~\bibnamefont {Lesgourgues}}, \bibinfo {author}
  {\bibfnamefont {L.}~\bibnamefont {Ludhova}}, \bibinfo {author} {\bibfnamefont
  {S.}~\bibnamefont {Mertens}}, \bibinfo {author} {\bibfnamefont
  {A.}~\bibnamefont {Schukraft}},\ and\ \bibinfo {author} {\bibfnamefont
  {M.}~\bibnamefont {Wurm}},\ }\bibfield  {title} {\bibinfo {title} {{Status of
  Light Sterile Neutrino Searches}},\ }\href
  {https://doi.org/10.1016/j.ppnp.2019.103736} {\bibfield  {journal} {\bibinfo
  {journal} {Prog. Part. Nucl. Phys.}\ }\textbf {\bibinfo {volume} {111}},\
  \bibinfo {pages} {103736} (\bibinfo {year} {2020})},\ \Eprint
  {https://arxiv.org/abs/1906.01739} {arXiv:1906.01739 [hep-ex]} \BibitemShut
  {NoStop}%
\bibitem [{\citenamefont {Dasgupta}\ and\ \citenamefont
  {Kopp}(2021)}]{Dasgupta:2021ies}%
  \BibitemOpen
  \bibfield  {author} {\bibinfo {author} {\bibfnamefont {B.}~\bibnamefont
  {Dasgupta}}\ and\ \bibinfo {author} {\bibfnamefont {J.}~\bibnamefont
  {Kopp}},\ }\bibfield  {title} {\bibinfo {title} {{Sterile Neutrinos}},\
  }\href {https://doi.org/10.1016/j.physrep.2021.06.002} {\bibfield  {journal}
  {\bibinfo  {journal} {Phys. Rept.}\ }\textbf {\bibinfo {volume} {928}},\
  \bibinfo {pages} {63} (\bibinfo {year} {2021})},\ \Eprint
  {https://arxiv.org/abs/2106.05913} {arXiv:2106.05913 [hep-ph]} \BibitemShut
  {NoStop}%
\bibitem [{\citenamefont {Hannestad}\ \emph {et~al.}(2014)\citenamefont
  {Hannestad}, \citenamefont {Hansen},\ and\ \citenamefont
  {Tram}}]{Hannestad:2013ana}%
  \BibitemOpen
  \bibfield  {author} {\bibinfo {author} {\bibfnamefont {S.}~\bibnamefont
  {Hannestad}}, \bibinfo {author} {\bibfnamefont {R.~S.}\ \bibnamefont
  {Hansen}},\ and\ \bibinfo {author} {\bibfnamefont {T.}~\bibnamefont {Tram}},\
  }\bibfield  {title} {\bibinfo {title} {{How Self-Interactions can Reconcile
  Sterile Neutrinos with Cosmology}},\ }\href
  {https://doi.org/10.1103/PhysRevLett.112.031802} {\bibfield  {journal}
  {\bibinfo  {journal} {Phys. Rev. Lett.}\ }\textbf {\bibinfo {volume} {112}},\
  \bibinfo {pages} {031802} (\bibinfo {year} {2014})},\ \Eprint
  {https://arxiv.org/abs/1310.5926} {arXiv:1310.5926 [astro-ph.CO]}
  \BibitemShut {NoStop}%
\bibitem [{\citenamefont {Dasgupta}\ and\ \citenamefont
  {Kopp}(2014)}]{Dasgupta:2013zpn}%
  \BibitemOpen
  \bibfield  {author} {\bibinfo {author} {\bibfnamefont {B.}~\bibnamefont
  {Dasgupta}}\ and\ \bibinfo {author} {\bibfnamefont {J.}~\bibnamefont
  {Kopp}},\ }\bibfield  {title} {\bibinfo {title} {{Cosmologically Safe
  eV-Scale Sterile Neutrinos and Improved Dark Matter Structure}},\ }\href
  {https://doi.org/10.1103/PhysRevLett.112.031803} {\bibfield  {journal}
  {\bibinfo  {journal} {Phys. Rev. Lett.}\ }\textbf {\bibinfo {volume} {112}},\
  \bibinfo {pages} {031803} (\bibinfo {year} {2014})},\ \Eprint
  {https://arxiv.org/abs/1310.6337} {arXiv:1310.6337 [hep-ph]} \BibitemShut
  {NoStop}%
\bibitem [{\citenamefont {Archidiacono}\ \emph {et~al.}(2015)\citenamefont
  {Archidiacono}, \citenamefont {Hannestad}, \citenamefont {Hansen},\ and\
  \citenamefont {Tram}}]{Archidiacono:2014nda}%
  \BibitemOpen
  \bibfield  {author} {\bibinfo {author} {\bibfnamefont {M.}~\bibnamefont
  {Archidiacono}}, \bibinfo {author} {\bibfnamefont {S.}~\bibnamefont
  {Hannestad}}, \bibinfo {author} {\bibfnamefont {R.~S.}\ \bibnamefont
  {Hansen}},\ and\ \bibinfo {author} {\bibfnamefont {T.}~\bibnamefont {Tram}},\
  }\bibfield  {title} {\bibinfo {title} {{Cosmology with self-interacting
  sterile neutrinos and dark matter - A pseudoscalar model}},\ }\href
  {https://doi.org/10.1103/PhysRevD.91.065021} {\bibfield  {journal} {\bibinfo
  {journal} {Phys. Rev. D}\ }\textbf {\bibinfo {volume} {91}},\ \bibinfo
  {pages} {065021} (\bibinfo {year} {2015})},\ \Eprint
  {https://arxiv.org/abs/1404.5915} {arXiv:1404.5915 [astro-ph.CO]}
  \BibitemShut {NoStop}%
\bibitem [{\citenamefont {Archidiacono}\ \emph
  {et~al.}(2016{\natexlab{a}})\citenamefont {Archidiacono}, \citenamefont
  {Hannestad}, \citenamefont {Hansen},\ and\ \citenamefont
  {Tram}}]{Archidiacono:2015oma}%
  \BibitemOpen
  \bibfield  {author} {\bibinfo {author} {\bibfnamefont {M.}~\bibnamefont
  {Archidiacono}}, \bibinfo {author} {\bibfnamefont {S.}~\bibnamefont
  {Hannestad}}, \bibinfo {author} {\bibfnamefont {R.~S.}\ \bibnamefont
  {Hansen}},\ and\ \bibinfo {author} {\bibfnamefont {T.}~\bibnamefont {Tram}},\
  }\bibfield  {title} {\bibinfo {title} {{Sterile neutrinos with pseudoscalar
  self-interactions and cosmology}},\ }\href
  {https://doi.org/10.1103/PhysRevD.93.045004} {\bibfield  {journal} {\bibinfo
  {journal} {Phys. Rev. D}\ }\textbf {\bibinfo {volume} {93}},\ \bibinfo
  {pages} {045004} (\bibinfo {year} {2016}{\natexlab{a}})},\ \Eprint
  {https://arxiv.org/abs/1508.02504} {arXiv:1508.02504 [astro-ph.CO]}
  \BibitemShut {NoStop}%
\bibitem [{\citenamefont {Archidiacono}\ \emph
  {et~al.}(2016{\natexlab{b}})\citenamefont {Archidiacono}, \citenamefont
  {Gariazzo}, \citenamefont {Giunti}, \citenamefont {Hannestad}, \citenamefont
  {Hansen}, \citenamefont {Laveder},\ and\ \citenamefont
  {Tram}}]{Archidiacono:2016kkh}%
  \BibitemOpen
  \bibfield  {author} {\bibinfo {author} {\bibfnamefont {M.}~\bibnamefont
  {Archidiacono}}, \bibinfo {author} {\bibfnamefont {S.}~\bibnamefont
  {Gariazzo}}, \bibinfo {author} {\bibfnamefont {C.}~\bibnamefont {Giunti}},
  \bibinfo {author} {\bibfnamefont {S.}~\bibnamefont {Hannestad}}, \bibinfo
  {author} {\bibfnamefont {R.}~\bibnamefont {Hansen}}, \bibinfo {author}
  {\bibfnamefont {M.}~\bibnamefont {Laveder}},\ and\ \bibinfo {author}
  {\bibfnamefont {T.}~\bibnamefont {Tram}},\ }\bibfield  {title} {\bibinfo
  {title} {{Pseudoscalar\textemdash{}sterile neutrino interactions: reconciling
  the cosmos with neutrino oscillations}},\ }\href
  {https://doi.org/10.1088/1475-7516/2016/08/067} {\bibfield  {journal}
  {\bibinfo  {journal} {JCAP}\ }\textbf {\bibinfo {volume} {08}},\ \bibinfo
  {pages} {067}},\ \Eprint {https://arxiv.org/abs/1606.07673} {arXiv:1606.07673
  [astro-ph.CO]} \BibitemShut {NoStop}%
\bibitem [{\citenamefont {Niedermann}\ and\ \citenamefont
  {Sloth}(2021{\natexlab{b}})}]{Letter}%
  \BibitemOpen
  \bibfield  {author} {\bibinfo {author} {\bibfnamefont {F.}~\bibnamefont
  {Niedermann}}\ and\ \bibinfo {author} {\bibfnamefont {M.~S.}\ \bibnamefont
  {Sloth}},\ }\bibfield  {title} {\bibinfo {title} {{Hot New Early Dark Energy:
  Towards a Unified Dark Sector of Neutrinos, Dark Energy and Dark Matter}},\
  }\href@noop {} {\  (\bibinfo {year} {2021}{\natexlab{b}})},\ \Eprint
  {https://arxiv.org/abs/2112.00759} {arXiv:2112.00759 [hep-ph]} \BibitemShut
  {NoStop}%
\bibitem [{\citenamefont {Linde}(1990)}]{Linde:1990gz}%
  \BibitemOpen
  \bibfield  {author} {\bibinfo {author} {\bibfnamefont {A.~D.}\ \bibnamefont
  {Linde}},\ }\bibfield  {title} {\bibinfo {title} {{Eternal extended inflation
  and graceful exit from old inflation without Jordan-Brans-Dicke}},\ }\href
  {https://doi.org/10.1016/0370-2693(90)90521-7} {\bibfield  {journal}
  {\bibinfo  {journal} {Phys. Lett. B}\ }\textbf {\bibinfo {volume} {249}},\
  \bibinfo {pages} {18} (\bibinfo {year} {1990})}\BibitemShut {NoStop}%
\bibitem [{\citenamefont {Adams}\ and\ \citenamefont
  {Freese}(1991)}]{Adams:1990ds}%
  \BibitemOpen
  \bibfield  {author} {\bibinfo {author} {\bibfnamefont {F.~C.}\ \bibnamefont
  {Adams}}\ and\ \bibinfo {author} {\bibfnamefont {K.}~\bibnamefont {Freese}},\
  }\bibfield  {title} {\bibinfo {title} {{Double field inflation}},\ }\href
  {https://doi.org/10.1103/PhysRevD.43.353} {\bibfield  {journal} {\bibinfo
  {journal} {Phys. Rev. D}\ }\textbf {\bibinfo {volume} {43}},\ \bibinfo
  {pages} {353} (\bibinfo {year} {1991})},\ \Eprint
  {https://arxiv.org/abs/hep-ph/0504135} {arXiv:hep-ph/0504135} \BibitemShut
  {NoStop}%
\bibitem [{\citenamefont {Copeland}\ \emph {et~al.}(1994)\citenamefont
  {Copeland}, \citenamefont {Liddle}, \citenamefont {Lyth}, \citenamefont
  {Stewart},\ and\ \citenamefont {Wands}}]{Copeland:1994vg}%
  \BibitemOpen
  \bibfield  {author} {\bibinfo {author} {\bibfnamefont {E.~J.}\ \bibnamefont
  {Copeland}}, \bibinfo {author} {\bibfnamefont {A.~R.}\ \bibnamefont
  {Liddle}}, \bibinfo {author} {\bibfnamefont {D.~H.}\ \bibnamefont {Lyth}},
  \bibinfo {author} {\bibfnamefont {E.~D.}\ \bibnamefont {Stewart}},\ and\
  \bibinfo {author} {\bibfnamefont {D.}~\bibnamefont {Wands}},\ }\bibfield
  {title} {\bibinfo {title} {{False vacuum inflation with Einstein gravity}},\
  }\href {https://doi.org/10.1103/PhysRevD.49.6410} {\bibfield  {journal}
  {\bibinfo  {journal} {Phys. Rev. D}\ }\textbf {\bibinfo {volume} {49}},\
  \bibinfo {pages} {6410} (\bibinfo {year} {1994})},\ \Eprint
  {https://arxiv.org/abs/astro-ph/9401011} {arXiv:astro-ph/9401011}
  \BibitemShut {NoStop}%
\bibitem [{\citenamefont {Adams}(1993)}]{Adams:1993zs}%
  \BibitemOpen
  \bibfield  {author} {\bibinfo {author} {\bibfnamefont {F.~C.}\ \bibnamefont
  {Adams}},\ }\bibfield  {title} {\bibinfo {title} {{General solutions for
  tunneling of scalar fields with quartic potentials}},\ }\href
  {https://doi.org/10.1103/PhysRevD.48.2800} {\bibfield  {journal} {\bibinfo
  {journal} {Phys. Rev.}\ }\textbf {\bibinfo {volume} {D48}},\ \bibinfo {pages}
  {2800} (\bibinfo {year} {1993})},\ \Eprint
  {https://arxiv.org/abs/hep-ph/9302321} {arXiv:hep-ph/9302321 [hep-ph]}
  \BibitemShut {NoStop}%
%%CITATION = HEP-PH/9302321;%%
\bibitem [{\citenamefont {Anderson}\ and\ \citenamefont
  {Hall}(1992)}]{Anderson:1991zb}%
  \BibitemOpen
  \bibfield  {author} {\bibinfo {author} {\bibfnamefont {G.~W.}\ \bibnamefont
  {Anderson}}\ and\ \bibinfo {author} {\bibfnamefont {L.~J.}\ \bibnamefont
  {Hall}},\ }\bibfield  {title} {\bibinfo {title} {{The Electroweak phase
  transition and baryogenesis}},\ }\href
  {https://doi.org/10.1103/PhysRevD.45.2685} {\bibfield  {journal} {\bibinfo
  {journal} {Phys. Rev. D}\ }\textbf {\bibinfo {volume} {45}},\ \bibinfo
  {pages} {2685} (\bibinfo {year} {1992})}\BibitemShut {NoStop}%
\bibitem [{\citenamefont {D'Amico}\ \emph {et~al.}(2020)\citenamefont
  {D'Amico}, \citenamefont {Gleyzes}, \citenamefont {Kokron}, \citenamefont
  {Markovic}, \citenamefont {Senatore}, \citenamefont {Zhang}, \citenamefont
  {Beutler},\ and\ \citenamefont {Gil-Marín}}]{DAmico:2019fhj}%
  \BibitemOpen
  \bibfield  {author} {\bibinfo {author} {\bibfnamefont {G.}~\bibnamefont
  {D'Amico}}, \bibinfo {author} {\bibfnamefont {J.}~\bibnamefont {Gleyzes}},
  \bibinfo {author} {\bibfnamefont {N.}~\bibnamefont {Kokron}}, \bibinfo
  {author} {\bibfnamefont {K.}~\bibnamefont {Markovic}}, \bibinfo {author}
  {\bibfnamefont {L.}~\bibnamefont {Senatore}}, \bibinfo {author}
  {\bibfnamefont {P.}~\bibnamefont {Zhang}}, \bibinfo {author} {\bibfnamefont
  {F.}~\bibnamefont {Beutler}},\ and\ \bibinfo {author} {\bibfnamefont
  {H.}~\bibnamefont {Gil-Marín}},\ }\bibfield  {title} {\bibinfo {title} {{The
  Cosmological Analysis of the SDSS/BOSS data from the Effective Field Theory
  of Large-Scale Structure}},\ }\href
  {https://doi.org/10.1088/1475-7516/2020/05/005} {\bibfield  {journal}
  {\bibinfo  {journal} {JCAP}\ }\textbf {\bibinfo {volume} {05}},\ \bibinfo
  {pages} {005}},\ \Eprint {https://arxiv.org/abs/1909.05271} {arXiv:1909.05271
  [astro-ph.CO]} \BibitemShut {NoStop}%
\bibitem [{\citenamefont {Beutler}\ \emph {et~al.}(2011)\citenamefont
  {Beutler}, \citenamefont {Blake}, \citenamefont {Colless}, \citenamefont
  {Jones}, \citenamefont {Staveley-Smith}, \citenamefont {Campbell},
  \citenamefont {Parker}, \citenamefont {Saunders},\ and\ \citenamefont
  {Watson}}]{6dF}%
  \BibitemOpen
  \bibfield  {author} {\bibinfo {author} {\bibfnamefont {F.}~\bibnamefont
  {Beutler}}, \bibinfo {author} {\bibfnamefont {C.}~\bibnamefont {Blake}},
  \bibinfo {author} {\bibfnamefont {M.}~\bibnamefont {Colless}}, \bibinfo
  {author} {\bibfnamefont {D.~H.}\ \bibnamefont {Jones}}, \bibinfo {author}
  {\bibfnamefont {L.}~\bibnamefont {Staveley-Smith}}, \bibinfo {author}
  {\bibfnamefont {L.}~\bibnamefont {Campbell}}, \bibinfo {author}
  {\bibfnamefont {Q.}~\bibnamefont {Parker}}, \bibinfo {author} {\bibfnamefont
  {W.}~\bibnamefont {Saunders}},\ and\ \bibinfo {author} {\bibfnamefont
  {F.}~\bibnamefont {Watson}},\ }\bibfield  {title} {\bibinfo {title} {{The 6dF
  Galaxy Survey: baryon acoustic oscillations and the local Hubble constant}},\
  }\href {https://doi.org/10.1111/j.1365-2966.2011.19250.x} {\bibfield
  {journal} {\bibinfo  {journal} {Monthly Notices of the Royal Astronomical
  Society}\ }\textbf {\bibinfo {volume} {416}},\ \bibinfo {pages} {3017}
  (\bibinfo {year} {2011})},\ \Eprint
  {https://arxiv.org/abs/http://oup.prod.sis.lan/mnras/article-pdf/416/4/3017/2985042/mnras0416-3017.pdf}
  {http://oup.prod.sis.lan/mnras/article-pdf/416/4/3017/2985042/mnras0416-3017.pdf}
  \BibitemShut {NoStop}%
\bibitem [{\citenamefont {Archidiacono}\ \emph {et~al.}(2019)\citenamefont
  {Archidiacono}, \citenamefont {Hooper}, \citenamefont {Murgia}, \citenamefont
  {Bohr}, \citenamefont {Lesgourgues},\ and\ \citenamefont
  {Viel}}]{Archidiacono:2019wdp}%
  \BibitemOpen
  \bibfield  {author} {\bibinfo {author} {\bibfnamefont {M.}~\bibnamefont
  {Archidiacono}}, \bibinfo {author} {\bibfnamefont {D.~C.}\ \bibnamefont
  {Hooper}}, \bibinfo {author} {\bibfnamefont {R.}~\bibnamefont {Murgia}},
  \bibinfo {author} {\bibfnamefont {S.}~\bibnamefont {Bohr}}, \bibinfo {author}
  {\bibfnamefont {J.}~\bibnamefont {Lesgourgues}},\ and\ \bibinfo {author}
  {\bibfnamefont {M.}~\bibnamefont {Viel}},\ }\bibfield  {title} {\bibinfo
  {title} {{Constraining Dark Matter-Dark Radiation interactions with CMB, BAO,
  and Lyman-$\alpha$}},\ }\href {https://doi.org/10.1088/1475-7516/2019/10/055}
  {\bibfield  {journal} {\bibinfo  {journal} {JCAP}\ }\textbf {\bibinfo
  {volume} {10}},\ \bibinfo {pages} {055}}\BibitemShut {NoStop}%
\bibitem [{\citenamefont {Linde}(1983)}]{Linde:1981zj}%
  \BibitemOpen
  \bibfield  {author} {\bibinfo {author} {\bibfnamefont {A.~D.}\ \bibnamefont
  {Linde}},\ }\bibfield  {title} {\bibinfo {title} {{Decay of the False Vacuum
  at Finite Temperature}},\ }\href
  {https://doi.org/10.1016/0550-3213(83)90072-X} {\bibfield  {journal}
  {\bibinfo  {journal} {Nucl. Phys. B}\ }\textbf {\bibinfo {volume} {216}},\
  \bibinfo {pages} {421} (\bibinfo {year} {1983})},\ \bibinfo {note} {[Erratum:
  Nucl.Phys.B 223, 544 (1983)]}\BibitemShut {NoStop}%
\bibitem [{\citenamefont {Dine}\ \emph {et~al.}(1992)\citenamefont {Dine},
  \citenamefont {Leigh}, \citenamefont {Huet}, \citenamefont {Linde},\ and\
  \citenamefont {Linde}}]{Dine:1992wr}%
  \BibitemOpen
  \bibfield  {author} {\bibinfo {author} {\bibfnamefont {M.}~\bibnamefont
  {Dine}}, \bibinfo {author} {\bibfnamefont {R.~G.}\ \bibnamefont {Leigh}},
  \bibinfo {author} {\bibfnamefont {P.~Y.}\ \bibnamefont {Huet}}, \bibinfo
  {author} {\bibfnamefont {A.~D.}\ \bibnamefont {Linde}},\ and\ \bibinfo
  {author} {\bibfnamefont {D.~A.}\ \bibnamefont {Linde}},\ }\bibfield  {title}
  {\bibinfo {title} {{Towards the theory of the electroweak phase
  transition}},\ }\href {https://doi.org/10.1103/PhysRevD.46.550} {\bibfield
  {journal} {\bibinfo  {journal} {Phys. Rev. D}\ }\textbf {\bibinfo {volume}
  {46}},\ \bibinfo {pages} {550} (\bibinfo {year} {1992})},\ \Eprint
  {https://arxiv.org/abs/hep-ph/9203203} {arXiv:hep-ph/9203203} \BibitemShut
  {NoStop}%
\bibitem [{\citenamefont {Arnold}\ and\ \citenamefont
  {Espinosa}(1993)}]{Arnold:1992rz}%
  \BibitemOpen
  \bibfield  {author} {\bibinfo {author} {\bibfnamefont {P.~B.}\ \bibnamefont
  {Arnold}}\ and\ \bibinfo {author} {\bibfnamefont {O.}~\bibnamefont
  {Espinosa}},\ }\bibfield  {title} {\bibinfo {title} {{The Effective potential
  and first order phase transitions: Beyond leading-order}},\ }\href
  {https://doi.org/10.1103/PhysRevD.47.3546} {\bibfield  {journal} {\bibinfo
  {journal} {Phys. Rev. D}\ }\textbf {\bibinfo {volume} {47}},\ \bibinfo
  {pages} {3546} (\bibinfo {year} {1993})},\ \bibinfo {note} {[Erratum:
  Phys.Rev.D 50, 6662 (1994)]},\ \Eprint {https://arxiv.org/abs/hep-ph/9212235}
  {arXiv:hep-ph/9212235} \BibitemShut {NoStop}%
\bibitem [{\citenamefont {Binetruy}\ \emph {et~al.}(2012)\citenamefont
  {Binetruy}, \citenamefont {Bohe}, \citenamefont {Caprini},\ and\
  \citenamefont {Dufaux}}]{Binetruy:2012ze}%
  \BibitemOpen
  \bibfield  {author} {\bibinfo {author} {\bibfnamefont {P.}~\bibnamefont
  {Binetruy}}, \bibinfo {author} {\bibfnamefont {A.}~\bibnamefont {Bohe}},
  \bibinfo {author} {\bibfnamefont {C.}~\bibnamefont {Caprini}},\ and\ \bibinfo
  {author} {\bibfnamefont {J.-F.}\ \bibnamefont {Dufaux}},\ }\bibfield  {title}
  {\bibinfo {title} {{Cosmological Backgrounds of Gravitational Waves and
  eLISA/NGO: Phase Transitions, Cosmic Strings and Other Sources}},\ }\href
  {https://doi.org/10.1088/1475-7516/2012/06/027} {\bibfield  {journal}
  {\bibinfo  {journal} {JCAP}\ }\textbf {\bibinfo {volume} {06}},\ \bibinfo
  {pages} {027}},\ \Eprint {https://arxiv.org/abs/1201.0983} {arXiv:1201.0983
  [gr-qc]} \BibitemShut {NoStop}%
\bibitem [{\citenamefont {Caprini}\ and\ \citenamefont
  {Figueroa}(2018)}]{Caprini:2018mtu}%
  \BibitemOpen
  \bibfield  {author} {\bibinfo {author} {\bibfnamefont {C.}~\bibnamefont
  {Caprini}}\ and\ \bibinfo {author} {\bibfnamefont {D.~G.}\ \bibnamefont
  {Figueroa}},\ }\bibfield  {title} {\bibinfo {title} {{Cosmological
  Backgrounds of Gravitational Waves}},\ }\href
  {https://doi.org/10.1088/1361-6382/aac608} {\bibfield  {journal} {\bibinfo
  {journal} {Class. Quant. Grav.}\ }\textbf {\bibinfo {volume} {35}},\ \bibinfo
  {pages} {163001} (\bibinfo {year} {2018})},\ \Eprint
  {https://arxiv.org/abs/1801.04268} {arXiv:1801.04268 [astro-ph.CO]}
  \BibitemShut {NoStop}%
\bibitem [{\citenamefont {Kajantie}\ \emph {et~al.}(1996)\citenamefont
  {Kajantie}, \citenamefont {Laine}, \citenamefont {Rummukainen},\ and\
  \citenamefont {Shaposhnikov}}]{Kajantie:1996mn}%
  \BibitemOpen
  \bibfield  {author} {\bibinfo {author} {\bibfnamefont {K.}~\bibnamefont
  {Kajantie}}, \bibinfo {author} {\bibfnamefont {M.}~\bibnamefont {Laine}},
  \bibinfo {author} {\bibfnamefont {K.}~\bibnamefont {Rummukainen}},\ and\
  \bibinfo {author} {\bibfnamefont {M.~E.}\ \bibnamefont {Shaposhnikov}},\
  }\bibfield  {title} {\bibinfo {title} {{Is there a hot electroweak phase
  transition at m(H) larger or equal to m(W)?}},\ }\href
  {https://doi.org/10.1103/PhysRevLett.77.2887} {\bibfield  {journal} {\bibinfo
   {journal} {Phys. Rev. Lett.}\ }\textbf {\bibinfo {volume} {77}},\ \bibinfo
  {pages} {2887} (\bibinfo {year} {1996})},\ \Eprint
  {https://arxiv.org/abs/hep-ph/9605288} {arXiv:hep-ph/9605288} \BibitemShut
  {NoStop}%
\bibitem [{\citenamefont {Laine}\ and\ \citenamefont
  {Rummukainen}(1999)}]{Laine:1998jb}%
  \BibitemOpen
  \bibfield  {author} {\bibinfo {author} {\bibfnamefont {M.}~\bibnamefont
  {Laine}}\ and\ \bibinfo {author} {\bibfnamefont {K.}~\bibnamefont
  {Rummukainen}},\ }\bibfield  {title} {\bibinfo {title} {{What's new with the
  electroweak phase transition?}},\ }\href
  {https://doi.org/10.1016/S0920-5632(99)85017-8} {\bibfield  {journal}
  {\bibinfo  {journal} {Nucl. Phys. B Proc. Suppl.}\ }\textbf {\bibinfo
  {volume} {73}},\ \bibinfo {pages} {180} (\bibinfo {year} {1999})},\ \Eprint
  {https://arxiv.org/abs/hep-lat/9809045} {arXiv:hep-lat/9809045} \BibitemShut
  {NoStop}%
\bibitem [{\citenamefont {Raveri}\ \emph {et~al.}(2017)\citenamefont {Raveri},
  \citenamefont {Hu}, \citenamefont {Hoffman},\ and\ \citenamefont
  {Wang}}]{Raveri:2017jto}%
  \BibitemOpen
  \bibfield  {author} {\bibinfo {author} {\bibfnamefont {M.}~\bibnamefont
  {Raveri}}, \bibinfo {author} {\bibfnamefont {W.}~\bibnamefont {Hu}}, \bibinfo
  {author} {\bibfnamefont {T.}~\bibnamefont {Hoffman}},\ and\ \bibinfo {author}
  {\bibfnamefont {L.-T.}\ \bibnamefont {Wang}},\ }\bibfield  {title} {\bibinfo
  {title} {{Partially Acoustic Dark Matter Cosmology and Cosmological
  Constraints}},\ }\href {https://doi.org/10.1103/PhysRevD.96.103501}
  {\bibfield  {journal} {\bibinfo  {journal} {Phys. Rev.}\ }\textbf {\bibinfo
  {volume} {D96}},\ \bibinfo {pages} {103501} (\bibinfo {year} {2017})},\
  \Eprint {https://arxiv.org/abs/1709.04877} {arXiv:1709.04877 [astro-ph.CO]}
  \BibitemShut {NoStop}%
%%CITATION = ARXIV:1709.04877;%%
\bibitem [{\citenamefont {Blinov}\ and\ \citenamefont
  {Marques-Tavares}(2020)}]{Blinov:2020hmc}%
  \BibitemOpen
  \bibfield  {author} {\bibinfo {author} {\bibfnamefont {N.}~\bibnamefont
  {Blinov}}\ and\ \bibinfo {author} {\bibfnamefont {G.}~\bibnamefont
  {Marques-Tavares}},\ }\bibfield  {title} {\bibinfo {title} {{Interacting
  radiation after Planck and its implications for the Hubble Tension}},\
  }\href@noop {} {\  (\bibinfo {year} {2020})},\ \Eprint
  {https://arxiv.org/abs/2003.08387} {arXiv:2003.08387 [astro-ph.CO]}
  \BibitemShut {NoStop}%
\bibitem [{\citenamefont {Blinov}\ \emph {et~al.}(2020)\citenamefont {Blinov},
  \citenamefont {Keith},\ and\ \citenamefont {Hooper}}]{Blinov:2020uvz}%
  \BibitemOpen
  \bibfield  {author} {\bibinfo {author} {\bibfnamefont {N.}~\bibnamefont
  {Blinov}}, \bibinfo {author} {\bibfnamefont {C.}~\bibnamefont {Keith}},\ and\
  \bibinfo {author} {\bibfnamefont {D.}~\bibnamefont {Hooper}},\ }\bibfield
  {title} {\bibinfo {title} {{Warm Decaying Dark Matter and the Hubble
  Tension}},\ }\href {https://doi.org/10.1088/1475-7516/2020/06/005} {\bibfield
   {journal} {\bibinfo  {journal} {JCAP}\ }\textbf {\bibinfo {volume} {06}},\
  \bibinfo {pages} {005}},\ \Eprint {https://arxiv.org/abs/2004.06114}
  {arXiv:2004.06114 [astro-ph.CO]} \BibitemShut {NoStop}%
\bibitem [{\citenamefont {Coleman}(1977)}]{Coleman:1977py}%
  \BibitemOpen
  \bibfield  {author} {\bibinfo {author} {\bibfnamefont {S.~R.}\ \bibnamefont
  {Coleman}},\ }\bibfield  {title} {\bibinfo {title} {{The Fate of the False
  Vacuum. 1. Semiclassical Theory}},\ }\href
  {https://doi.org/10.1103/PhysRevD.16.1248} {\bibfield  {journal} {\bibinfo
  {journal} {Phys. Rev. D}\ }\textbf {\bibinfo {volume} {15}},\ \bibinfo
  {pages} {2929} (\bibinfo {year} {1977})},\ \bibinfo {note} {[Erratum:
  Phys.Rev.D 16, 1248 (1977)]}\BibitemShut {NoStop}%
\bibitem [{\citenamefont {Cirelli}\ \emph {et~al.}(2006)\citenamefont
  {Cirelli}, \citenamefont {Fornengo},\ and\ \citenamefont
  {Strumia}}]{Cirelli:2005uq}%
  \BibitemOpen
  \bibfield  {author} {\bibinfo {author} {\bibfnamefont {M.}~\bibnamefont
  {Cirelli}}, \bibinfo {author} {\bibfnamefont {N.}~\bibnamefont {Fornengo}},\
  and\ \bibinfo {author} {\bibfnamefont {A.}~\bibnamefont {Strumia}},\
  }\bibfield  {title} {\bibinfo {title} {{Minimal dark matter}},\ }\href
  {https://doi.org/10.1016/j.nuclphysb.2006.07.012} {\bibfield  {journal}
  {\bibinfo  {journal} {Nucl. Phys. B}\ }\textbf {\bibinfo {volume} {753}},\
  \bibinfo {pages} {178} (\bibinfo {year} {2006})},\ \Eprint
  {https://arxiv.org/abs/hep-ph/0512090} {arXiv:hep-ph/0512090} \BibitemShut
  {NoStop}%
\bibitem [{\citenamefont {Mohapatra}\ and\ \citenamefont
  {Valle}(1986)}]{Mohapatra:1986bd}%
  \BibitemOpen
  \bibfield  {author} {\bibinfo {author} {\bibfnamefont {R.~N.}\ \bibnamefont
  {Mohapatra}}\ and\ \bibinfo {author} {\bibfnamefont {J.~W.~F.}\ \bibnamefont
  {Valle}},\ }\bibfield  {title} {\bibinfo {title} {{Neutrino Mass and Baryon
  Number Nonconservation in Superstring Models}},\ }\href
  {https://doi.org/10.1103/PhysRevD.34.1642} {\bibfield  {journal} {\bibinfo
  {journal} {Phys. Rev. D}\ }\textbf {\bibinfo {volume} {34}},\ \bibinfo
  {pages} {1642} (\bibinfo {year} {1986})}\BibitemShut {NoStop}%
\bibitem [{\citenamefont {Gonzalez-Garcia}\ and\ \citenamefont
  {Valle}(1989)}]{Gonzalez-Garcia:1988okv}%
  \BibitemOpen
  \bibfield  {author} {\bibinfo {author} {\bibfnamefont {M.~C.}\ \bibnamefont
  {Gonzalez-Garcia}}\ and\ \bibinfo {author} {\bibfnamefont {J.~W.~F.}\
  \bibnamefont {Valle}},\ }\bibfield  {title} {\bibinfo {title} {{Fast Decaying
  Neutrinos and Observable Flavor Violation in a New Class of Majoron
  Models}},\ }\href {https://doi.org/10.1016/0370-2693(89)91131-3} {\bibfield
  {journal} {\bibinfo  {journal} {Phys. Lett. B}\ }\textbf {\bibinfo {volume}
  {216}},\ \bibinfo {pages} {360} (\bibinfo {year} {1989})}\BibitemShut
  {NoStop}%
\bibitem [{\citenamefont {Deppisch}\ and\ \citenamefont
  {Valle}(2005)}]{Deppisch:2004fa}%
  \BibitemOpen
  \bibfield  {author} {\bibinfo {author} {\bibfnamefont {F.}~\bibnamefont
  {Deppisch}}\ and\ \bibinfo {author} {\bibfnamefont {J.~W.~F.}\ \bibnamefont
  {Valle}},\ }\bibfield  {title} {\bibinfo {title} {{Enhanced lepton flavor
  violation in the supersymmetric inverse seesaw model}},\ }\href
  {https://doi.org/10.1103/PhysRevD.72.036001} {\bibfield  {journal} {\bibinfo
  {journal} {Phys. Rev. D}\ }\textbf {\bibinfo {volume} {72}},\ \bibinfo
  {pages} {036001} (\bibinfo {year} {2005})},\ \Eprint
  {https://arxiv.org/abs/hep-ph/0406040} {arXiv:hep-ph/0406040} \BibitemShut
  {NoStop}%
\bibitem [{\citenamefont {Archidiacono}\ and\ \citenamefont
  {Hannestad}(2014)}]{Archidiacono:2013dua}%
  \BibitemOpen
  \bibfield  {author} {\bibinfo {author} {\bibfnamefont {M.}~\bibnamefont
  {Archidiacono}}\ and\ \bibinfo {author} {\bibfnamefont {S.}~\bibnamefont
  {Hannestad}},\ }\bibfield  {title} {\bibinfo {title} {{Updated constraints on
  non-standard neutrino interactions from Planck}},\ }\href
  {https://doi.org/10.1088/1475-7516/2014/07/046} {\bibfield  {journal}
  {\bibinfo  {journal} {JCAP}\ }\textbf {\bibinfo {volume} {07}},\ \bibinfo
  {pages} {046}},\ \Eprint {https://arxiv.org/abs/1311.3873} {arXiv:1311.3873
  [astro-ph.CO]} \BibitemShut {NoStop}%
\bibitem [{\citenamefont {Barenboim}\ \emph {et~al.}(2021)\citenamefont
  {Barenboim}, \citenamefont {Chen}, \citenamefont {Hannestad}, \citenamefont
  {Oldengott}, \citenamefont {Tram},\ and\ \citenamefont
  {Wong}}]{Barenboim:2020vrr}%
  \BibitemOpen
  \bibfield  {author} {\bibinfo {author} {\bibfnamefont {G.}~\bibnamefont
  {Barenboim}}, \bibinfo {author} {\bibfnamefont {J.~Z.}\ \bibnamefont {Chen}},
  \bibinfo {author} {\bibfnamefont {S.}~\bibnamefont {Hannestad}}, \bibinfo
  {author} {\bibfnamefont {I.~M.}\ \bibnamefont {Oldengott}}, \bibinfo {author}
  {\bibfnamefont {T.}~\bibnamefont {Tram}},\ and\ \bibinfo {author}
  {\bibfnamefont {Y.~Y.~Y.}\ \bibnamefont {Wong}},\ }\bibfield  {title}
  {\bibinfo {title} {{Invisible neutrino decay in precision cosmology}},\
  }\href {https://doi.org/10.1088/1475-7516/2021/03/087} {\bibfield  {journal}
  {\bibinfo  {journal} {JCAP}\ }\textbf {\bibinfo {volume} {03}},\ \bibinfo
  {pages} {087}},\ \Eprint {https://arxiv.org/abs/2011.01502} {arXiv:2011.01502
  [astro-ph.CO]} \BibitemShut {NoStop}%
\bibitem [{\citenamefont {Mention}\ \emph {et~al.}(2011)\citenamefont
  {Mention}, \citenamefont {Fechner}, \citenamefont {Lasserre}, \citenamefont
  {Mueller}, \citenamefont {Lhuillier}, \citenamefont {Cribier},\ and\
  \citenamefont {Letourneau}}]{Mention:2011rk}%
  \BibitemOpen
  \bibfield  {author} {\bibinfo {author} {\bibfnamefont {G.}~\bibnamefont
  {Mention}}, \bibinfo {author} {\bibfnamefont {M.}~\bibnamefont {Fechner}},
  \bibinfo {author} {\bibfnamefont {T.}~\bibnamefont {Lasserre}}, \bibinfo
  {author} {\bibfnamefont {T.~A.}\ \bibnamefont {Mueller}}, \bibinfo {author}
  {\bibfnamefont {D.}~\bibnamefont {Lhuillier}}, \bibinfo {author}
  {\bibfnamefont {M.}~\bibnamefont {Cribier}},\ and\ \bibinfo {author}
  {\bibfnamefont {A.}~\bibnamefont {Letourneau}},\ }\bibfield  {title}
  {\bibinfo {title} {{The Reactor Antineutrino Anomaly}},\ }\href
  {https://doi.org/10.1103/PhysRevD.83.073006} {\bibfield  {journal} {\bibinfo
  {journal} {Phys. Rev. D}\ }\textbf {\bibinfo {volume} {83}},\ \bibinfo
  {pages} {073006} (\bibinfo {year} {2011})},\ \Eprint
  {https://arxiv.org/abs/1101.2755} {arXiv:1101.2755 [hep-ex]} \BibitemShut
  {NoStop}%
\bibitem [{\citenamefont {Gariazzo}\ \emph {et~al.}(2016)\citenamefont
  {Gariazzo}, \citenamefont {Giunti}, \citenamefont {Laveder}, \citenamefont
  {Li},\ and\ \citenamefont {Zavanin}}]{Gariazzo:2015rra}%
  \BibitemOpen
  \bibfield  {author} {\bibinfo {author} {\bibfnamefont {S.}~\bibnamefont
  {Gariazzo}}, \bibinfo {author} {\bibfnamefont {C.}~\bibnamefont {Giunti}},
  \bibinfo {author} {\bibfnamefont {M.}~\bibnamefont {Laveder}}, \bibinfo
  {author} {\bibfnamefont {Y.~F.}\ \bibnamefont {Li}},\ and\ \bibinfo {author}
  {\bibfnamefont {E.~M.}\ \bibnamefont {Zavanin}},\ }\bibfield  {title}
  {\bibinfo {title} {{Light sterile neutrinos}},\ }\href
  {https://doi.org/10.1088/0954-3899/43/3/033001} {\bibfield  {journal}
  {\bibinfo  {journal} {J. Phys. G}\ }\textbf {\bibinfo {volume} {43}},\
  \bibinfo {pages} {033001} (\bibinfo {year} {2016})},\ \Eprint
  {https://arxiv.org/abs/1507.08204} {arXiv:1507.08204 [hep-ph]} \BibitemShut
  {NoStop}%
\bibitem [{\citenamefont {Gonzalez-Garcia}\ \emph {et~al.}(2016)\citenamefont
  {Gonzalez-Garcia}, \citenamefont {Maltoni},\ and\ \citenamefont
  {Schwetz}}]{Gonzalez-Garcia:2015qrr}%
  \BibitemOpen
  \bibfield  {author} {\bibinfo {author} {\bibfnamefont {M.~C.}\ \bibnamefont
  {Gonzalez-Garcia}}, \bibinfo {author} {\bibfnamefont {M.}~\bibnamefont
  {Maltoni}},\ and\ \bibinfo {author} {\bibfnamefont {T.}~\bibnamefont
  {Schwetz}},\ }\bibfield  {title} {\bibinfo {title} {{Global Analyses of
  Neutrino Oscillation Experiments}},\ }\href
  {https://doi.org/10.1016/j.nuclphysb.2016.02.033} {\bibfield  {journal}
  {\bibinfo  {journal} {Nucl. Phys. B}\ }\textbf {\bibinfo {volume} {908}},\
  \bibinfo {pages} {199} (\bibinfo {year} {2016})},\ \Eprint
  {https://arxiv.org/abs/1512.06856} {arXiv:1512.06856 [hep-ph]} \BibitemShut
  {NoStop}%
\bibitem [{\citenamefont {Acero}\ \emph {et~al.}(2008)\citenamefont {Acero},
  \citenamefont {Giunti},\ and\ \citenamefont {Laveder}}]{Acero:2007su}%
  \BibitemOpen
  \bibfield  {author} {\bibinfo {author} {\bibfnamefont {M.~A.}\ \bibnamefont
  {Acero}}, \bibinfo {author} {\bibfnamefont {C.}~\bibnamefont {Giunti}},\ and\
  \bibinfo {author} {\bibfnamefont {M.}~\bibnamefont {Laveder}},\ }\bibfield
  {title} {\bibinfo {title} {{Limits on nu(e) and anti-nu(e) disappearance from
  Gallium and reactor experiments}},\ }\href
  {https://doi.org/10.1103/PhysRevD.78.073009} {\bibfield  {journal} {\bibinfo
  {journal} {Phys. Rev. D}\ }\textbf {\bibinfo {volume} {78}},\ \bibinfo
  {pages} {073009} (\bibinfo {year} {2008})},\ \Eprint
  {https://arxiv.org/abs/0711.4222} {arXiv:0711.4222 [hep-ph]} \BibitemShut
  {NoStop}%
\bibitem [{\citenamefont {Giunti}\ and\ \citenamefont
  {Laveder}(2011)}]{Giunti:2010zu}%
  \BibitemOpen
  \bibfield  {author} {\bibinfo {author} {\bibfnamefont {C.}~\bibnamefont
  {Giunti}}\ and\ \bibinfo {author} {\bibfnamefont {M.}~\bibnamefont
  {Laveder}},\ }\bibfield  {title} {\bibinfo {title} {{Statistical Significance
  of the Gallium Anomaly}},\ }\href
  {https://doi.org/10.1103/PhysRevC.83.065504} {\bibfield  {journal} {\bibinfo
  {journal} {Phys. Rev. C}\ }\textbf {\bibinfo {volume} {83}},\ \bibinfo
  {pages} {065504} (\bibinfo {year} {2011})},\ \Eprint
  {https://arxiv.org/abs/1006.3244} {arXiv:1006.3244 [hep-ph]} \BibitemShut
  {NoStop}%
\bibitem [{\citenamefont {Abratenko}\ \emph
  {et~al.}(2021{\natexlab{a}})\citenamefont {Abratenko} \emph
  {et~al.}}]{MicroBooNE:2021rmx}%
  \BibitemOpen
  \bibfield  {author} {\bibinfo {author} {\bibfnamefont {P.}~\bibnamefont
  {Abratenko}} \emph {et~al.} (\bibinfo {collaboration} {MicroBooNE}),\
  }\bibfield  {title} {\bibinfo {title} {{Search for an Excess of Electron
  Neutrino Interactions in MicroBooNE Using Multiple Final State Topologies}},\
  }\href@noop {} {\  (\bibinfo {year} {2021}{\natexlab{a}})},\ \Eprint
  {https://arxiv.org/abs/2110.14054} {arXiv:2110.14054 [hep-ex]} \BibitemShut
  {NoStop}%
\bibitem [{\citenamefont {Abratenko}\ \emph
  {et~al.}(2021{\natexlab{b}})\citenamefont {Abratenko} \emph
  {et~al.}}]{MicroBooNE:2021sne}%
  \BibitemOpen
  \bibfield  {author} {\bibinfo {author} {\bibfnamefont {P.}~\bibnamefont
  {Abratenko}} \emph {et~al.} (\bibinfo {collaboration} {MicroBooNE}),\
  }\bibfield  {title} {\bibinfo {title} {{Search for an anomalous excess of
  charged-current $\nu_e$ interactions without pions in the final state with
  the MicroBooNE experiment}},\ }\href@noop {} {\  (\bibinfo {year}
  {2021}{\natexlab{b}})},\ \Eprint {https://arxiv.org/abs/2110.14065}
  {arXiv:2110.14065 [hep-ex]} \BibitemShut {NoStop}%
\bibitem [{\citenamefont {Arg\"uelles}\ \emph {et~al.}(2021)\citenamefont
  {Arg\"uelles}, \citenamefont {Esteban}, \citenamefont {Hostert},
  \citenamefont {Kelly}, \citenamefont {Kopp}, \citenamefont {Machado},
  \citenamefont {Martinez-Soler},\ and\ \citenamefont
  {Perez-Gonzalez}}]{Arguelles:2021meu}%
  \BibitemOpen
  \bibfield  {author} {\bibinfo {author} {\bibfnamefont {C.~A.}\ \bibnamefont
  {Arg\"uelles}}, \bibinfo {author} {\bibfnamefont {I.}~\bibnamefont
  {Esteban}}, \bibinfo {author} {\bibfnamefont {M.}~\bibnamefont {Hostert}},
  \bibinfo {author} {\bibfnamefont {K.~J.}\ \bibnamefont {Kelly}}, \bibinfo
  {author} {\bibfnamefont {J.}~\bibnamefont {Kopp}}, \bibinfo {author}
  {\bibfnamefont {P.~A.~N.}\ \bibnamefont {Machado}}, \bibinfo {author}
  {\bibfnamefont {I.}~\bibnamefont {Martinez-Soler}},\ and\ \bibinfo {author}
  {\bibfnamefont {Y.~F.}\ \bibnamefont {Perez-Gonzalez}},\ }\bibfield  {title}
  {\bibinfo {title} {{MicroBooNE and the $\nu_e$ Interpretation of the
  MiniBooNE Low-Energy Excess}},\ }\href@noop {} {\  (\bibinfo {year}
  {2021})},\ \Eprint {https://arxiv.org/abs/2111.10359} {arXiv:2111.10359
  [hep-ph]} \BibitemShut {NoStop}%
\bibitem [{\citenamefont {Dodelson}\ and\ \citenamefont
  {Widrow}(1994)}]{Dodelson:1993je}%
  \BibitemOpen
  \bibfield  {author} {\bibinfo {author} {\bibfnamefont {S.}~\bibnamefont
  {Dodelson}}\ and\ \bibinfo {author} {\bibfnamefont {L.~M.}\ \bibnamefont
  {Widrow}},\ }\bibfield  {title} {\bibinfo {title} {{Sterile-neutrinos as dark
  matter}},\ }\href {https://doi.org/10.1103/PhysRevLett.72.17} {\bibfield
  {journal} {\bibinfo  {journal} {Phys. Rev. Lett.}\ }\textbf {\bibinfo
  {volume} {72}},\ \bibinfo {pages} {17} (\bibinfo {year} {1994})},\ \Eprint
  {https://arxiv.org/abs/hep-ph/9303287} {arXiv:hep-ph/9303287} \BibitemShut
  {NoStop}%
\bibitem [{\citenamefont {Abada}\ \emph {et~al.}(2014)\citenamefont {Abada},
  \citenamefont {Arcadi},\ and\ \citenamefont {Lucente}}]{Abada:2014zra}%
  \BibitemOpen
  \bibfield  {author} {\bibinfo {author} {\bibfnamefont {A.}~\bibnamefont
  {Abada}}, \bibinfo {author} {\bibfnamefont {G.}~\bibnamefont {Arcadi}},\ and\
  \bibinfo {author} {\bibfnamefont {M.}~\bibnamefont {Lucente}},\ }\bibfield
  {title} {\bibinfo {title} {{Dark Matter in the minimal Inverse Seesaw
  mechanism}},\ }\href {https://doi.org/10.1088/1475-7516/2014/10/001}
  {\bibfield  {journal} {\bibinfo  {journal} {JCAP}\ }\textbf {\bibinfo
  {volume} {10}},\ \bibinfo {pages} {001}},\ \Eprint
  {https://arxiv.org/abs/1406.6556} {arXiv:1406.6556 [hep-ph]} \BibitemShut
  {NoStop}%
\bibitem [{\citenamefont {Gelmini}\ and\ \citenamefont
  {Roncadelli}(1981)}]{Gelmini:1980re}%
  \BibitemOpen
  \bibfield  {author} {\bibinfo {author} {\bibfnamefont {G.~B.}\ \bibnamefont
  {Gelmini}}\ and\ \bibinfo {author} {\bibfnamefont {M.}~\bibnamefont
  {Roncadelli}},\ }\bibfield  {title} {\bibinfo {title} {{Left-Handed Neutrino
  Mass Scale and Spontaneously Broken Lepton Number}},\ }\href
  {https://doi.org/10.1016/0370-2693(81)90559-1} {\bibfield  {journal}
  {\bibinfo  {journal} {Phys. Lett. B}\ }\textbf {\bibinfo {volume} {99}},\
  \bibinfo {pages} {411} (\bibinfo {year} {1981})}\BibitemShut {NoStop}%
\bibitem [{\citenamefont {Grimus}\ \emph {et~al.}(2000)\citenamefont {Grimus},
  \citenamefont {Pfeiffer},\ and\ \citenamefont {Schwetz}}]{Grimus:1999fz}%
  \BibitemOpen
  \bibfield  {author} {\bibinfo {author} {\bibfnamefont {W.}~\bibnamefont
  {Grimus}}, \bibinfo {author} {\bibfnamefont {R.}~\bibnamefont {Pfeiffer}},\
  and\ \bibinfo {author} {\bibfnamefont {T.}~\bibnamefont {Schwetz}},\
  }\bibfield  {title} {\bibinfo {title} {{A Four neutrino model with a Higgs
  triplet}},\ }\href {https://doi.org/10.1007/s100520000255} {\bibfield
  {journal} {\bibinfo  {journal} {Eur. Phys. J. C}\ }\textbf {\bibinfo {volume}
  {13}},\ \bibinfo {pages} {125} (\bibinfo {year} {2000})},\ \Eprint
  {https://arxiv.org/abs/hep-ph/9905320} {arXiv:hep-ph/9905320} \BibitemShut
  {NoStop}%
\bibitem [{\citenamefont {Lusignoli}\ \emph {et~al.}(1990)\citenamefont
  {Lusignoli}, \citenamefont {Masiero},\ and\ \citenamefont
  {Roncadelli}}]{Lusignoli:1990yk}%
  \BibitemOpen
  \bibfield  {author} {\bibinfo {author} {\bibfnamefont {M.}~\bibnamefont
  {Lusignoli}}, \bibinfo {author} {\bibfnamefont {A.}~\bibnamefont {Masiero}},\
  and\ \bibinfo {author} {\bibfnamefont {M.}~\bibnamefont {Roncadelli}},\
  }\bibfield  {title} {\bibinfo {title} {{Spontaneous versus explicit breaking
  of a continuous global symmetry}},\ }\href
  {https://doi.org/10.1016/0370-2693(90)90868-7} {\bibfield  {journal}
  {\bibinfo  {journal} {Phys. Lett. B}\ }\textbf {\bibinfo {volume} {252}},\
  \bibinfo {pages} {247} (\bibinfo {year} {1990})}\BibitemShut {NoStop}%
\bibitem [{\citenamefont {Garny}\ \emph {et~al.}(2016)\citenamefont {Garny},
  \citenamefont {Sandora},\ and\ \citenamefont {Sloth}}]{Garny:2015sjg}%
  \BibitemOpen
  \bibfield  {author} {\bibinfo {author} {\bibfnamefont {M.}~\bibnamefont
  {Garny}}, \bibinfo {author} {\bibfnamefont {M.}~\bibnamefont {Sandora}},\
  and\ \bibinfo {author} {\bibfnamefont {M.~S.}\ \bibnamefont {Sloth}},\
  }\bibfield  {title} {\bibinfo {title} {{Planckian Interacting Massive
  Particles as Dark Matter}},\ }\href
  {https://doi.org/10.1103/PhysRevLett.116.101302} {\bibfield  {journal}
  {\bibinfo  {journal} {Phys. Rev. Lett.}\ }\textbf {\bibinfo {volume} {116}},\
  \bibinfo {pages} {101302} (\bibinfo {year} {2016})},\ \Eprint
  {https://arxiv.org/abs/1511.03278} {arXiv:1511.03278 [hep-ph]} \BibitemShut
  {NoStop}%
\bibitem [{\citenamefont {Garny}\ \emph {et~al.}(2018)\citenamefont {Garny},
  \citenamefont {Palessandro}, \citenamefont {Sandora},\ and\ \citenamefont
  {Sloth}}]{Garny:2017kha}%
  \BibitemOpen
  \bibfield  {author} {\bibinfo {author} {\bibfnamefont {M.}~\bibnamefont
  {Garny}}, \bibinfo {author} {\bibfnamefont {A.}~\bibnamefont {Palessandro}},
  \bibinfo {author} {\bibfnamefont {M.}~\bibnamefont {Sandora}},\ and\ \bibinfo
  {author} {\bibfnamefont {M.~S.}\ \bibnamefont {Sloth}},\ }\bibfield  {title}
  {\bibinfo {title} {{Theory and Phenomenology of Planckian Interacting Massive
  Particles as Dark Matter}},\ }\href
  {https://doi.org/10.1088/1475-7516/2018/02/027} {\bibfield  {journal}
  {\bibinfo  {journal} {JCAP}\ }\textbf {\bibinfo {volume} {02}},\ \bibinfo
  {pages} {027}},\ \Eprint {https://arxiv.org/abs/1709.09688} {arXiv:1709.09688
  [hep-ph]} \BibitemShut {NoStop}%
\bibitem [{\citenamefont {Esteban}\ \emph {et~al.}(2020)\citenamefont
  {Esteban}, \citenamefont {Gonzalez-Garcia}, \citenamefont {Maltoni},
  \citenamefont {Schwetz},\ and\ \citenamefont {Zhou}}]{Esteban:2020cvm}%
  \BibitemOpen
  \bibfield  {author} {\bibinfo {author} {\bibfnamefont {I.}~\bibnamefont
  {Esteban}}, \bibinfo {author} {\bibfnamefont {M.~C.}\ \bibnamefont
  {Gonzalez-Garcia}}, \bibinfo {author} {\bibfnamefont {M.}~\bibnamefont
  {Maltoni}}, \bibinfo {author} {\bibfnamefont {T.}~\bibnamefont {Schwetz}},\
  and\ \bibinfo {author} {\bibfnamefont {A.}~\bibnamefont {Zhou}},\ }\bibfield
  {title} {\bibinfo {title} {{The fate of hints: updated global analysis of
  three-flavor neutrino oscillations}},\ }\href
  {https://doi.org/10.1007/JHEP09(2020)178} {\bibfield  {journal} {\bibinfo
  {journal} {JHEP}\ }\textbf {\bibinfo {volume} {09}},\ \bibinfo {pages}
  {178}},\ \Eprint {https://arxiv.org/abs/2007.14792} {arXiv:2007.14792
  [hep-ph]} \BibitemShut {NoStop}%
\bibitem [{\citenamefont {Raghuram}\ and\ \citenamefont
  {Taylor}(2018)}]{Raghuram:2018hjn}%
  \BibitemOpen
  \bibfield  {author} {\bibinfo {author} {\bibfnamefont {N.}~\bibnamefont
  {Raghuram}}\ and\ \bibinfo {author} {\bibfnamefont {W.}~\bibnamefont
  {Taylor}},\ }\bibfield  {title} {\bibinfo {title} {{Large U(1) charges in
  F-theory}},\ }\href {https://doi.org/10.1007/JHEP10(2018)182} {\bibfield
  {journal} {\bibinfo  {journal} {JHEP}\ }\textbf {\bibinfo {volume} {10}},\
  \bibinfo {pages} {182}},\ \Eprint {https://arxiv.org/abs/1809.01666}
  {arXiv:1809.01666 [hep-th]} \BibitemShut {NoStop}%
\bibitem [{\citenamefont {Kallosh}\ \emph {et~al.}(1995)\citenamefont
  {Kallosh}, \citenamefont {Linde}, \citenamefont {Linde},\ and\ \citenamefont
  {Susskind}}]{Kallosh:1995hi}%
  \BibitemOpen
  \bibfield  {author} {\bibinfo {author} {\bibfnamefont {R.}~\bibnamefont
  {Kallosh}}, \bibinfo {author} {\bibfnamefont {A.~D.}\ \bibnamefont {Linde}},
  \bibinfo {author} {\bibfnamefont {D.~A.}\ \bibnamefont {Linde}},\ and\
  \bibinfo {author} {\bibfnamefont {L.}~\bibnamefont {Susskind}},\ }\bibfield
  {title} {\bibinfo {title} {{Gravity and global symmetries}},\ }\href
  {https://doi.org/10.1103/PhysRevD.52.912} {\bibfield  {journal} {\bibinfo
  {journal} {Phys. Rev. D}\ }\textbf {\bibinfo {volume} {52}},\ \bibinfo
  {pages} {912} (\bibinfo {year} {1995})},\ \Eprint
  {https://arxiv.org/abs/hep-th/9502069} {arXiv:hep-th/9502069} \BibitemShut
  {NoStop}%
\bibitem [{\citenamefont {Delle~Rose}\ \emph {et~al.}(2020)\citenamefont
  {Delle~Rose}, \citenamefont {Khalil}, \citenamefont {King},\ and\
  \citenamefont {Moretti}}]{DelleRose:2019ukt}%
  \BibitemOpen
  \bibfield  {author} {\bibinfo {author} {\bibfnamefont {L.}~\bibnamefont
  {Delle~Rose}}, \bibinfo {author} {\bibfnamefont {S.}~\bibnamefont {Khalil}},
  \bibinfo {author} {\bibfnamefont {S.~J.~D.}\ \bibnamefont {King}},\ and\
  \bibinfo {author} {\bibfnamefont {S.}~\bibnamefont {Moretti}},\ }\bibfield
  {title} {\bibinfo {title} {{$R_K$ and $R_{K^*}$ in an Aligned 2HDM with
  Right-Handed Neutrinos}},\ }\href
  {https://doi.org/10.1103/PhysRevD.101.115009} {\bibfield  {journal} {\bibinfo
   {journal} {Phys. Rev. D}\ }\textbf {\bibinfo {volume} {101}},\ \bibinfo
  {pages} {115009} (\bibinfo {year} {2020})},\ \Eprint
  {https://arxiv.org/abs/1903.11146} {arXiv:1903.11146 [hep-ph]} \BibitemShut
  {NoStop}%
\bibitem [{\citenamefont {Delle~Rose}\ \emph {et~al.}(2021)\citenamefont
  {Delle~Rose}, \citenamefont {Khalil},\ and\ \citenamefont
  {Moretti}}]{DelleRose:2020oaa}%
  \BibitemOpen
  \bibfield  {author} {\bibinfo {author} {\bibfnamefont {L.}~\bibnamefont
  {Delle~Rose}}, \bibinfo {author} {\bibfnamefont {S.}~\bibnamefont {Khalil}},\
  and\ \bibinfo {author} {\bibfnamefont {S.}~\bibnamefont {Moretti}},\
  }\bibfield  {title} {\bibinfo {title} {{Explaining electron and muon $g$
  \ensuremath{-} 2 anomalies in an Aligned 2-Higgs Doublet Model with
  right-handed neutrinos}},\ }\href
  {https://doi.org/10.1016/j.physletb.2021.136216} {\bibfield  {journal}
  {\bibinfo  {journal} {Phys. Lett. B}\ }\textbf {\bibinfo {volume} {816}},\
  \bibinfo {pages} {136216} (\bibinfo {year} {2021})},\ \Eprint
  {https://arxiv.org/abs/2012.06911} {arXiv:2012.06911 [hep-ph]} \BibitemShut
  {NoStop}%
\bibitem [{\citenamefont {Dolan}\ and\ \citenamefont
  {Jackiw}(1974)}]{Dolan:1973qd}%
  \BibitemOpen
  \bibfield  {author} {\bibinfo {author} {\bibfnamefont {L.}~\bibnamefont
  {Dolan}}\ and\ \bibinfo {author} {\bibfnamefont {R.}~\bibnamefont {Jackiw}},\
  }\bibfield  {title} {\bibinfo {title} {{Symmetry Behavior at Finite
  Temperature}},\ }\href {https://doi.org/10.1103/PhysRevD.9.3320} {\bibfield
  {journal} {\bibinfo  {journal} {Phys. Rev. D}\ }\textbf {\bibinfo {volume}
  {9}},\ \bibinfo {pages} {3320} (\bibinfo {year} {1974})}\BibitemShut
  {NoStop}%
\bibitem [{\citenamefont {Croon}\ \emph {et~al.}(2021)\citenamefont {Croon},
  \citenamefont {Gould}, \citenamefont {Schicho}, \citenamefont {Tenkanen},\
  and\ \citenamefont {White}}]{Croon:2020cgk}%
  \BibitemOpen
  \bibfield  {author} {\bibinfo {author} {\bibfnamefont {D.}~\bibnamefont
  {Croon}}, \bibinfo {author} {\bibfnamefont {O.}~\bibnamefont {Gould}},
  \bibinfo {author} {\bibfnamefont {P.}~\bibnamefont {Schicho}}, \bibinfo
  {author} {\bibfnamefont {T.~V.~I.}\ \bibnamefont {Tenkanen}},\ and\ \bibinfo
  {author} {\bibfnamefont {G.}~\bibnamefont {White}},\ }\bibfield  {title}
  {\bibinfo {title} {{Theoretical uncertainties for cosmological first-order
  phase transitions}},\ }\href {https://doi.org/10.1007/JHEP04(2021)055}
  {\bibfield  {journal} {\bibinfo  {journal} {JHEP}\ }\textbf {\bibinfo
  {volume} {04}},\ \bibinfo {pages} {055}},\ \Eprint
  {https://arxiv.org/abs/2009.10080} {arXiv:2009.10080 [hep-ph]} \BibitemShut
  {NoStop}%
\bibitem [{\citenamefont {Gould}\ and\ \citenamefont
  {Tenkanen}(2021)}]{Gould:2021oba}%
  \BibitemOpen
  \bibfield  {author} {\bibinfo {author} {\bibfnamefont {O.}~\bibnamefont
  {Gould}}\ and\ \bibinfo {author} {\bibfnamefont {T.~V.~I.}\ \bibnamefont
  {Tenkanen}},\ }\bibfield  {title} {\bibinfo {title} {{On the perturbative
  expansion at high temperature and implications for cosmological phase
  transitions}},\ }\href@noop {} {\  (\bibinfo {year} {2021})},\ \Eprint
  {https://arxiv.org/abs/2104.04399} {arXiv:2104.04399 [hep-ph]} \BibitemShut
  {NoStop}%
\bibitem [{\citenamefont {Weinberg}\ and\ \citenamefont
  {Wu}(1987)}]{Weinberg:1987vp}%
  \BibitemOpen
  \bibfield  {author} {\bibinfo {author} {\bibfnamefont {E.~J.}\ \bibnamefont
  {Weinberg}}\ and\ \bibinfo {author} {\bibfnamefont {A.-q.}\ \bibnamefont
  {Wu}},\ }\bibfield  {title} {\bibinfo {title} {{Understanding complex
  perturbative effective potentials}},\ }\href
  {https://doi.org/10.1103/PhysRevD.36.2474} {\bibfield  {journal} {\bibinfo
  {journal} {Phys. Rev. D}\ }\textbf {\bibinfo {volume} {36}},\ \bibinfo
  {pages} {2474} (\bibinfo {year} {1987})}\BibitemShut {NoStop}%
\end{thebibliography}%

\end{document}